\documentstyle[prl,aps,epsf,epsfig,amsmath]{revtex}  

\font\cmss=cmss12   
\def\1{\hbox{{1}\kern-.25em\hbox{l}}}  
\def\bfZ{\relax{\hbox{\cmss Z\kern-.4em Z}}}  

\newcommand{\cP}{{\cal P}}

\newcommand{\pz}{\bar P \cdot z}  
\newcommand{\rz}{\Delta \cdot z}  
\newcommand{\beq}{\begin{equation}}  
\newcommand{\eeq}{\end{equation}}  
\newcommand{\bea}{\begin{eqnarray}}  
\newcommand{\eea}{\end{eqnarray}}

\begin{document}  
                 
\title{A detailed next-to-leading 
order QCD analysis of deeply virtual Compton scattering observables}  
 
\author{Andreas Freund\thanks{andreas.freund@physik.uni-regensburg.de}} 
\address{Institut f{\"u}r Theoretische Physik, 
Univ. Regensburg, Universit{\"a}tstr. 31, 93053 Regensburg 
, Germany}  
\author{Martin McDermott\thanks{martinmc@amtp.liv.ac.uk}} 
\address{Division of Theoretical Physics, Dept. Math. Sciences, Univ. of Liverpool, Liverpool, L69 3BX, UK}    
\maketitle  
\begin{abstract} 
We present a detailed next-to-leading order (NLO) leading twist QCD analysis of 
deeply virtual Compton scattering (DVCS) observables, for several 
different input scenarios, in the $\overline{\mbox{MS}}$ scheme. We discuss 
the size of the NLO effects and the behavior of the observables in skewedness 
$\zeta$, momentum transfer, $t$, and photon virtuality, $q^2=-Q^2$. We present results  
on the amplitude level for unpolarized and longitudinally polarized lepton probes, and unpolarized and 
longitudinally polarized proton targets. We make predictions for various asymmetries and for the DVCS cross section and compare with the available data. 
\vspace{1pc}  
\end{abstract}    
 
\section{Introduction} 
 
In the quest for understanding the structure of hadrons, hard, 
exclusive lepton-nucleon processes have emerged as very 
promising candidates to further constrain the dynamical degrees of freedom of hadronic matter. 
Experimentally, such processes are typified by a clear spatial separation of 
the scattered final state nucleon and the diffractively-produced, exclusive system, 
$X$, i.e. by the presence of a large rapidity gap. The hard scale required for a  
perturbative analysis is either provided by the spacelike 
virtuality, $Q^2 = -q^2 \gg \Lambda^2_{\mbox{\footnotesize{QCD}}}$, of the exchanged photon, a heavy quark mass, or by a large momentum transfer 
to the hadron in the t-channel, $t \ll 0$. Deeply virtual Compton scattering (DVCS) \cite{mul,rad,ji,diehl,van,jcaf,jios,ffs,zeus,h1,herm,clas,zeus2}, $\gamma^* (q) + p (P) \to \gamma (q') + p (P') $, is the most promising\footnote{The fact that the produced real photon 
is an {\it elementary} quantum state eliminates the need for further 
non-perturbative information, which is required for example in exclusive vector meson 
production \cite{cfs}. This simplifies the theoretical treatment considerably.} 
of these processes. 
One reason for this is that on the lepton level it interferes with a competing QED process, known as the Bethe-Heitler (BH) process, 
in which the final state photon is radiated from either the initial or final state lepton. 
The associated interference term offers the unique possibility to directly measure both the imaginary and real 
parts of QCD amplitudes, via various angular asymmetries.   
A factorization theorem has been proven for the DVCS process \cite{jcaf,jios} 
which relates the experimentally-accessible 
amplitudes to a new class of fundamental functions, called generalized parton distributions (GPDs) 
\cite{mul,rad,ji,bfm,pet,ffgs}, which encode detailed information about the partonic structure of hadrons. 
 
GPDs are an extension of the well-known parton distribution functions 
(PDFs) appearing in inclusive processes such as deep inelastic 
scattering (DIS), or Drell-Yan, and encode additional information 
about the partonic structure of hadrons, above and beyond that of 
conventional PDFs. They are defined as the Fourier transforms of  
{\it non-local} light-cone operators\footnote{compared to local 
  operators in inclusive reactions.} sandwiched between nucleon states of  
{\it different} momenta\footnote{in inclusive reactions the momenta 
  are the same.}, commensurate with a finite momentum transfer in the t-channel 
to the final state proton. As such, the GPDs depend on {\it four} variables ($X,\zeta,Q^2,t$) rather than just 
{\it two} ($X,Q^2$) as is the case for regular PDFs. 
This allows an extended mapping of the dynamical behavior of a nucleon in the two 
extra variables, {\it skewedness} $\zeta$, and momentum transfer, $t$. 
In fact, knowledge of the 
behavior of the GPDs in these two extra variables would allow one to obtain, 
for the first time, a three dimensional map of the proton in terms of its partonic constituents. 
The GPDs are true two-particle correlation functions, whereas the PDFs are effectively only one-particle 
distributions. They contain, in addition to the usual PDF-type information 
residing in the so-called ``DGLAP region'' \cite{dglap} (for which the momentum fraction variable is larger than the skewedness parameter, $X>\zeta$),  
supplementary information about the distribution amplitudes of virtual 
``meson-like'' states in the nucleon in the so-called ``ERBL region''
\cite{erbl} ($X<\zeta$). 
 
A good knowledge of GPDs is required to establish the boundary conditions 
for a large class of exclusive processes calculable in QCD. 
Unfortunately, reliable perturbative calculations can only be made if $t$ 
is either small or large, confining a sensible comparison between theory and experiment 
to restricted kinematical regions where either the $t$-dependence is a 
purely nonperturbative function (small $t$, large $Q^2$) as in DVCS or 
the $Q^2$-dependence is mainly non perturbative (small $Q^2$, large 
$t$) as in, for example, large-$t$ photoproduction of a real photon 
(wide angle DVCS) \cite{rad98}. 
In light of this observation, one might question the practicality of studying and measuring these exclusive distributions, given that inclusive PDFs 
may only be constrained well via a global analysis of a large number of data points from numerous experiments. 

It turns out that the GPDs are rather more tightly constrained than one might naively assume \cite{frmc1}. 
Firstly, they are obliged to reproduce the regular PDFs in the forward 
limit $\zeta\to 0$ \cite{mul,rad,ji}. 
Secondly, they are each required to be either symmetric or 
antisymmetric about the point $X=\zeta/2$ in the ERBL region, and they 
have to obey a polynomiality constraint (see for example 
\cite{poly}). These are properties which need to be preserved under evolution. 
Lastly, the DVCS amplitudes, and thus certain observables, seem to be very sensitive to the shape 
of the GPD in the small $\zeta$ region, especially the real part of DVCS amplitudes \cite{frmc2,frmc3}. 
Hence, experimental measurements of DVCS observables, even of only moderate statistics, 
appear to give a good opportunity to pin-down the GPDs, given these theoretical restrictions. 
Therefore a careful, thorough and accurate theoretical analysis of DVCS is necessary to 
understand how varying the input GPDs affects the physical observables and the quantitative 
and qualitative changes in going from leading order (LO) to next-to-leading order (NLO) accuracy in 
perturbation theory. In this paper, we present such a NLO analysis, 
in the $\overline{\mbox{MS}}$-scheme, for both polarized and unpolarized scattering, and explore 
some of the necessary issues required to make an optimal extraction of the GPDs from current and 
future data. 
 
The physical picture emerging for DVCS is also very interesting in its own right. Several important questions immediately arise. 
What does the energy and $Q^2$-dependence of DVCS reveal about the nature of diffractive exchange and how does 
it compare to other diffractive processes ? Why does DVCS appear to have significant 
probability in the valance region at larger $x$, i.e. outside of the region in which generic 
diffraction usually occurs ? Is the physical picture for DVCS the same in both regions ?  
We will attempt a partial answer to these questions in the following. 
 
This paper gives a comprehensive analysis of DVCS observables at NLO accuracy. Complementary information may be found in \cite{frmc1,frmc2,frmc3}. To bring our analysis right up to date, we introduce unpolarized input GPDs based on two of the most recent PDF sets, 
CTEQ5M \cite{cteq5m} and MRST99 \cite{mrst99} (in addition to GRV98 \cite{grv98}, and the older MRSA' \cite{mrsap} used in our earlier publications). This allows us to push our input scale for skewed evolution down to $Q_0=1$~GeV. 
For the purposes of comparison we also present the results for DVCS observables obtained 
using our earlier input models. Various computer codes used in our analysis our available 
from the HEPDATA website \cite{website}.
 
This paper is structured as follows. In section \ref{sec:dvcskin} we reiterate 
the kinematics of DVCS and BH and define the DVCS observables, i.e. various measurable angular asymmetries 
and cross sections. In section \ref{sec:gpds} we describe the various input GPDs and present numerical results for the NLO evolution \cite{bfm,nloevol} of the new sets. 
Section \ref{sec:dvcsamp} discusses how the various DVCS amplitudes are produced via convolution integrals of GPDs with coefficient functions. We present the $Q^2$ and $\zeta$-dependence of the unpolarized amplitudes for the new sets graphically. 
In section \ref{sec:res} we give our predictions for the DVCS observables in $\zeta$, $Q^2$ and $t$, for various polarizations of probe and target, as well as discussing their implications. We compare our results with the currently available data and 
with other theoretical predictions in section \ref{sec:dat}. 
Finally, we briefly conclude in section \ref{sec:con}. 
\begin{figure} 
\centering 
\mbox{\epsfig{file=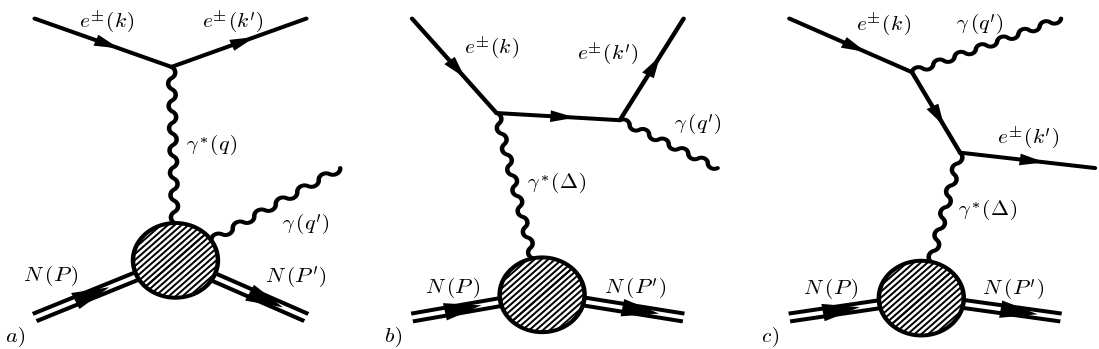,width=14.5cm,height=4.5cm}} 
\caption{a) DVCS graph, b) BH with photon from final state lepton and
  c) with photon from initial state lepton.} 
\label{dvcspic} 
\end{figure}

\section{Kinematics and observables for DVCS and BH} 
\label{sec:dvcskin} 
 
\subsection{Kinematics and frame definition} 

The lepton level process, $e^{\pm} (k,\kappa) ~N(P,S) \to e^{\pm}(k',\kappa') ~N(P',S')~\gamma(q',\epsilon')$,  
receives contributions from each of the graphs shown in Fig.\ \ref{dvcspic}.
The corresponding differential cross section is given by\footnote{In this section we follow closely the notation of \cite{bemu1}.}: 
\begin{align} 
&d\sigma^{DVCS+BH} = \frac{1}{4k \cdot P}|{\cal T}^{\pm}|^2 
(2\pi)^4\delta^{(4)} (k+P -k'-P'-q')\frac{d^3{\bf 
k'}}{2k'_{0}(2\pi)^3} \frac{d^3{\bf P'}}{2 P'_{0} (2\pi)^3}\frac{d^3{\bf 
q'}}{2 q'_{0} (2\pi)^3} \, , 
\label{dvcscross} 
\end{align} 
\noindent where the square of the amplitude receives contributions from pure DVCS (Fig.\ 1a), from pure BH (Figs. 1b, 1c) and from their interference 
(with a sign governed by lepton charge), 
\bea 
&|{\cal T}^{\pm}|^2 =  
\sum_{\kappa',S',\epsilon'}\Big[|{\cal T}^{\pm}_{DVCS}|^2 + ({\cal T}^{\pm *}_{DVCS}{\cal T}_{BH} + {\cal T}^{\pm}_{DVCS} {\cal T}^*_{BH}) + |{\cal T}_{BH}|^2\Big] \, .   
\label{tdef} 
\eea 
The DVCS amplitude is given by 
\begin{eqnarray} 
\label{VCS-amplitude} 
{\cal T}_{DVCS}^{\pm} 
= \pm \frac{e^3}{q^2} \epsilon^{' \ast}_\mu T^{\mu\nu} 
\bar u (k^\prime) \gamma_\nu u(k) \quad 
\left\{ {+ \mbox{\ for\ } e^+ \atop - \mbox{\ for\ } e^- } \right. \, , 
\end{eqnarray} 
where $q = k - k'$, $\epsilon^{' *}_\mu$ is the polarisation vector of the outgoing 
real photon, and the hadronic tensor, $T_{\mu \nu}$, is defined by a time ordered product of two electromagnetic currents: 
\begin{eqnarray} 
T_{\mu\nu} ({\bar q}, P, P') = 
i \int dx e^{i x \cdot {\bar q}} 
\langle P', S' | T j_\mu (x/2) j_\nu (-x/2) | P, S \rangle \, , 
\end{eqnarray} 
where ${\bar q} = (q + q')/2$. This hadronic tensor contains twelve\footnote{$12 
= \frac{1}{2} \times 3 \mbox{\ (virtual photon)} \times 2 \mbox{\ (final 
photon)} \times 2 \mbox{\ (initial nucleon)} \times 2 \mbox{\ (final 
nucleon)}$. The reduction factor $1/2$ is a result of parity invariance.} 
independent kinematical structures \cite{KroSchGui96} for a spin-1/2 target. 
Here, we restrict ourselves to the twist-2 part of $T_{\mu\nu}$ and drop all 
other terms of either kinematical or dynamical higher twist. From the 
structure of the operator product expansion (OPE) one can immediately 
conclude that, to twist-2 accuracy\footnote{We drop a twist-2 
contribution arising from a double helicity flip of the photon, 
i.e. going from helicity 
$+1$ to helicity $-1$ or vice versa, which is suppressed in $\alpha_s$ 
since this double flip can only be mediated by gluons. Thus when we 
speak of twist-2 contributions we really mean twist-2 modulo this 
double flip contribution.}, one has the following form factor decomposition: 
\begin{eqnarray} 
\label{decom-T} 
T_{\mu\nu} (q,P,\Delta) 
&=& 
- \tilde{g}_{\mu\nu} \frac{{\bar q} \cdot V_1}{2 {\bar P}\cdot {\bar q}} 
- i \tilde{\epsilon}_{\mu \nu \rho \sigma} 
\frac{{\bar q}^{\rho} A_1^\sigma}{2 {\bar P} \cdot {\bar q}} \, , 
\end{eqnarray} 
where ${\bar P} = (P + P')/2$ and the gauge invariant tensors $\tilde g_{\mu\nu} 
= \cP_{\mu\rho} g_{\rho\sigma} \cP_{\sigma\nu}$ and $\tilde \epsilon_{\mu\nu\alpha\beta} 
= \cP_{\mu\rho} \epsilon_{\rho\sigma\alpha\beta} \cP_{\sigma\nu}$ are 
constructed through the projection tensor 
$\cP_{\mu\nu} \equiv g_{\mu\nu} - q_{\mu} q'_{\nu}/q \cdot q'$. 
The vector $V^{\mu}_1$ and axial-vector $A^{\nu}_1$ are expressed, 
again to twist-2 accuracy, through the following form factor decomposition, 
\begin{eqnarray} 
\label{dec-FF-V} 
V_{1\mu} 
= \bar U (P', S') 
\left( {\cal H}_1 \gamma_\mu 
- {\cal E}_1 \frac{i\sigma_{\mu\nu} \Delta^\nu}{2M} 
\right) U (P, S) \, , \\ 
\label{dec-FF-A} 
A_{1\mu} 
= \bar U (P',S') 
\left( \widetilde{\cal H}_1 \gamma_\mu\gamma_5 
- \widetilde{\cal E}_1 \frac{\Delta_\mu \gamma_5}{2M} 
\right) U (P,S) \, , 
\end{eqnarray} 
where $U, {\bar U}$ are spinors for the incoming and outgoing hadron state, 
$\Delta = P - P'$ is the momentum transfered from the hadron\footnote{Note that there is a relative minus sign between our definition of $\Delta$ and that of 
\cite{bemu1} (our definition is the same as $r$ in \cite{poly}). } 
and $M$ is the hadron mass. 
The various Lorentz structures have associated amplitudes:   
${\cal H}_1, {\cal E}_1$ are unpolarized helicity non-flip and helicity flip 
amplitudes, respectively, and ${\tilde {\cal H}}_1,{\tilde {\cal E}}_1$ are their polarized counterparts. These amplitudes are expressed, 
via the DVCS factorization theorem \cite{jcaf,jios}, as  
convolutions of a hard scattering coefficient function and a GPD. They depend on the following 
Lorentz-invariant variables: 
\begin{eqnarray} 
\xi = \frac{Q^2}{2 {\bar P} \cdot {\bar q}} \, , 
\qquad 
{\bar {\cal Q}}^2 = -{\bar q}^2, 
\qquad 
t = \Delta^2 = (P-P')^2 \, , 
\nonumber 
\end{eqnarray} 
which are related to the experimentally accessible variables,  
$\zeta \equiv x_{bj} = -q^2/(2P \cdot q)$ and $Q^2 = - q^2$, used throughout this paper, via 
\begin{align} 
&{\bar {\cal Q}}^2 = \frac{1}{2}Q^2\left(1+\frac{t}{Q^2}\right)\approx \frac{1}{2}Q^2, \quad ~\mbox{and} \quad 
\xi = \frac{\zeta\left(1+\frac{t}{2 Q^2}\right)}{2-\zeta\left(1-\frac{t}{Q^2}\right)}\approx 
\frac{\zeta}{2-\zeta} \, . \label{qbar} 
\end{align}   
The BH amplitude is purely real and is given by the sum of the graphs in Fig.\ 1b and Fig.\ 1c:
\begin{align} 
&{\cal T}_{BH} = -\frac{e^3}{t}\epsilon^{' *}_{\mu} L^{\mu\nu} J_{\nu} \, , 
\end{align} 
with the leptonic tensor 
\begin{align} 
&L_{\mu\nu} = {\bar u}(k',\kappa')\left[\gamma_{\mu}({\not\!k} + 
{\not\!\Delta})^{-1}\gamma_{\nu} + \gamma_{\nu}({\not\!k'} - 
{\not\!\Delta})^{-1}\gamma_{\mu}\right]u(k,\kappa) \, , 
\label{emtensor} 
\end{align} 
and the hadronic current 
\begin{align} 
J_{\nu} = {\bar U}(P',S')\left(F_1(t)\gamma_{\nu} - iF_2(t)\sigma_{\nu\tau}\frac{\Delta^{\tau}}{2M}\right)U(P,S) \, , 
\end{align} 
where $F_1, F_2$ are the Dirac and Pauli form factors, respectively, 
normalized such that $F^p_1(0) = 1$, $F^p_2(0) \equiv \kappa_p = -1.79$, and $F^n_1(0) = 0$, 
$F^n_2(0)\equiv \kappa_n = -1.91$ for proton, $p$, and neutron, $n$. 
These are known from low energy exclusive scattering and have been parametrised, using  
dipole formulas for small $t$, as linear combinations of the electric and magnetic form factors: 
\begin{align} 
&F^i_1 = \frac{G^i_E(t)+\frac{t}{4M^2}G^i_M(t)}{1+\frac{t}{4M^2}},~ 
\quad ~F^i_2(t) = \frac{G^i_M(t)-G^i_E(t)}{1+\frac{t}{4M^2}} \, , 
\label{fipn}
\end{align} 
\noindent with 
\begin{align} 
&G^p_E(t) = \frac{G^p_M(t)}{1+\kappa_p} = \frac{G^n_M(t)}{\kappa_n} = 
\frac{1}{\left(1-\frac{t}{m^2_V}\right)^2},~G^n_E(t) = 0 \, , 
\label{gipn}
\end{align} 
and $m_V = 0.84~\mbox{GeV}$ \cite{Oku90}.

Following \cite{bemu2}, we chose to work in the $\gamma^* P$ frame with the proton at rest such that the direction of the four vector $q = k - k'$ (i.e. the virtual photon direction for the DVCS graph) defines the negative $z$-axis 
\footnote{This frame is related to the center-of-mass system of 
\cite{diehl} by a boost of the hadron in the $z$-direction.}. Without loss of generality we can choose the 
incoming electron to only have a non-zero component along the positive $x$-axis in the transverse ($x-y$) plane: 
$k = (k_0, k_0 \sin \theta_e, 0 , k_0 \cos\theta_e )$, $q = (q_0, 0,-|q_3|)$, $P = 
(M, 0, 0, 0)$ and $P' = (P'_0, |\mbox{\boldmath$P'$}| \cos\phi \sin\theta_H, 
|\mbox{\boldmath$P'$}| \sin\phi \sin\theta_H, |\mbox{\boldmath$P'$}| \cos\theta_H)$, 
where $\phi$ is the azimuthal angle between the lepton ($x-z$) and hadron scattering planes. A dependence on the angle $\phi$ can be either induced kinematically, or through the phase of hadronic amplitudes as will be shown below. 
The spin vectors of the nucleon target for longitudinal and transverse 
polarizations are given by 
\begin{align} 
&S^{\pm}_L = 
\frac{\Lambda}{Q\sqrt{1+\frac{Q^2}{4M^2\zeta^2}}}\left(q_0-\frac{Q^2}{2M\zeta}, 0, 0, -|q_3|)  \right) \quad ,  \quad~S^T = (0, \cos \Phi, \sin \Phi,0) \, ,
\label{targetpol} 
\end{align} 
\noindent where $\Lambda = \pm 1$. Note that $t = (P - P')^2$ has a kinematical minimum value which is given by 
\begin{align} 
&t_{\mbox{min}} = - \frac{M^2x_{bj}^2}{1-x_{bj}+\frac{x_{bj}M^2}{Q^2}} (1 + {\cal O} (\frac{M^2}{Q^2}) ) \, .
\end{align} 
The motivation for using this frame is that the frame-dependent expressions for the $s$- and $u$-channel BH propagators 
appearing in eq.(\ref{emtensor}) (cf. Figs. 1b, 1c, respectively) have a particularly simple Fourier expansion\footnote{In \cite{bemu1} a different frame is used which induces a more complicated $\phi$-dependence than one 
finds in our frame.} 
in the angle $\phi$. 
The dimensionless forms ${\cal P}_1 = 
\frac{(k+\Delta)^2}{Q^2}=\frac{t + 2 k\cdot \Delta}{Q^2}$ and 
${\cal P}_2 = \frac{(k-q')^2}{Q^2}=1-\frac{2k\cdot \Delta}{Q^2}$ 
are (up to corrections of order 
${\cal O} ((-t/Q^2),(M^2/Q^2))$):  
\begin{align} 
&{\cal P}_1 = \frac{1}{y}\Big\{1+ 
2\sqrt{\frac{(t_{\mbox{min}}-t)(1-x)(1-y)}{Q^2}}\cos \phi - 
\frac{t}{Q^2}(1-y-x(2-y)) - 2\frac{M^2x^2}{Q^2}(2-y)\Big\} \, , \nonumber\\ 
&{\cal P}_2 = -\frac{1}{y}\Big\{(1-y) + 
2\sqrt{\frac{(t_{\mbox{min}}-t)(1-x)(1-y)}{Q^2}}\cos \phi - 
\frac{t}{Q^2}(1-x(2-y)) - 2\frac{M^2x^2}{Q^2}(2-y)\Big\} \, . 
\label{bhprop} 
\end{align} 
The product ${\cal P}_1{\cal P}_2$ appears in the BH and interference 
expressions and thus induces an additional kinematical (rather 
than hadronic) $\phi$-dependence. In certain kinematical regions, 
this additional $\phi$-dependence can fake certain 
hadronic $\phi$-dependences \cite{bemu2}, it can also 
lead to unwanted contributions in certain $\phi$-asymmetries, as 
discussed below. 
One way out of this dilemma is to weight DVCS 
observables with ${\cal P}_1{\cal P}_2$, leaving only the pure 
hadronic $\phi$-dependence (exploiting the orthogonality of $\cos \phi$ and $\cos m' \phi$ for integer $m' \neq 1$). 
We will not do this here because such a weighting requires a good $\phi$ resolution not available for the present data  
and also because the pure twist-2 contributions may well explain most of the observed 
data without needing to take twist-3 or higher contributions into account. 
Nevertheless, studying weighted DVCS observables should be 
done as soon as good experimental $\phi$-resolution is available. 
   
\subsection{DVCS observables: differential cross section and asymmetries} 

\label{subsec:obs}
 
After performing the phase space integration in eq.(\ref{dvcscross}), the triple differential 
cross section on the lepton level is given by 
\begin{align} 
&\frac{d\sigma^{(3)} (e^{\pm} p \to e^{\pm} \gamma p)}{dx_{bj} dQ^2d|t|} = \int^{2\pi}_0d\phi \frac{d\sigma^{(4)}}{dx_{bj} dQ^2d|t|d\phi} \, =  
\frac{\alpha_{e.m.}^3 x_{bj} y^2}{8\pi Q^4}\left(1+\frac{4M^2 x_{bj}^2}{Q^2}\right)^{-1/2} \int^{2\pi}_0d\phi|{\cal T}^{\pm}|^2 \, .  
\label{crossx} 
\end{align}             
The twist two expressions for the DVCS squared, interference and BH squared terms, for all probe and target polarizations, 
required for eq.(\ref{crossx}) are very similar to eqs.(24-32) 
of \cite{bemu1} but with the full expressions for the BH propagators of eq.(\ref{bhprop}) included to 
reinstate the correct $y$- and $\phi$-dependence (a correction factor of 
$-(1-y)/(y^2 {\cal P}_1 {\cal P}_2)$ should be applied to eqs.(27-32) of \cite{bemu1}).  

From the pure DVCS piece (following the usual single photon exchange flux factor convention adopted in \cite{h1rho}), changing variable from $x_{bj}$ to $y$ 
and integrating over $t$, one may define the virtual-photon proton cross section via
\begin{align} 
&\frac{d\sigma^{(2)} (e^{\pm} p \to e^{\pm} \gamma p)}{dy dQ^2} =   
\frac{\alpha_{e.m.} (1 + (1-y)^2) }{2 \pi Q^2 y} \sigma(\gamma^{*} P \to \gamma P)
 \, .  
\label{crossx2} 
\end{align}             
\noindent We give predictions for $\sigma(\gamma^{*} P \to \gamma P)$ and compare with the recent experimental data from the H1 Collab. \cite{h1} in section \ref{sec:dat}.
 
We will now define various DVCS observables, in terms of a list of asymmetries in the 
azimuthal angle $\phi$: 
 
\begin{itemize} 
\item The (unpolarized) azimuthal angle asymmetry (AAA), measured in 
the scattering of an unpolarized probe on an unpolarized target, is 
defined by 
\begin{center} 
\bea 
\mbox{AAA} =\frac{\Big.\int^{\pi/2}_{-\pi/2} d\phi 
(d\sigma^{DVCS+BH}-d\sigma^{BH}) - \Big.\int^{3\pi/2}_{\pi/2} d\phi (d\sigma^{DVCS+BH}-d\sigma^{BH})}{\Big.\int^{2\pi}_{0} d\phi d\sigma^{DVCS+BH}} \, , 
\label{aaadef} 
\eea 
\end{center} 
where $d \sigma^{BH}$ is the pure BH term. 
 
\item The single spin asymmetry (SSA), measured in the scattering of a 
longitudinally polarized probe on an unpolarized target, is defined by 
\begin{align} 
&\mbox{SSA} = \frac{\int^{\pi}_{0} d\phi \Delta\sigma^{DVCS+BH} - \int^{2\pi}_{\pi} d\phi \Delta\sigma^{DVCS+BH}}{\int^{2\pi}_{0} d\phi (d\sigma^{DVCS+BH,\uparrow}+d\sigma^{DVCS+BH,\downarrow})} \label{eq:ssa} \, , 
\end{align} 
where $\Delta\sigma = d\sigma^{\uparrow} - d\sigma^{\downarrow}$ and 
$\uparrow$ and $\downarrow$ signify that the lepton is polarized along or against its direction, respectively. 
 
\item The asymmetry of an unpolarized probe on a longitudinally 
polarized target (UPLT) is given by: 
\begin{align} 
&\mbox{UPLT} = \frac{\int^{\pi}_{0} d\phi \Delta\sigma_{LT}^{DVCS+BH} 
  - \int^{2\pi}_{\pi} d\phi 
  \Delta\sigma_{LT}^{DVCS+BH}}{\int^{2\pi}_{0} d\phi 
  (d\sigma^{DVCS+BH}_{\uparrow} +d\sigma^{DVCS+BH}_{\downarrow})} \, , 
\end{align} 
where $\Delta\sigma_{LT} = d\sigma_{\uparrow} - d\sigma_{\downarrow}$ with 
$\uparrow$ and $\downarrow$ signifying that the target is polarized
along or against the $+z$-direction, respectively, corresponding to $\Lambda = \mp 1$ in eq.(\ref{targetpol}). 
 
\item The asymmetry of an unpolarized probe on a transversely 
polarized target (UPTT) is given by: 
\begin{align} 
&\mbox{UPTT} = \frac{\int^{\pi}_{0} d\phi \Delta\sigma_{TT}^{DVCS+BH} 
  - \int^{2\pi}_{\pi} d\phi 
  \Delta\sigma_{TT}^{DVCS+BH}}{\int^{2\pi}_{0} d\phi (d\sigma^{DVCS+BH}_{\to}+d\sigma^{DVCS+BH}_{\gets})} \, , 
\end{align} 
where $\Delta\sigma_{TT} = d\sigma_{\to} - d\sigma_{\gets}$ with 
$\to$ and $\gets$ signify that the target transverse polarization vector, $S^T$,   
points along the $+x$ and $-x$ directions (i.e. $\Phi = 0,\pi$), respectively. 
 
\item The charge asymmetry (CA) in the scattering of an unpolarized probe 
on an unpolarized target: 
\begin{align} 
&\mbox{CA} =\frac{\Big.\int^{\pi/2}_{-\pi/2} d\phi 
\Delta d^C\sigma^{DVCS+BH} - \Big.\int^{3\pi/2}_{\pi/2} d\phi \Delta d^C\sigma^{DVCS+BH}}{\Big.\int^{2\pi}_{0} d\phi (d^+\sigma^{DVCS+BH}+d^-\sigma^{DVCS+BH})} \, , 
\label{cadef} 
\end{align} 
where $\Delta d^C\sigma = d^{+}\sigma - d^{-}\sigma$ corresponds to the difference 
of the scattering with a positron probe and an electron probe. 
 
\item The charge asymmetry with a double spin flip of a longitudinally polarized probe on a longitudinally polarized target (CADSFL): 
\begin{align} 
&\mbox{CADSFL} = \frac{\Big.\int^{\pi/2}_{-\pi/2} d\phi 
\Delta d^C\sigma_{LT}^{DVCS+BH,LP} - \Big.\int^{3\pi/2}_{\pi/2} d\phi 
\Delta d^C\sigma_{LT}^{DVCS+BH,LP}}{\Big.\int^{2\pi}_{0} d\phi 
(d^+\sigma^{DVCS+BH}_{\uparrow} +d^-\sigma^{DVCS+BH}_{\downarrow})} \, , 
\label{cadsfldef}
\end{align} 
where $\Delta d^C\sigma_{LT}^{LP} = d^+\sigma^{\uparrow}_{\uparrow} - 
d^-\sigma^{\downarrow}_{\downarrow} - \Delta d^C\sigma$ with $d^+\sigma^{\uparrow}_{\uparrow}$ corresponding to a positron beam polarized along its own direction scattering with a target with its polarization vector having a positive $z$-component, 
$d^-\sigma^{\downarrow}_{\downarrow}$ corresponds to an electron beam polarized against its own direction scattering on 
target with its polarization vector having a negative $z$-component and with $\Delta d^C\sigma$ having the same meaning as for CA. 
\item  
The charge asymmetry with a double spin flip of a longitudinally polarized probe on a transversally polarized target (CADSFT): 
\begin{align} 
&\mbox{CADSFT} = \frac{\Big.\int^{\pi/2}_{-\pi/2} d\phi 
\Delta d^C\sigma_{TT}^{DVCS+BH,LP} - \Big.\int^{3\pi/2}_{\pi/2} d\phi \Delta d^C\sigma_{TT}^{DVCS+BH,LP}}{\Big.\int^{2\pi}_{0} d\phi (d^+\sigma^{DVCS+BH}_{\to}+d^-\sigma^{DVCS+BH}_{\gets})} \, , 
\end{align} 
where $\Delta d^C\sigma_{TT}^{LP} = d^+\sigma^{\uparrow}_{\to} - 
d^-\sigma^{\downarrow}_{\gets} - \Delta d^C\sigma$, with $d^+\sigma^{\uparrow}_{\to}$ corresponding to a 
positron beam polarized along its own direction scattering on a target with a polarization vector pointing in the $+x$ direction, $d^-\sigma^{\downarrow}_{\gets}$ corresponds to an electron beam polarized against its own direction scattering on a target with a polarization vector pointing in the $-x$ direction, and $\Delta d^C\sigma$ having the same meaning as in CA. 
   
\end{itemize} 
 
The definitions above make the asymmetries directly proportional to the 
real part of a combination of DVCS amplitudes, in the case of the 
AAA, CA, CADSFT and CADSFL, and to the imaginary part of a combination of DVCS amplitudes for SSA, UPTT and UPLT. 
If one forms the proper combinations from eqs.(24-32) of \cite{bemu1}, with the 
correction factor included, one observes that 
for small-$x$ DVCS observables the information from transversally 
polarized targets is redundant to the information from longitudinally polarized targets 
as far as the information on the real and imaginary part of DVCS amplitudes is concerned.
For large $x$, this is, strictly speaking, no longer true! However, higher twist corrections, 
especially in the normalization of the asymmetries, will make extraction of information on 
individual amplitudes virtually impossible.
For this reason we will focus only on DVCS observables which may be obtained using an
unpolarized or longitudinally polarized target. 
Note that for small $x$ and $t$, these combinations of amplitudes reduce to just 
the unpolarized (for AAA, CA and SSA) or 
polarized (for UPLT and CADSFL) helicity non-flip amplitudes \cite{bemu1}. 
Note also that the definition of the AAA is different from the usual one 
(see e.g. the first reference of \cite{ffs}) and is designed to ensure that 
the numerator contains only the interference term and is thus directly proportional 
to a DVCS amplitude. 
This slight change was necessary since, on inspection, it was realised that 
the pure BH contribution to the numerator does not vanish when the $\phi$ integrations in the numerator of 
eq.(\ref{aaadef}) are carried out (due to the correction factor, $\propto 1/{\cal P}_1 {\cal P}_2$, applied to eq.(27) of \cite{bemu1}). 
Hence the BH contribution needs to be subtracted from the differential cross section in  order to have an asymmetry which is 
directly proportional to the real part of a hadronic DVCS amplitude.
 
Note that in the following we will always assume a positron probe, except in the case of charge asymmetries 
where one needs both positron and electron. Thus for the corresponding electron observables the 
overall sign in the results we will quote below has to be reversed. 

At this point, we wish to make a few general comments on the $\phi$-behavior 
of the lepton level expressions given in eqs.(30-32) of \cite{bemu1} 
(modified by the  propagator factors).
Firstly, one can cleanly separate the real and imaginary parts of DVCS amplitudes using their different $\phi$ behavior either by taking moments with respect to a function in $\phi$ (usually either sine or cosine) thereby 
projecting out the unwanted contributions, or, equivalently, by forming asymmetries in the angle $\phi$, as we 
did above. Secondly, as was observed in \cite{diehl,bemu2}, the $\phi$-dependence of 
the expressions also allows a clean separation into different twist contributions. 
For example (see \cite{bemu2}), the real and imaginary parts of twist-2 can be separated from one another by using $\phi$-moments,  
since the real and imaginary parts of twist-3 amplitudes have a $\phi$-dependence which is very different from that 
of twist-2: e.g. $\cos \phi$ (twist-2) and $\cos 2\phi$ (twist-3) \cite{diehl,bemu2}. 
Thus DVCS allows the real and imaginary parts of hadronic twist-2 and 
twist-3 amplitudes to be isolated for the first time within the same experiment, simply by using 
different moments or asymmetries in the azimuthal angle $\phi$. 
Therefore, experimental upgrades which enhance the instrumentation in the forward region 
(see e.g. \cite{favart} for a H1 Collaboration proposal) are vital to 
maximise the physics scope of these experiments. 
We note the asymmetries as defined above do not necessarily require a good resolution in the angle $\phi$, 
it is sufficient to specify which hemisphere in $\phi$ a given event corresponds to. 
A reasonable sample of events will then be sufficient to measure the asymmetries.
If one wishes to produce asymmetries weighted for example with a sine or cosine, then a good phi 
resolution is of course required. 
This concludes our comments on DVCS/BH kinematics and DVCS observables. 
 
\section{GPDs and input models}  

\label{sec:gpds}  
  
\subsection{Symmetries and representations of GPDs}  

For the definition of our input GPDs we follow precisely the prescription 
given in \cite{frmc1}. GPDs result from matrix elements for quark and gluon 
correlators of unequal momentum nucleon states and may be defined in a number of ways. Following \cite{rad,ji}, we initially  
chose a definition which treated the initial and final state nucleon momentum ($P,P'$, respectively) symmetrically by involving parton light-cone fractions  
with respect to the momentum transfer, $\Delta = P - P'$, and the average  
momentum, $\bar P = (P + P')/2$. The inherent symmetries of the matrix elements are clearly manifest in associated symmetries of the GPDs. We then shifted to a definition \cite{golbier} based on light-cone fractions of the incoming hadron, for the purposes of evolution and a direct comparison with conventional PDFs and with experiment. We discussed the manifestation of the symmetries in this representation and explicitly illustrated their preservation under evolution.
  
Matrix elements of the non-local operators are defined on the light cone and involve 
a light-like vector $z^{\mu}$ ($z^2=0$). They can be most generally represented by a 
double spectral representation with respect to $\pz$ and $\rz$ 
\cite{mul,rad,poly} (see eq.(4) of \cite{frmc1}).    
In accordance with the associated Lorentz structures, the non-singlet, singlet and gluon matrix elements involve functions   
corresponding to proton helicity conservation (labelled with $F$) and to proton helicity flip (labelled with $K$) which are collectively known as double distributions. Henceforth, for brevity, we shall only discuss the helicity non-flip parts explicitly. However, the helicity flip case is exactly analogous. 
The $D$-terms in eq.(4) of \cite{frmc1} correspond to  
resonance-like exchange\cite{poly,vanderhagen1} and permit  
non-zero values for the singlet and gluon matrix elements in the limit 
$\pz \to 0$  and $\rz \neq 0$, which is allowed by their evenness in $\pz$.  
  
By making a particular choice of the light-cone vector, $z^{\mu}$, as  
a light-ray vector (so that in light-cone variables, $z_{\pm} = z_0  
\pm z_3 $, only its minus component is   
non-zero: $z^{\mu} = (0,z_{-},{\vec 0})$) one may reduce the double spectral  
representation of eq.(4) of \cite{frmc1}, defined on the entire light-cone, to a one dimensional spectral representation, defined along a light ray,  
depending on the skewedness parameter, $\xi$, defined by   
\begin{align}  
\xi = \rz /2 \pz = \Delta_{+}/ 2 {\bar P}_{+} \, ,  
\end{align}  
\noindent which is equivalent to our definition in Sec.\ \ref{sec:dvcskin} (cf. eq.(\ref{qbar})). The resultant GPDs  
are the off-forward parton distribution functions (OFPDFs) introduced in \cite{mul,diehl}:  
\begin{align}  
&H(v,\xi,t) =    
\int^1_{-1}dx' \int^{1-|x'|}_{-1+|x'|}dy' \delta(x'+ \xi y' - v) F(x',y',t) \, ,   
\label{1dim}   
\end{align}   
\noindent where $v \in [-1,1]$. In terms of individual flavor  
decompositions, the singlet, non-singlet and gluon distributions are  
given through
   
\begin{align}  
&H^S(v,\xi) = \sum_a H^{q,a} (v,\xi) \mp H^{q,a} (-v,\xi) \, ,\nonumber\\   
&H^{NS,a}(v,\xi) = H^{q,a} (v,\xi) \pm H^{q,a} (-v,\xi)   \, , \nonumber\\   
&H^G(v,\xi) = H^g(v,\xi) \pm H^g(-v,\xi) \, ,  
\end{align}  
where the upper (lower) signs corresponds to the unpolarized  
(polarized) case. Note that the symmetries which hold for the matrix  
elements can change for the $H^i$s,  due to the influence of the $\pz$, $\hat z$ factors in eq.(4) of \cite{frmc1}. 
In particular, the unpolarized quark singlet is   
antisymmetric about $v=0$, as are both $D$-terms, whereas the unpolarized quark non-singlet and the gluon are symmetric.  
The opposite symmetries hold for the  
polarized distributions. The helicity flip GPDs, are found analogously  
(double integrals with respect to the $K$s) and  
similar reasoning establishes their symmetry properties.   
   
As in \cite{frmc1}, we shall make the usual assumption that the 
$t$-dependence  of all of these functions factorizes into implicit 
form factors. One should bear in mind that in order to make predictions 
for physical amplitudes (for $t \neq 0$) these form factors must be specified.   
Note that the assumption of a factorized $t$-dependence, as a general  
statement, must be justified within the kinematic regime concerned.  
It appears to be valid at small $x$ and small $t$, from the HERA data  
on a variety of diffractive measurements. However, it appears not to hold   
for moderate to large $t$ and larger $x$ \cite{ppg}.  
  
For the purposes of comparing to experiment it is natural to define  
GPDs in terms of momentum fractions, $X \in [0,1]$, of the incoming  
proton momentum, $P$, carried by the outgoing   
parton. To this end we adapt the notation and definitions of  
\cite{golbier} introducing two non-diagonal parton distribution  
functions (NDPDFs), ${\cal F}^q$ and ${\cal F}^{\bar q}$, for flavor, $a$:   
\begin{align}  
&{\cal F}^{q,a} \left (X_1=\frac{v_1 + \xi}{1+\xi},\zeta\right ) =  
\frac{H^{q,a} (v_1,\xi)}{(1-\zeta/2)}, \qquad  
{\cal F}^{{\bar q},a} \left (X_2=\frac{\xi - v_2}{1+\xi},\zeta\right ) = -\frac{H^{q,a} (v_2,\xi)}{(1-\zeta/2)} \, , \label{fqbar}   
\end{align}   
where $v_1 \in [-\xi,1], \, v_2 \in [-1,\xi]$ (see Fig.\ 4 of  
\cite{golbier}), $\zeta \equiv \Delta^+/P^+$ is the skewedness  
defined on the domain $\zeta \in [0,1]$ such that $\xi \approx \zeta/(2-\zeta)$ and $\zeta = x_{bj}$ for DVCS (this definition is equivalent and the  
relations are the same as those in Sec.\ \ref{sec:dvcskin}).    
The transformations between the $v_1, v_2$ and $X_1,X_2$ are given implicitly in eq.(\ref{fqbar}), the inverse transformations are:   
\bea   
v_1 = \frac{X_1 - \zeta/2}{1-\zeta/2} \, , & \quad & ~v_2 = \frac{\zeta/2 - X_2}{1-\zeta/2} \, .  
\eea   
For the gluon one may use either transformation, e.g.   
\bea   
{\cal F}^{g} (X,\zeta) &=& \frac{H^g (v_1, \xi)}{(1-\zeta/2)} \, .  
\eea   
     
There are two distinct kinematic regions for the GPDs, with different  
physical interpretations.   
In the DGLAP \cite{dglap} region, $X > \zeta$ ($|v| > \xi$),  
${\cal F}^q (X,\zeta)$ and  ${\cal F}^{\bar q}(X,\zeta)$ are  
independent functions, corresponding to quark or anti-quark   
fields leaving the nucleon with momentum fraction $X$ and returning  
with positive momentum  fraction $X-\zeta$. As such they correspond to  
a generalization of regular DGLAP PDFs  (which have equal outgoing and  
returning fractions). In the ERBL \cite{erbl} region,   
$X<\zeta$ ($|v| < \xi$), both quark and anti-quark carry positive momentum   
fractions ($X,\zeta-X$) {\it away} from the nucleon in a meson-like  
configuration, and the GPDs behave like ERBL \cite{erbl}  
distributional amplitudes characterising mesons.   
This implies that ${\cal F}^q $ and ${\cal F}^{\bar q}$ are not  
independent in the ERBL region and indeed a symmetry is observed:  
${\cal F}^q (\zeta-X,\zeta) = {\cal F}^{\bar q} (X,\zeta)$  
(which directly reflects the symmetry of $H^q(v,\xi)$ about  
$v=0$). Similarly, the gluon distribution, ${\cal F}^g$, is DGLAP-like for   
$X>\zeta$ and ERBL-like for $X<\zeta$. This leads to unpolarized non-singlet,   
${\cal F}^{NS,a} = {\cal F}^{q,a} - {\cal F}^{{\bar q},a}$,  and gluon  
GPDs which are symmetric, and a singlet quark distribution  
${\cal F}^S = \sum_a {\cal F}^{q,a} + {\cal F}^{{\bar q},a}$   
which is antisymmetric, about the point $X = \zeta/2$ in the ERBL region.   
Again the opposite symmetries hold for the polarized distributions.  
  
\subsection{GPD input models}  

\label{subsec:gpdsinp}

For our input models we we follow precisely the procedure given in section III of \cite{frmc1} which is based on Radyushkin's ansatz \cite{rad2} for GPDs. Here we 
briefly describe some salient features which are required for the discussion and give various technical details not included in \cite{frmc1}.  
The input distributions, ${\cal F}^{q, {\bar q}, g} (X,\zeta, Q_0) $, at the input scale,  
$Q_0$, have the correct symmetries   
and properties and are built from conventional PDFs in the DGLAP region, for both  
the unpolarized and polarized cases. These input NDPDFs then serve as  
the boundary conditions for our numerical evolution.   
   
Factoring out the overall $t$-dependence we have the following integral  
relations between the double distributions and NDPDFs for the quark and  
antiquark:  
\begin{align}   
&{\cal F}^{q,a} (X,\zeta) = \frac{H^{q,a} (v_1,\xi)}{1-\zeta/2} =   
\int^{1}_{-1} dx' \int^{1-|x'|}_{-1+|x'|} dy' \delta \left( x' + \xi y' - v_1 \right)   
\frac{F^{q,a} (x',y')}{\left(1-\zeta/2\right)} \, , \nonumber \\   
&{\cal F}^{{\bar q},a} (X,\zeta) = -\frac{H^{q,a} (v_2,\xi)}{1-\zeta/2} =   
\int^{1}_{-1} dx' \int^{1-|x'|}_{-1+|x'|} dy' \delta \left( x' + \xi y' - v_2 \right)   
\frac{F^{q,a} (x',y')}{\left(1-\zeta/2\right)} \, .   
\label{fqinp}   
\end{align}   
\noindent with $1> v_1 > -\xi$, $-1< v_2 < \xi$ and a similar relation for the gluon (for which one can use either $v_1$ or $v_2$).     
   
Following \cite{bemu1,rad2} we employ a factorized ansatz for the   
double distribution where they are given by a product of a profile function,  
$\pi^{i}$, and a conventional PDF, $f^{i}$, $(i=q,g$).   
The profile functions are chosen to guarantee the correct symmetry properties   
in the ERBL region and their normalization is specified by demanding that the   
conventional distributions are reproduced in the forward limit: e.g.     
${\cal F}^{g} (X,\zeta \to 0) \to f^g (X)$. The exact  
$t$-dependence will be specified in Sec.\ \ref{sec:dvcsamp}, since it  
depends on whether one is dealing with helicity-flip or helicity  
non-flip amplitudes, unpolarized or polarized in origin.  
 
In \cite{frmc1} we specified two particular forward input distributions for the GPDs by using two consistent sets of  
inclusive unpolarized and polarized PDFs, i.e. GRV98 \cite{grv98} and GRSV00\footnote{the ``standard'' scenario with an unbroken flavor sea.} \cite{grsv00} with $\Lambda^{(4,NLO)}_{{\mbox{\tiny QCD}}} = 246~\mbox{MeV}$, and MRSA' \cite{mrsap} and GS(gluon 'A')  
\cite{gehrst} with $\Lambda^{(4,NLO)}_{{\mbox{\tiny QCD}}} = 231~\mbox{MeV}$ at the common input scale  $Q^2_0 = 4~\mbox{GeV}^2$ and $\Lambda^{(4,LO)}_{{\mbox{\tiny QCD}}} = 174~\mbox{MeV}$ for both sets. These pairs of unpolarized and polarized sets were consistent in the sense that the unpolarized PDFs were  
used to constrain the respective polarized PDFs and both use the same  
choices for $\Lambda_{\mbox{{\tiny QCD}}}$, etc in the evolution.  
 
In order to bring our analysis more up-to-date\footnote{In particular the MRSA'/GS(A)  set from 1995 is based on rather old data.} and to further  
investigate the input model dependence, we relax our  
(rather weak) consistency requirement in this paper and use two additional contemporary $\overline{MS}$-scheme unpolarized sets. We use CTEQ5M \cite{cteq5m} and MRST99 \cite{mrst99} in conjunction with the GRSV00\footnote{the ``valence'' or broken flavor sea scenario.} polarized set at the common input scale of $Q^2_0=1~\mbox{GeV}^2$ (with $\Lambda^{(4,NLO)}_{{\mbox{\tiny QCD}}} = 326~\mbox{MeV}$ for CTEQ5M\footnote{we use the FORTRAN code supplied by Pumplin \cite{pumplin} at the input scale of $Q_0 = 1~\mbox{GeV}$, to feed into our double distribution code.
Despite the fact that this code is not recommended for use at such a low scale we found that the results matched very smoothly onto the results at higher scales.
}, and $\Lambda^{(4,NLO)}_{{\mbox{\tiny QCD}}} = 300~\mbox{MeV}$ for MRST99). For the LO evolution of these sets, we have for both models $\Lambda^{(4,LO)}_{{\mbox{\tiny QCD}}} =  192~\mbox{MeV}$.  
  
Having defined this particular input model for the double distribution  
one may then  
perform the $y'$-integration in eq.(\ref{fqinp}) using the delta  
function. This modifies the limits on the $x'$ integration according  
to the region concerned: for the DGLAP region $X> \zeta$ one has:   
\begin{align}   
&{\cal F}^{q,a} (X,\zeta) = \frac{2}{\zeta}  
\int^{\frac{v_1+\xi}{1+\xi}}_{\frac{v_1-\xi}{1-\xi}} dx' \pi^q \left (x', \frac{v_1 - x'}{\xi} \right) q^a (x') \, \label{dglapq}.  
\end{align}   
For the anti-quark, since $v_2 = -v_1$ one may use  
eqs.(\ref{fqbar},\ref{fqinp}) with $v_2 \to -v_1$, and,   
exploiting the fact that $f^{q} (x) = - {\bar q} (|x|) $ for $ x < 0  
$, one arrives at   
\begin{align}   
&{\cal F}^{\bar q,a} (X,\zeta) =  
\frac{2}{\zeta}  
\int^{\frac{-v_1+\xi}{1-\xi}}_{\frac{-v_1-\xi}{1+\xi}} dx' \pi^q \left   
(x',\frac{-v_1 - x'}{\xi} \right ){\bar q}^a(|x'|) \label{dglapbq}.   
\end{align}  
  
\noindent In the ERBL region ($X < \zeta, |v| < \xi$) integration over $y'$ leads to:   
\begin{align}  
&{\cal F}^{q,a} (X,\zeta) = \frac{2}{\zeta}   
\left[ \int^{\frac{v_1+\xi}{1+\xi}}_{0} dx' \pi^q \left(x', \frac{v_1  
- x'}{\xi} \right) q^a (x') - \int^{0}_{\frac{-(\xi-v_1)}{1+\xi}} dx'  
\pi^q \left(x', \frac{v_1 - x'}{\xi} \right) {\bar q}^a (|x'|) \right]  
\, , \label{erblq}  
\end{align}  
\begin{align}   
&{\cal F}^{{\bar q},a} (X,\zeta) = -\frac{2}{\zeta}   
\left[ \int^{\frac{\xi-v_1}{1+\xi}}_{0} dx' \pi^q \left(x', \frac{-v_1  
- x'}{\xi} \right) q^a (x') -  \int^{0}_{\frac{-(\xi+v_1)}{1+\xi}} dx'  
\pi^q \left(x', \frac{-v_1 - x'}{\xi} \right) {\bar q}^a (|x'|)  
\right] \, \label{erblbq}.   
\end{align}  
  
\noindent The non-singlet (valence) and singlet quark combinations are given by:   
\bea   
{\cal F}^{NS,a} &\equiv  {\cal F}^{q,a} + {\cal F}^{{\bar q},a}  &\equiv  \frac{[H^{q,a} (v_1,\xi) - H^{q,a} (-v_1,\xi)]}{1-\zeta/2} \,  , \label{fnsing} \\   
{\cal F}^{S}    &\equiv  \sum_a {\cal F}^{q,a} - {\cal F}^{{\bar q},a}  &\equiv \sum_a \frac{[H^{q,a} (v_1,\xi) + H^{q,a} (-v_1,\xi)]}{1-\zeta/2} \, . \label{fsing}   
\eea   
  
We implemented eqs.(\ref{fnsing},\ref{fsing}) using eqs.(\ref{dglapq},\ref{dglapbq}) and 
eqs.(\ref{erblq},\ref{erblbq}) for the DGLAP and ERBL regions respectively,  
employing an adaptive Gaussian numerical integration routine on a non-equidistant  
grid \footnote{This uses up most of the computing time for an evolution  
  run.}. We comment further on our usage of grids and integration routines below.  
  
The integration ranges in eqs.(\ref{dglapq},\ref{dglapbq}) and  eqs.(\ref{erblq},\ref{erblbq})
sample the input PDFs all the way down to zero in $x'$. Generally
speaking, the providers of PDF sets issue programs that only allow
their PDFs to be called for $x'$ greater than some minimum value. This
is partly for technical reasons but also partly because the PDFs have
not yet been well constrained by inclusive data in the very small $x$
region, which corresponds to very high centre-of-mass energies. 
For the implementation of MRST, MRSA' and GS we are fortunate to have access to analytic forms themselves at the input scale\footnote{for CTEQ5M we use Pumplin's code \cite{pumplin}} and we simply extrapolate these into the very small $x$ region. For GRV98 and GRSV00 this is not the case and it was necessary to perform fits, at the $Q_0$-scale concerned, to the small $x$ behavior of these sets for values of $x$ where they are available  
and then extrapolate these fits into the very small $x$ regime\footnote{Having tried several forms we eventually settled on fits of the type $f(x) = f(x_1) ~(x/x_1)^a ~(1 + b~\log (x/x_1) + c~(x-x_1))$, with $x_1$ being the minimum value of $x$ available and with the power $a$ constrained to be greater than zero to allow convergence as $x \to 0$.}.  
In investigating this issue we noticed that if PDFs (and the quarks in
particular) were too singular the integrals for the individual $q$ and
${\bar q}$ were divergent (although this divergence is cancelled in forming the singlet, some regulation of the limit $x' \to 0$ was required). 
This issue is not merely of technical interest.  
The physics message is clear: the GPDs defined above, 
using Radyushkin's ansatz \cite{rad2}, 
and hence the DVCS observables are very sensitive to the behavior of the PDFs in the small $x$ region and also to their extrapolation to extremely small $x$. To the best of our knowledge this has not been explicitly pointed out before.
 
The unpolarized singlet also includes the D-term on the right hand side of 
eq.(\ref{fsing})  (in principle there is also an analogous term  
in the unpolarized gluon ($D^G$ in eq.(4) of \cite{frmc1}), but we choose to  
set it to zero, since nothing is known for the gluon D-term  
except its symmetry.). We adapt the model introduced in  
\cite{vanderhagen1} for the unpolarized singlet D-term, which is based on the  
chiral-quark-soliton model.  This D-term is antisymmetric in its argument, i.\ e.\ about the point $X = \zeta/2$ (in keeping with the anti-symmetry of ${\cal F}^S$, and  
$H^S$ about $v=0$). It is non-zero only in the ERBL region and hence  
vanishes entirely in the forward limit. In practice it only  
assumes numerical significance for large $\zeta$ (see Fig.\ 6 of \cite{frmc1}). 
  
\subsection{GPD evolution} 
\label{sec:gpdevol} 
 
The input GPDs must now be evolved in $Q^2$, using renormalization group equations, in order to make predictions for DVCS amplitudes at evolved scales. 
The input GPDs, as previously defined, are continuous functions which span the 
DGLAP and ERBL regions, and evolve in scale appropriately according to 
generalised versions of the DGLAP or ERBL evolution equations.  
Note that the evolution in the ERBL region depends on the DGLAP region, 
i.e. there is a convolution integral in the ERBL equations spanning the 
DGLAP region $[\zeta,1]$, whereas the 
DGLAP evolution is independent of the ERBL region. As the scale increases, 
partons are pushed from the DGLAP into the ERBL region simply 
through momentum degradation, but not vice versa. 
The ERBL region thus acts as a sink for the partons. 
Hence, in the asymptotic limit of infinite $Q^2$, we recover the simple 
asymptotic pion-type distribution amplitude in the ERBL region and a 
completely empty DGLAP region. This will have strong 
implications for the DVCS amplitudes and cross section. 
Note that in performing the evolution 
we assumed that the $t$-dependences of the quarks and gluons (which mix 
under evolution) are the same and factorize such that they do not influence the degree of mixing under evolution. Otherwise, any assumed $t$-dependence will be modified by the QCD evolution, complicating calculations. 
In fact the $t$-dependence of quarks and 
gluons should be different, however, since we study DVCS only at small 
$t$, the differences in their $t$-dependence should be 
small. This unresolved problem of $t$-dependence mixing will be 
addressed in another paper. 
      
The renormalization group equations (RGEs), or evolution 
equations, for the DGLAP region and ERBL regions, involve convolutions
of GPDs with generalised kernels. They are implemented 
in a FORTRAN numerical evolution code. 
In the DGLAP region the quark flavor singlet and the gluon 
distributions mix under evolution according to   
generalized DGLAP kernels\footnote{Note that in the second reference of \cite{bfm} there was a typographical error in Eq.\ 
(188) where the overall sign of the polarized pure singlet term in the 
QQ sector should be $-$ so as to be consistent with Eq.\ 
(178). In another typographical error, the factor $3$ in the first term of the square 
bracket in the second line of Eq.\ (194) of the same reference, 
i.e. the equation for the unpolarized GQ kernel, should be replaced by 
$3\zeta$. These mistakes were properly corrected in the 
implementation of the kernels in the GPD evolution code. 
} taken from \cite{bfm}. 
The flavor non-singlet (NS) quark combinations 
do not mix under evolution. 
Note that in order to do the full evolution and afterwards extract 
the various quark species separately one needs to solve two separate 
evolution equations. One for a symmetric combination $q_{+}^{a} = q^a + {\bar 
  q}^a - 1/N_f\sum_a(q^a+{\bar q}^a)$ and 
one for an antisymmetric combination $q_{-}^{a} = q^{a} - {\bar q}^{a}$. 
A single quark species, i.e. quark or anti-quark in the DGLAP region, 
or just a singlet or non-singlet quark combination in the ERBL region, 
can be extracted the following way: 
\begin{align} 
&\left(q^a \atop {\bar q}^a\right) = \frac{1}{2}(q_{+}^{a} \pm q_{-}^{a} + q^S) \, , \quad \mbox{with} \qquad 
q^S=\frac{1}{N_F}\sum_a(q^{a} + {\bar q}^{a}) \, ,
\end{align} 
in the DGLAP region and 
\begin{align} 
&q^{S,a} = (q_{+}^{a} + q^S) \, , \quad \mbox{and} \quad q^{NS,a} = q_{-}^{a} \, ,
\end{align} 
in the ERBL region. This procedure was adopted in our FORTRAN code. 
  
A numerical implementation of the convolution integrals of the RNG equations 
involves specifying a treatment of the integrable endpoint singularities. This 
is achieved via the following definition of the $+$-distributions (in this case we 
chose to apply it to the ``whole kernel'') in the DGLAP region:  
\begin{align}  
&\int^1_y \frac{dz}{z} P\left(\frac{y}{z},\frac{\zeta}{z}\right)_+{\cal F}(z,\zeta) =  
\int^1_y \frac{dz}{z} P\left(\frac{y}{z},\frac{\zeta}{z}\right)\left( {\cal F} (z,\zeta) - {\cal F}(\zeta,\zeta)\right) - 
{\cal F}(\zeta,\zeta)\left[\int^1_{\frac{\zeta}{y}} dz P\left(z,\frac{\zeta}{y}\right)  -\int^1_y dz P\left(z,z\frac{\zeta}{y}\right)\right] \, ,  
\end{align} 
and accordingly implemented in our code. Note that the lower limit of 
the first integral in the last bracket is $\zeta/y$, which is not necessarily a grid point. Since we 
initially used an equidistant grid in the integration 
variable\footnote{Our method of integration is the following: we first 
  introduce an equidistant grid which we then stretch both in the ERBL 
  and DGLAP regions with particular transformation functions in order 
  to be able to treat the important region around $\zeta$ and $0$ more 
  accurately. We then compute the Jacobian of this transformation for 
  the inverse transformation we need. On the non-equidistant grid we 
  compute the input distributions and then the kernels. Using the 
  Jacobian we transform back onto the equidistant grid and perform the 
  convolution integrals using an equidistant grid integration routine 
  like a semi-open Simpson to account for the remaining integrable 
  singularities at $y$ and $\zeta$.},  we needed to resort to a slower integration routine for 
a non-equidistant grid such as an adaptive Gaussian integration routine. 
 
In the ERBL region, the quark singlet and gluon again mix under evolution. The generalized ERBL kernels may also be found in \cite{bfm}.  
The $+$-distribution, again applied to the whole kernel, takes the following form in the ERBL region:  
\begin{align} 
\int^1_y dz V\left(\frac{y}{\zeta},\frac{z}{\zeta}\right)_+ {\cal F}(z,\zeta) &= \int^1_y dz V\left(\frac{y}{\zeta},\frac{z}{\zeta}\right)\left[ {\cal F}(z,\zeta)-{\cal F}(\zeta,\zeta)\right] + {\cal F}(\zeta,\zeta)\Big[\int^{\zeta}_y dz \left(V\left(\frac{y}{\zeta},\frac{z}{\zeta}\right) -  
V\left(\overline {\frac{z}{\zeta}},\overline {\frac{y}{\zeta}}\right)\right)\nonumber\\
&+\int^1_{\zeta} dz V\left(\frac{y}{\zeta},\frac{z}{\zeta}\right)\Big]
\, , \nonumber\\  
\int^y_0 dz V\left(\overline {\frac{y}{\zeta}},\overline 
  {\frac{z}{\zeta}}\right)_+{\cal F}(z,\zeta) &= \int^y_0 dz 
V\left(\overline {\frac{y}{\zeta}},\overline {\frac{z}{\zeta}}\right) \left[{\cal F} (z,\zeta)-{\cal F}(\zeta,\zeta)\right] 
+ {\cal F}(\zeta,\zeta)\int^y_0 dz \left(V\left(\overline 
    {\frac{y}{\zeta}},\overline {\frac{z}{\zeta}}\right) - V\left(\frac{z}{\zeta},\frac{y}{\zeta}\right)\right) \, ,  
\end{align} 
where the terms have been arranged in such away that all divergences 
explicitly cancel in each term  separately and only integrable 
divergences, as in the case of forward evolution at the point 
$y=x_{bj}$, remain. 
The bar notation means, for example, $\overline{z/\zeta} = 1 - z/\zeta$. 
In the code, the integrations were dealt with in an analogous 
way to the DGLAP region. Note that for the solution of the differential equation 
in $Q^2$, we adopted the CTEQ-routines, which are based on a Runge-Kutta 
predictor-corrector algorithm. 

\begin{figure}  
\centering  
\mbox{\epsfig{file=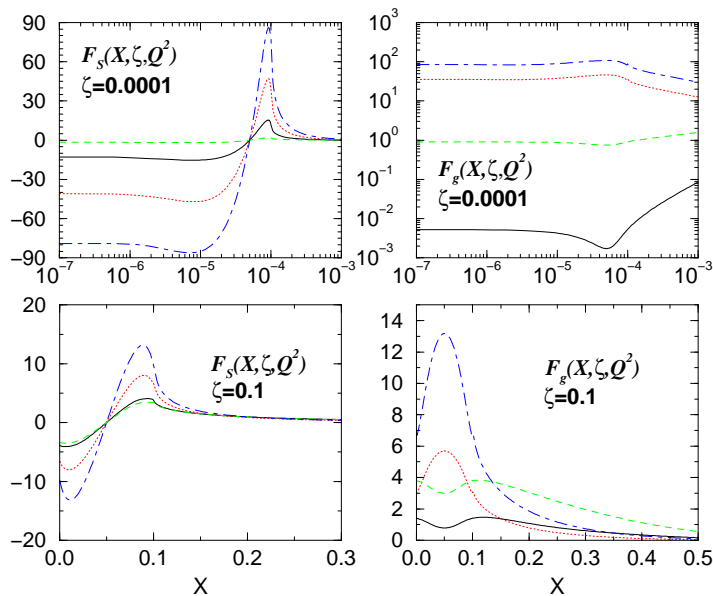,width=9.5cm,height=8.0cm}}  
\caption{Unpolarized NLO input and evolved singlet quark and gluon GPDs at small and large skewedness. The solid curves are the input MRST99 GPDs at $Q_0 = 1~\mbox{GeV}$, the dotted ones show them evolved to $Q =10~\mbox{GeV}$.  
The dashed curves are input CTEQ5M GPDs at $Q_0 = 1~\mbox{GeV}$ and the dashed-dotted ones show them evolved to $Q = 10~\mbox{GeV}$. 
The quark singlet is scaled by a factor of $10^{-4}$ at $\zeta = 0.0001$ and by $10^{-2}$ at $\zeta = 0.1$. 
For $\zeta=0.1$ the symmetry of the gluon GPD and the anti-symmetry of the singlet quark GPD are apparent about the point $X=\zeta/2 = 0.05$.} 
\label{fig:evolvealln} 
\end{figure} 
 
In Fig.(\ref{fig:evolvealln}) we plot the input distributions for the ``new'' unpolarized inputs (CTEQ5M and MRST99) at their input scale, and at an evolved scale to demonstrate the evolution effects in NLO. 

\section{DVCS amplitudes} 
\label{sec:dvcsamp} 

\subsection{Convolution formulae}
\label{sec:dvcsconv} 

In \cite{frmc3} we investigated unpolarised and polarised DVCS amplitudes in detail. 
For convenience in the following equations we introduce the notation $V$ for the unpolarized case and $A$ for the polarized case, where applicable, they 
take the upper and lower signs, respectively\footnote{refering to vector and axial-vector currents for the quarks}.
The factorization theorem \cite{jcaf,jios} for DVCS proves that the amplitude takes the following factorized form in the non-diagonal representation
(up to power-suppressed corrections of ${\cal O} (1/Q) $):
  
\begin{align}  
{\cal T}^{S,V/A}_{DVCS} (\zeta,\mu^2,Q^2,t) = \sum_a e^2_a \left(\frac{2 - \zeta}{\zeta} \right) \Big[  
&\int^1_0 dX~T^{S(a),V/A} \left(\frac{2X}{\zeta} - 1+i\epsilon, \frac{Q^2}{\mu^2} \right) ~{\cal F}^{S(a),V/A} (X,\zeta,\mu^2,t)  \nonumber\\  
\Big. \mp &\int^1_{\zeta} dX~T^{S(a),V/A} \left(1 - \frac{2X}{\zeta},\frac{Q^2}{\mu^2}\right)~{\cal F}^{S(a),V/A} (X,\zeta,\mu^2,t) \Big] ,\nonumber\\  
{\cal T}^{g,V/A}_{DVCS} (\zeta,\mu^2,Q^2,t) = \frac{1}{N_f}\left (\frac{2 - \zeta}{\zeta}\right )^2  \Big[ 
&\int^1_0 dX~T^{g,V/A} \left(\frac{2X}{\zeta} - 1+i\epsilon, \frac{Q^2}{\mu^2} \right) ~{\cal F}^{g,V/A} (X,\zeta,\mu^2,t)  \nonumber\\ 
\pm \Big. &\int^1_{\zeta} dX~T^{g,V/A} \left(1 - \frac{2X}{\zeta}, \frac{Q^2}{\mu^2}\right) ~{\cal F}^{g,V/A}(X,\zeta,\mu^2,t) \Big] \, .  
\label{tdvcs}  
\end{align} 
\noindent 
Note that the second integral is purely real and does not need a $+i\epsilon$ prescription since there is no divergence of the coefficient function in the integration interval. Also note the opposite sign structure in the quark 
singlet and gluon case due to opposite symmetries of the quark singlet and 
gluon. 
For our numerical calculations, we set the factorization scale, $\mu^2$, 
equal to the photon virtuality, $Q^2$ 
(in \cite{frmc3} we studied the effects of its variation and 
found them to be rather mild; the associated uncertainties are less than those due to 
differences in the input model GPDs so we neglect them).
Henceforth, we  suppress the factorized $t$-dependence and give predictions for $t=0$.
We will specify it for each case later.
The LO and NLO coefficient functions, $T^{i,V/A}$, are  
taken from eqs.(14-17) of \cite{bemu3} and are summarized in 
appendix A of \cite{frmc3}.
They contain logarithms of the type $\log(1-X/\zeta)^n/(1-X/\zeta)^{n_1}$, with $n,n_1 = 0,1,2,3$. Hence, 
depending on the region of integration, they can have both real and imaginary parts
(i.e. if the argument of the log is positive or negative), 
which in turn generate real and imaginary parts of the DVCS amplitudes. 

In eq.(\ref{tdvcs}), we employ the $+i\epsilon$ prescription through the Cauchy principal value prescription (``P.V.'') which we implemented through the following algorithm:
\begin{align} 
&\Big. P.V. \int^1_0 dX~T\left(\frac{2X}{\zeta} - 1\right) {\cal F}(X,\zeta,Q^2) = 
\int^{\zeta}_0 dX~T\left(\frac{2X}{\zeta} - 1\right)\left({\cal F}(X,\zeta,Q^2)-{\cal F}(\zeta,\zeta,Q^2)\right) + \nonumber\\ 
& \quad \quad \int^1_{\zeta} dX~T\left(\frac{2X}{\zeta} - 1\right) \left({\cal F}(X,\zeta,Q^2) -{\cal F}(\zeta,\zeta,Q^2)\right) + 
{\cal F}(\zeta,\zeta,Q^2)\int^1_0 dX~T \left(\frac{2X}{\zeta} - 1\right) \, . 
\label{subtraction} 
\end{align} 
Each term in eq.~(\ref{subtraction}) is now either separately finite 
or only contains an integrable logarithmic singularity.  
This algorithm closely resembles the implementation of the $+$ regularization in the evolution of PDFs and GPDs. 
We note that the first integral (in the ERBL region) is strictly real. 
The second and third terms contain both real 
and imaginary parts (which are generated in the DGLAP region). Explicit expressions for the real and imaginary parts of the DVCS amplitudes are given in Sec.(II) of \cite{frmc3}. They involve integrals over the coefficient functions which 
may be calculated explicitly and are given in Appendix B of \cite{frmc3}. 
The real and imaginary parts of the unpolarized and polarized DVCS 
amplitudes were computed using a FORTRAN code based on numerical 
integration routines. We implemented the exact solution to the RNG 
equation for $\alpha_s$ in LO or NLO in our calculation, as appropriate,  
to be consistent throughout our analysis.

\subsection{Numerical results: unpolarized case}

\begin{figure} 
\centering 
\mbox{\epsfig{file=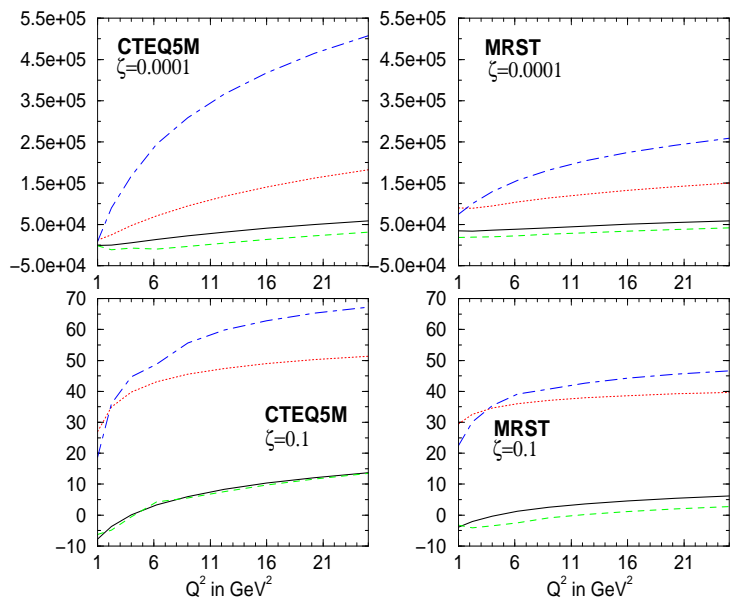,width=9.5cm,height=8.5cm}} 
\caption{The $Q^2$-dependence of the real and imaginary parts of the quark singlet DVCS 
amplitude. The solid (dashed) curve is the real part in LO (NLO) and 
the dotted (dashed-dotted) curve is the imaginary part in LO (NLO).} 
\label{ampunpolsqn} 
\end{figure} 

\begin{figure} 
\centering 
\mbox{\epsfig{file=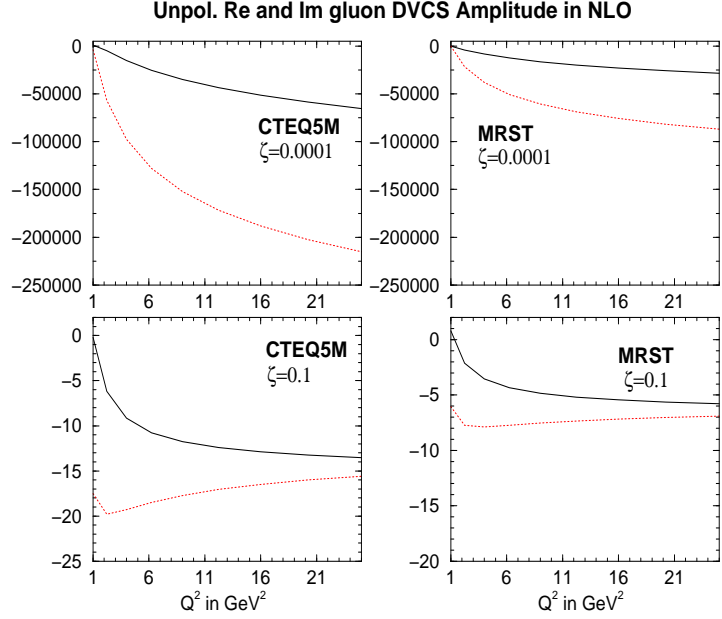,width=9.5cm,height=8.5cm}} 
\caption{The $Q^2$-dependence of the real (solid line) and imaginary (dotted line) parts of the unpolarized gluon DVCS 
amplitude, which starts at NLO, for two representative values of $\zeta$.} 
\label{ampunpolgqn} 
\end{figure}

In this subsection we illustrate the $Q^2$- and $\zeta$-dependence, as well as the size, 
of the real and imaginary parts of the unpolarized
DVCS amplitudes for the new input distributions (MRST99, CTEQ5M), calculated at LO and NLO accuracy. 
Fig. \ref{ampunpolsqn} shows the $Q^2$-dependence of the real and imaginary 
parts of the quark singlet contribution, at LO and NLO,   
for two values of $\zeta = 0.1, 0.0001$, representative of HERMES and HERA kinematics, respectively. Correspondingly, Fig.\ \ref{ampunpolgqn} shows the gluon contributions, which start at NLO.
Note the strong $Q^2$-dependence in NLO of the imaginary part of the quark singlet and gluon amplitude at small $\zeta$. This might raise concerns about the convergence of the perturbative expansion, especially when comparing NLO with LO in the quark singlet, where NLO grows much stronger with $Q^2$ than LO. This is due to the same type of logarithmic divergences as $X\to \zeta$ in both the evolution kernels and NLO coefficient functions. These divergences enhance the region around $\zeta$, which is important for the value of the imaginary part \cite{frmc2}, more quickly in $Q^2$ than $\alpha_s$ drops as $Q^2$ increases. However, the quark singlet amplitude itself is not an observable quantity at NLO but rather the sum of quark singlet and gluon. When comparing the physical amplitudes at LO and NLO, the relative NLO corrections decrease as $Q^2$ increases, as they indeed should \cite{frmc2}.   

\begin{figure} 
\centering 
\mbox{\epsfig{file=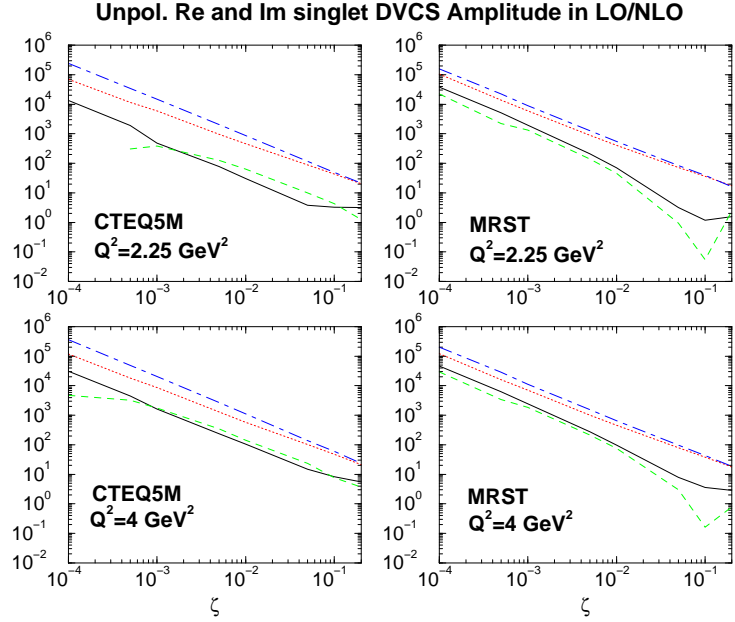,width=9.5cm,height=8.5cm}} 
\caption{The real and imaginary parts of the unpolarized  quark singlet DVCS 
amplitude, as a function of $\zeta$. The solid (dashed) curve is the real part 
in LO (NLO) and the dotted (dashed-dotted) curve is the imaginary part in LO 
(NLO). To be able to plot the NLO real part of the quark singlet amplitude in 
a viewable manner, we removed the first point for CTEQ5M in the upper 
left plot (since it was negative and thus not easily handled in a log-log plot) and shifted the NLO curve of the real part in the upper right plot 
upward by an amount of ${\cal O}(1)$, again to avoid negative numbers.} 
\label{ampunpolsxn} 
\end{figure}

\begin{figure} 
\centering 
\mbox{\epsfig{file=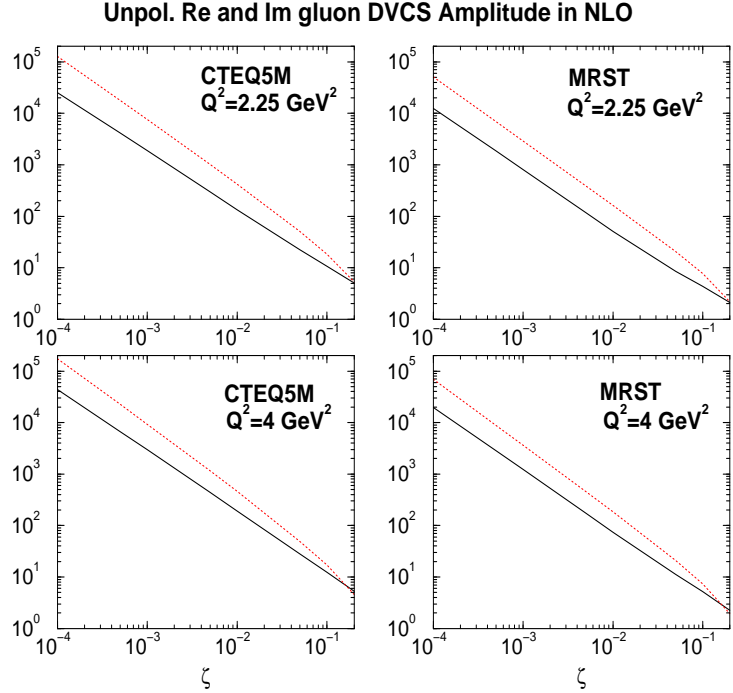,width=9.5cm,height=9.5cm}} 
\caption{The real and imaginary parts of the unpolarized gluon DVCS 
amplitude as  functions of $\zeta$, for fixed $Q^2$.  
The solid curve is the modulus of the real part (which is actually 
negative) and the dotted curve is the modulus of the imaginary part of the gluon amplitude, which is also negative!} 
\label{ampunpolgxn} 
\end{figure}

The above figures are complemented by Figs.\ \ref{ampunpolsxn} and Figs.\ \ref{ampunpolgxn}, 
which show the $\zeta$-dependence at fixed $Q^2$ for the quark singlet and gluon 
contributions, respectively. Again we would like to point out the remarkable power-like 
behavior of the unpolarized amplitudes in $\zeta$ for fixed $Q^2$ already remarked upon 
and explained in \cite{frmc3} (with the exception of the NLO real parts of the quark 
singlet amplitude for CTEQ5M for the $Q^2$ values plotted due to a somewhat strange 
combination of small quark input and comparatively large gluon input. 
At higher $Q^2$, the CTEQ5M distribution also displays the characteristic power-like 
behavior in $\zeta$). 

\subsection{Specification of $t$-dependence and helicity flip amplitudes}
\label{sec:dvcst} 

In this subsection we specify the $t$-dependence of the various unpolarized/polarized 
helicity non-flip/flip DVCS amplitudes, 
${\cal H}_1, \widetilde{{\cal H}_1}, {\cal E}_1, \widetilde{{\cal E}}_1$, 
for each parton species $a = S(a=u,d,s), g $. These are required in 
eqs.(\ref{decom-T}, \ref{dec-FF-V}, \ref{dec-FF-A}) to fully specify ${\cal T}_{\mbox{{\scriptsize DVCS}}}$  (our choices follow those of \cite{bemu1} closely) \footnote{In fact the GPDs themselves may 
already be considered to have a given helicity specification. The assumption of 
a factorized $t$-dependence for the GPDs allows the specification of the $t$-dependence 
to be moved to the amplitude level.}. The various form factors are specified as follows:
\begin{align} 
&{\cal H}_1^a(\zeta,Q^2,t) = F^a_1(t) {\cal T}^{a,V}(\zeta,Q^2) \, , ~\qquad 
 ~{\cal {\tilde H}}_1^a(\zeta,Q^2,t) = G^a_1(t) {\cal T}^{a,A}(\zeta,Q^2) \, , \nonumber \\ 
&{\cal E}_1^a(\zeta,Q^2,t) = F^a_2(t){\cal 
  T}_{{\cal E}_1}^{a,V}(\zeta,Q^2) \, , ~\qquad  ~{\cal {\tilde E}}_1^a(\zeta,Q^2,t) = G^a_2(t) {\cal T}_{{\tilde {\cal E}}_1}^{a,A}(\zeta,Q^2) \, ,  
\label{forms}
\end{align}
\noindent where, in the helicity flip case on the second line, we have introduced additional subscripts, ${\cal E}_1, {\tilde {\cal E}}_1$,
on the right hand side to distinguish this case from the helicity non-flip one considered explicitly above.
For the up and down quark flavors we exploit 
the fact that proton and neutron form an iso-spin doublet to arrive at:
\begin{align} 
2 F^u_i (t) = 2 F^p_i (t) + F^n_i (t),~\qquad ~2 F^d_i (t) = F^p_i (t) + 2 F^n_i (t) \, , ~\qquad~ \mbox{for} \, ~i=1,2 \, , 
\label{f12ud}
\end{align} 
\noindent corresponding to the Dirac and Pauli form factors, respectively (see eqs.(\ref{fipn}, \ref{gipn})). 
For the helicity non-flip polarized case we choose \cite{bemu1}: 
\begin{align} 
G^{u,d}_1 (t) = \left(1-\frac{t}{m^2_A}\right)^{-2} \, , \qquad G^{s,g}_1 (t) = (1-\frac{t}{m_A^2})^{-3} \, ,
\label{tdepsga}
\end{align} 
where $m_A = 0.9~\mbox{GeV}$ \cite{Oku90}. The strange quark and the gluon sea-like form factors were chosen in \cite{bemu1} using the counting rules for elastic form factors which give $1/t^3$, for large $t$. Hence, for the unpolarized case the electric and magnetic form factors were chosen to be
\begin{align} 
G^{s,g}_E (t) = \frac{G^{s,g}_M(t)}{1+\kappa^{s,g}} = \left(1-\frac{t}{m^2_V}\right)^{-3} \, ,
\label{tdepsgv} 
\end{align} 
\noindent where we further assume $\kappa^{s,g} = 0$. This gives (cf. eq.(\ref{fipn}))
\begin{align} 
F^{s,g}_1 (t) = G_E^{s,g} (t) = \left(1-\frac{t}{m_V^2}\right)^{-3} \, , ~\qquad ~F^{s,g}_2 (t) = 0 \, ,
\label{f12sg}
\end{align} 
\noindent with $m_V = 0.84$~GeV.

For the helicity-flip case, we still have to specify the polarized and unpolarized GPDs. 
Unfortunately there are no inclusive analogs to guide us. However, we do know that the unpolarized helicity flip GPD contains a D-term  with a 
relative minus sign as compared to the helicity non-flip GPD, such 
that when added they cancel \footnote{This has to be the case since 
the n-th.\ moment in $X$ of the sum is a polynomial of degree 
n-1 in $\zeta$, whereas the n-th.\ moment of the non-flip and flip GPDs separately 
are polynomials of degree n. The highest power of the polynomial in 
each case is generated by the D-term and thus they must cancel.}. In contrast to \cite{bemu1} we 
set the GPD equal to this D-term, which is non-zero only in the ERBL 
region\footnote{Strictly speaking this violates the polynomiality condition since 
the helicity flip GPD is multiplied by the Pauli form factor rather than the Dirac 
form factor. However since the helicity flip term is numerically insignificant
for DVCS observables this model is sufficient for phenomenological purposes.}. We then evolve this 
distribution in $Q^2$ and use it as an input to calculate ${\cal T}_{{\cal E}_1}^{a,V}$ for use 
in eq.(\ref{forms}), with the $F_2^a (t)$ for various $a$ specified in eqs.(\ref{f12ud}, \ref{f12sg}).

For the polarized helicity flip amplitude, ${\cal T}_{\cal {\tilde E}}$, it 
was observed that, in a similar fashion to the effective pseudo scalar 
form factors in $\beta$-decay \cite{Oku90}, one can approximate it at small 
$t$ by the pion 
pole (see for example \cite{rad2}). Thus we chose the corresponding GPD to only contain the asymptotic pion 
distribution amplitude given by
\begin{align} 
\phi_{\pi}(X/\zeta) = 
\frac{8}{3}\frac{2-\zeta}{\zeta}\frac{X}{\zeta}\left(1-\frac{X}{\zeta}\right) \, ,
\end{align} 
in the ERBL region (and zero for $X > \zeta$). Since we use the 
asymptotic form we do not evolve the GPD and use it directly in the 
computation of the DVCS amplitude, ${\cal T}_{{\tilde {\cal E}}_1}^{a,A} (\zeta,Q^2)$.
The $t$-dependence is given by the pion pole and thus we find 
\begin{align} 
{\cal {\tilde E}}_1^a (\zeta,Q^2,t) = G_2^a (t) ~{\cal T}_{{\tilde {\cal E}}_1}^{a,A} (\zeta,Q^2) = \frac{4g^{(3)}_AM^2}{m_{\pi}^2-t} 
{\cal T}_{{\tilde {\cal E}}_1}^{a,A} (\zeta,Q^2) \, , ~\qquad \, \mbox{for} \, ~a = u,d \, ,
\end{align} 
\noindent where $g_A^{(3)} = 1.267$, $M$ is the nucleon mass and 
${\cal T}_{{\tilde {\cal E}}_1}^{a,A} = \phi_{\pi}$, for all $Q^2$.
For the s-quark and the gluon we set $\tilde {\cal E}$ to zero.
This completes the specification of the $t$-dependence of DVCS amplitudes, which are then 
used to compute the various DVCS observables defined in Sec.\ \ref{sec:dvcskin}. 
 
We close this section with a few comments. We note 
that ${\cal E}_1, {\cal {\tilde E}}_1$ only have real parts since the GPD is zero for $X > \zeta$.
Furthermore, in the asymptotic limit $Q^2 \to \infty$ for finite $\zeta$, when 
all the partons have accumulated in the ERBL region, all amplitudes 
have zero imaginary part and a real part which is given by an 
asymptotic distribution amplitude in the ERBL region, thus the overall 
amplitude, and hence the appropriately scaled triple 
differential cross section, remain non-zero, in contrast to inclusive DIS. 
 
\section{LO and NLO results for DVCS observables} 
 
\label{sec:res} 

In this section we present results for the triple differential cross section and for 
various asymmetries (AAA, SSA, CA, UPLT, CADSFL) defined in subsection \ref{subsec:obs},    
in kinematics appropriate for the H1, ZEUS and HERMES experiments. 
We show the results as functions of $t$ (for fixed $\zeta, Q^2$), of $\zeta$ (for fixed $t, Q^2$) and of $Q^2$ (for fixed $t, \zeta$) using the various input distributions defined in subsection \ref{subsec:gpdsinp}.
For our predictions in HERA kinematics we assume an $e^+ P$ scattering with a proton energy of $920$~GeV 
and a positron energy of $27.5$~GeV (except of course for the charge asymmetries, which also use 
an electron probe of $27.5$~GeV). 
 
\subsection{The triple differential cross section} 

\begin{figure} 
\centering 
\mbox{\epsfig{file=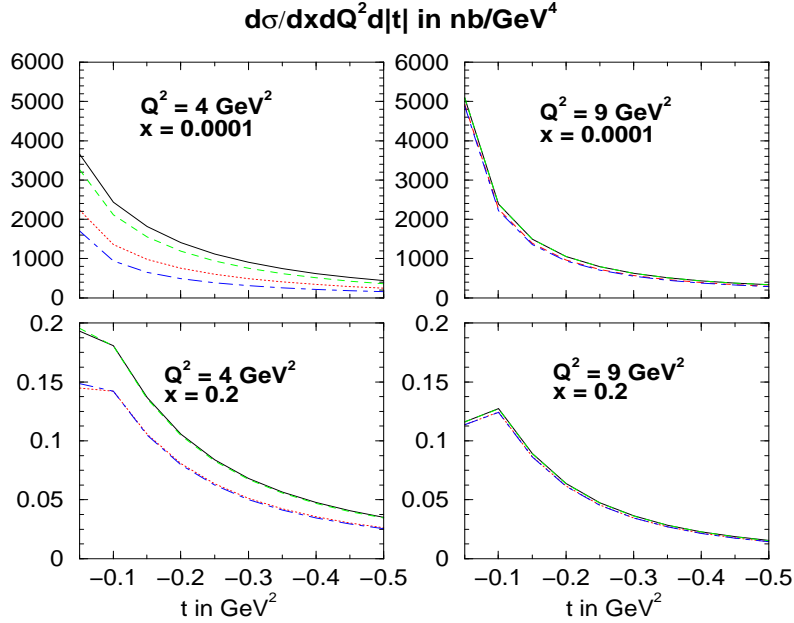,width=10.5cm,height=8.5cm}} 
\caption{The triple differential cross section in $t$ for fixed 
  $x = \zeta$ and $Q^2$. The solid (dotted) curve is the MRSA$^{'}$ set in LO (NLO) 
  and the dashed (dashed-dotted) curve is the GRV98 set in LO (NLO).} 
\label{td1} 
\end{figure}

\begin{figure} 
\centering 
\mbox{\epsfig{file=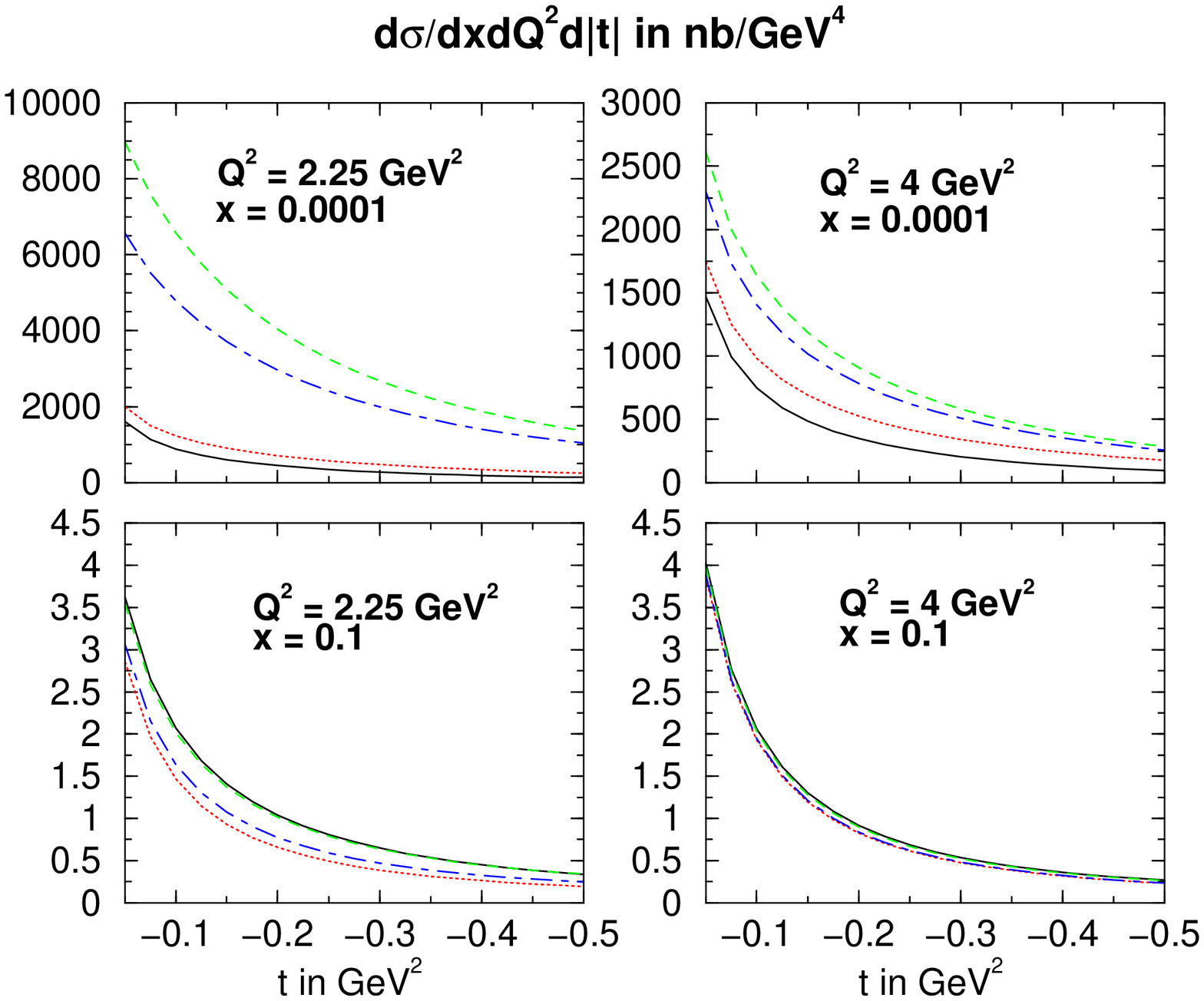,width=10.5cm,height=8.5cm}} 
\caption{The triple differential cross section in $t$ for fixed 
  $x=\zeta$ and $Q^2$. The solid (dotted) curve is the CTEQ5M set in LO (NLO) 
  and the dashed (dashed-dotted) curve is the MRST99 set in LO (NLO).} 
\label{td2} 
\end{figure}

In Fig.\ \ref{td1} and Fig.\ \ref{td2} we show the triple differential cross section of eq.(\ref{crossx}), as a function of $t$
at fixed $\zeta$ and $Q^2$, for our four input distributions. We note that at the common point $Q^2=4$~GeV$^2$, 
GRV98, MRSA' and MRST99 are in close agreement, for small $\zeta=x$. In general CTEQ5M and MRST99 
experience only moderate changes in going from LO to NLO. 

At small $\zeta=x$, for GRV98 and MRSA' input distributions, the NLO corrections are much larger. 
As $Q^2$ increases, at fixed $\zeta$, the spread of the predictions is seen to decrease. This is partly due to the evolution washing out the differences between the various input distributions, and partly due to an increased significance of the BH process.

\begin{figure} 
\centering 
\mbox{\epsfig{file=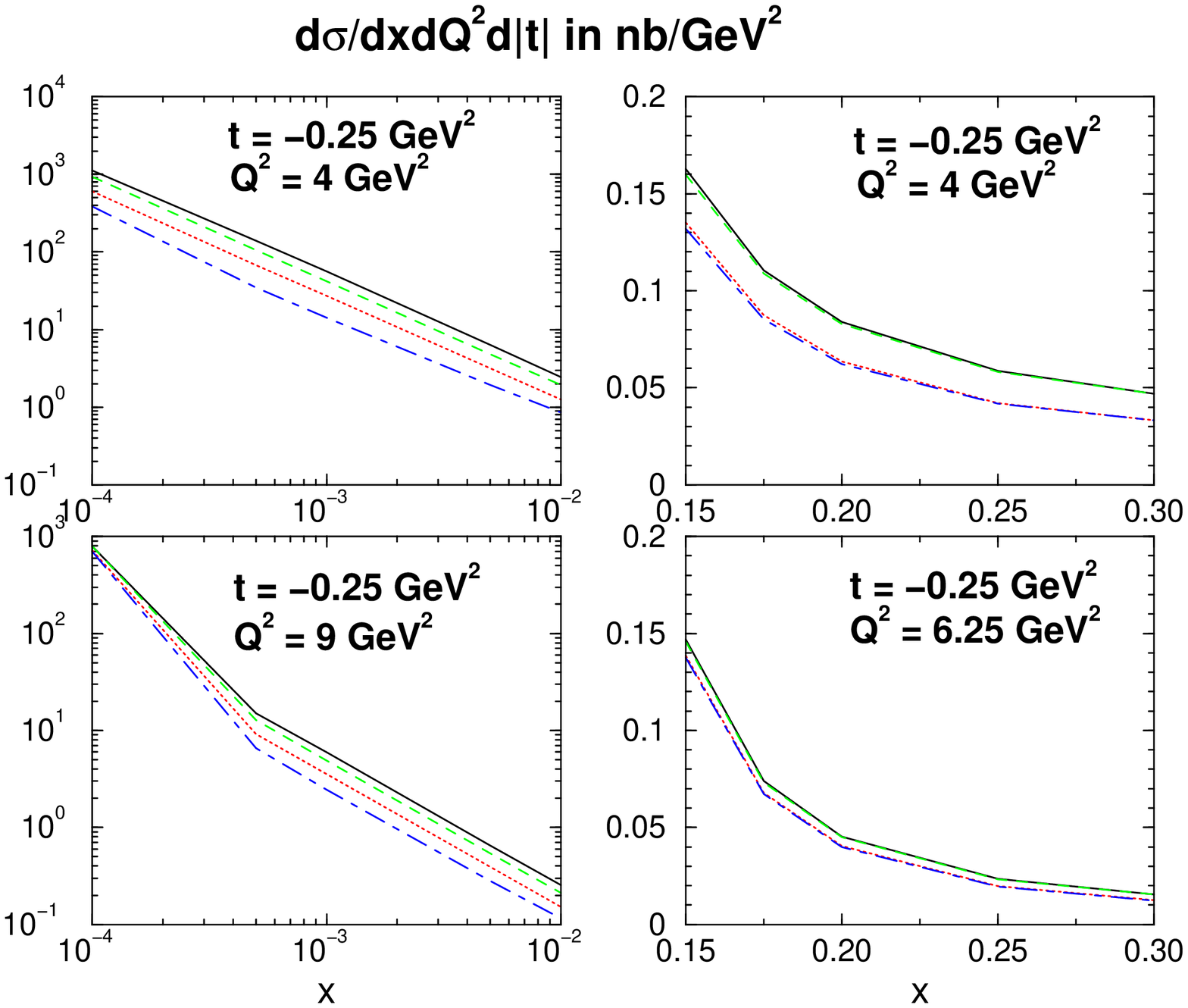,width=10.5cm,height=8.5cm}} 
\caption{The triple differential cross section in $x=\zeta$ for fixed 
  $t$ and $Q^2$. The solid (dotted) curve is the MRSA$^{'}$ set in LO (NLO)  
  and the dashed (dashed-dotted) curve is the GRV98 set in LO (NLO).} 
\label{td3} 
\end{figure} 
 
\begin{figure} 
\centering 
\mbox{\epsfig{file=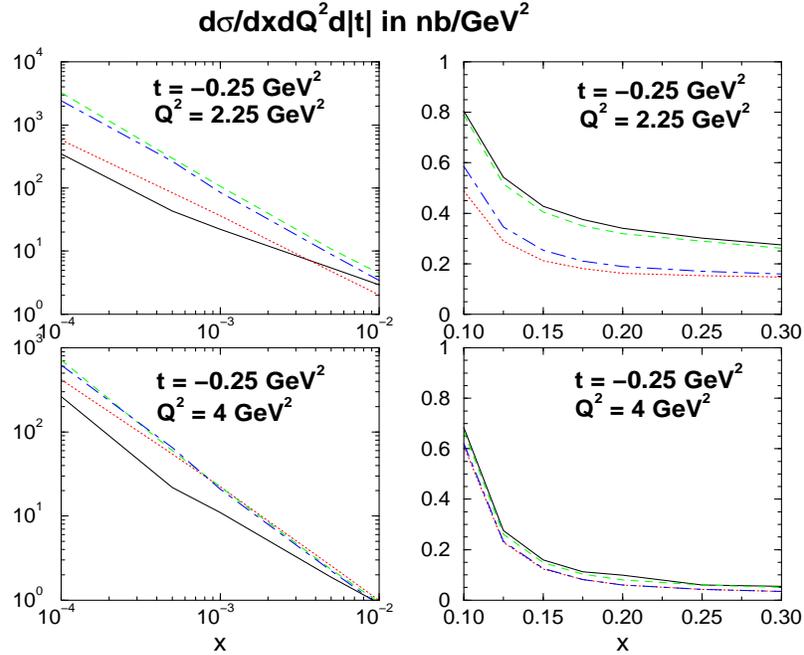,width=10.5cm,height=8.5cm}} 
\caption{The triple differential cross section in $x=\zeta$ for fixed 
  $t$ and $Q^2$. The solid (dotted) curve is the CTEQ5M set in LO (NLO) 
  and the dashed (dashed-dotted) curve is the MRST99 set in LO (NLO).} 
\label{td4} 
\end{figure} 

In Figs.\ \ref{td3}, \ref{td4}, we employ logarithmic scales for HERA kinematics ($x < 10^{-2}$) to illustrate the $\zeta$-behavior for fixed $Q^2$ and $t$. 
It is interesting to note that we find the same type of power law behavior, $\zeta^n$, for the triple differential cross section, which also includes the BH process, as we found previously for the  
unpolarized DVCS amplitudes (see Figs.\ \ref{ampunpolsxn} -\ref{ampunpolgxn}).
Note that the kink for $Q^2=9~\mbox{GeV}^2$ is due to going from a BH dominated region at $x=0.0001$ ($y>0.8$) to a region where DVCS dominates ($x > 0.0005$ and $y < 0.2$).  

\begin{figure} 
\centering 
\mbox{\epsfig{file=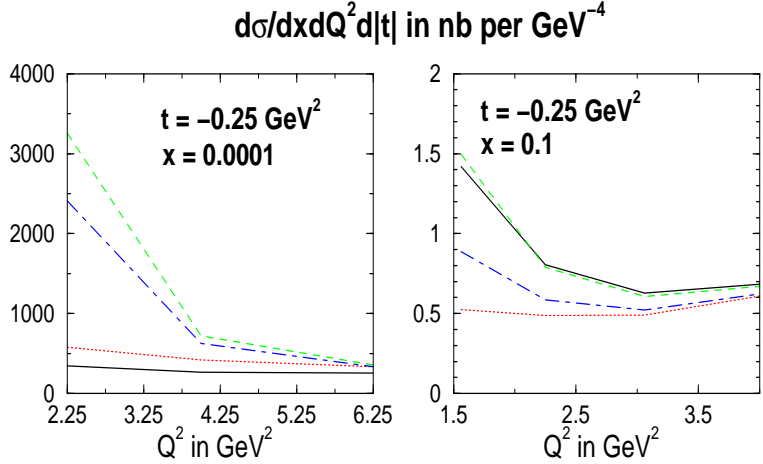,width=10cm,height=6.5cm}} 
\caption{The triple differential cross section in $Q^2$ for fixed 
  $t$ and $x=\zeta$. The solid (dotted) curve is the CTEQ5M set in LO (NLO) 
  and the dashed (dashed-dotted) curve is the MRST99 set in LO (NLO).} 
\label{td5} 
\end{figure}

The fact that the GRV98 and MRSA' input scale is $Q_0^2=4$~GeV$^2$, implies that 
the available $Q^2$-range at small $x$ is rather limited for these distributions. 
Hence, in Fig.\ \ref{td5} we show only the $Q^2$-dependence of the MRST99 and CTEQ5M sets, which start at $Q_0^2=1$~GeV$^2$.
For smaller $Q^2$, where DVCS dominates BH, the 
cross section falls quickly with $Q^2$ and one is sensitive to the details of 
the choice of input distribution. As $Q^2$ increases and BH starts to dominate 
over DVCS, one observes that the curves begin to converge and one loses 
sensitivity to the details of the input GPD.

We would like to point out that at large $x$ all of the distributions produce 
fairly similar results and that the NLO corrections are tame. This is expected
because each of the distributions have been strongly constrained by global 
fits to high statistics data in this region and hence behave similarly.
Any observed differences may result in part from the holistic nature of the 
GPDs, which requires a continuous function for all $X$, both at the input 
scale and upon evolution. 
The real part of the amplitude is sensitive to an integral over the ERBL 
region $X < \zeta$ so one may have some residual sensitivity to the behavior 
at very small $X \ll \zeta$, particularly if this behavior is extreme. 
The imaginary part is strongly influenced by the behavior at $X=\zeta$ which 
in turn is constrained by a symmetry in the ERBL region.
This sensitivity is further enhanced at smaller $\zeta$, where the input PDFs 
are not yet so well constrained by direct (mainly inclusive) measurements.
Hence, we optimistically speculate that a detailed measurement of the 
differential cross section, at HERA, for relatively low 
$Q^2 \approx 1-4$~GeV$^2$, could help to discriminate between different 
standard input PDFs.

\subsection{The single spin asymmetries (SSA,UPLT)} 
 
Having found a large spread of predictions for the triple differential cross 
section for different input GPDs, we now turn to two single spin asymmetries 
which are directly sensitive to the imaginary part of DVCS amplitudes. Let us 
first discuss the SSA, defined in eq.(\ref{eq:ssa}), 
which can be measured using a polarized lepton probe on an unpolarized target. 
At small $x$ the numerator in eq.(\ref{eq:ssa}) is directly proportional to 
$Im~{\cal H}_1$ (see the second term of eq.(30) of \cite{bemu1}).

\begin{figure} 
\centering 
\mbox{\epsfig{file=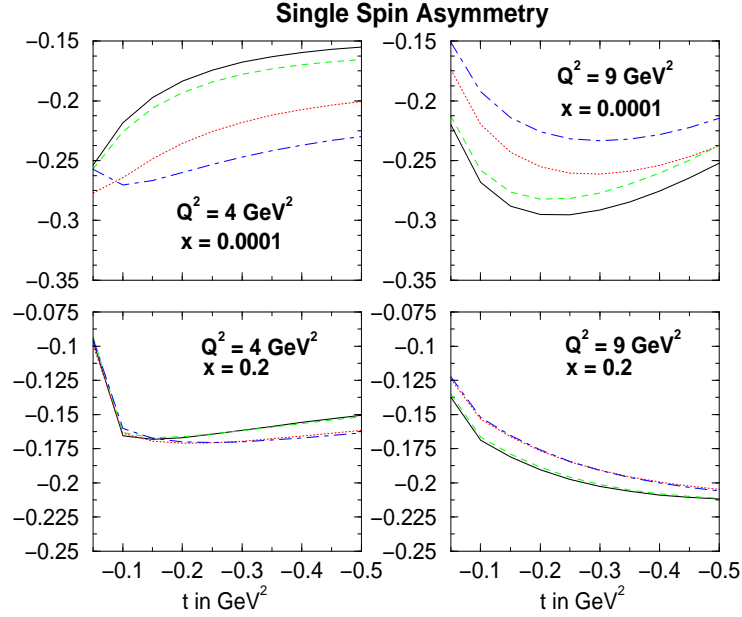,width=9.5cm,height=8.5cm}} 
\caption{The single spin asymmetry as a function of $t$, for fixed 
  $x=\zeta$ and $Q^2$. The solid (dotted) curve is the MRSA$^{'}$ set in LO (NLO) 
  and the dashed (dashed-dotted) curve is the GRV98 set in LO (NLO).} 
\label{ssa1} 
\end{figure}

\begin{figure} 
\centering 
\mbox{\epsfig{file=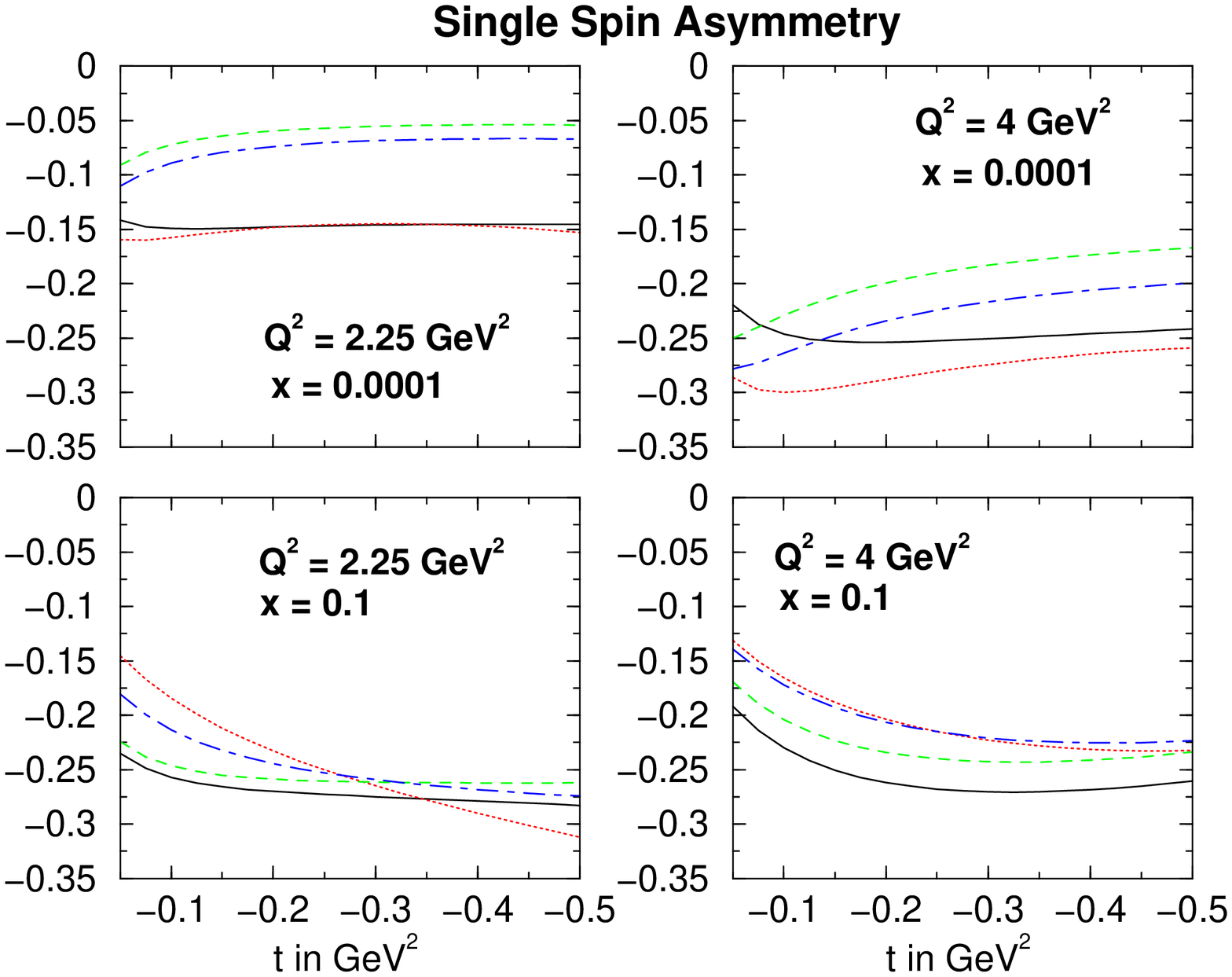,width=9.5cm,height=8.5cm}} 
\caption{The SSA in $t$ for fixed 
  $x$ and $Q^2$. The solid (dotted) curve is the CTEQ5M set in LO (NLO) 
  and the dashed (dashed-dotted) curve is the MRST99 set in LO (NLO).} 
\label{ssa2} 
\end{figure}

In Figs.\ \ref{ssa1}, \ref{ssa2} we illustrate the $t$-dependence at fixed $\zeta$ and $Q^2$, and 
note that the predictions range from as little as $5\%$ to as much as $30\%$. 
We note that in general the $t$-dependence is rather flat for $t>-0.1~\mbox{GeV}^2$, indicating that, to a certain extent, the $t$-dependence 
cancels in the ratio of eq.(\ref{eq:ssa}). This means that an experimental measurement of this asymmetry, even with 
a rather coarse binning in $t$, would be able to distinguish between different input scenarios, 
especially at larger $Q^2$ values.
At the common scale of $Q^2 = 4$~GeV$^2$, the GRV98 and MRSA' 
sets produce very similar numbers within LO and NLO, whereas CTEQ5M and MRST99 are more different within LO and NLO, at least at small $x$. 
The NLO to LO corrections are generally speaking small to moderate ($5-30\%$). 

\begin{figure} 
\centering 
\mbox{\epsfig{file=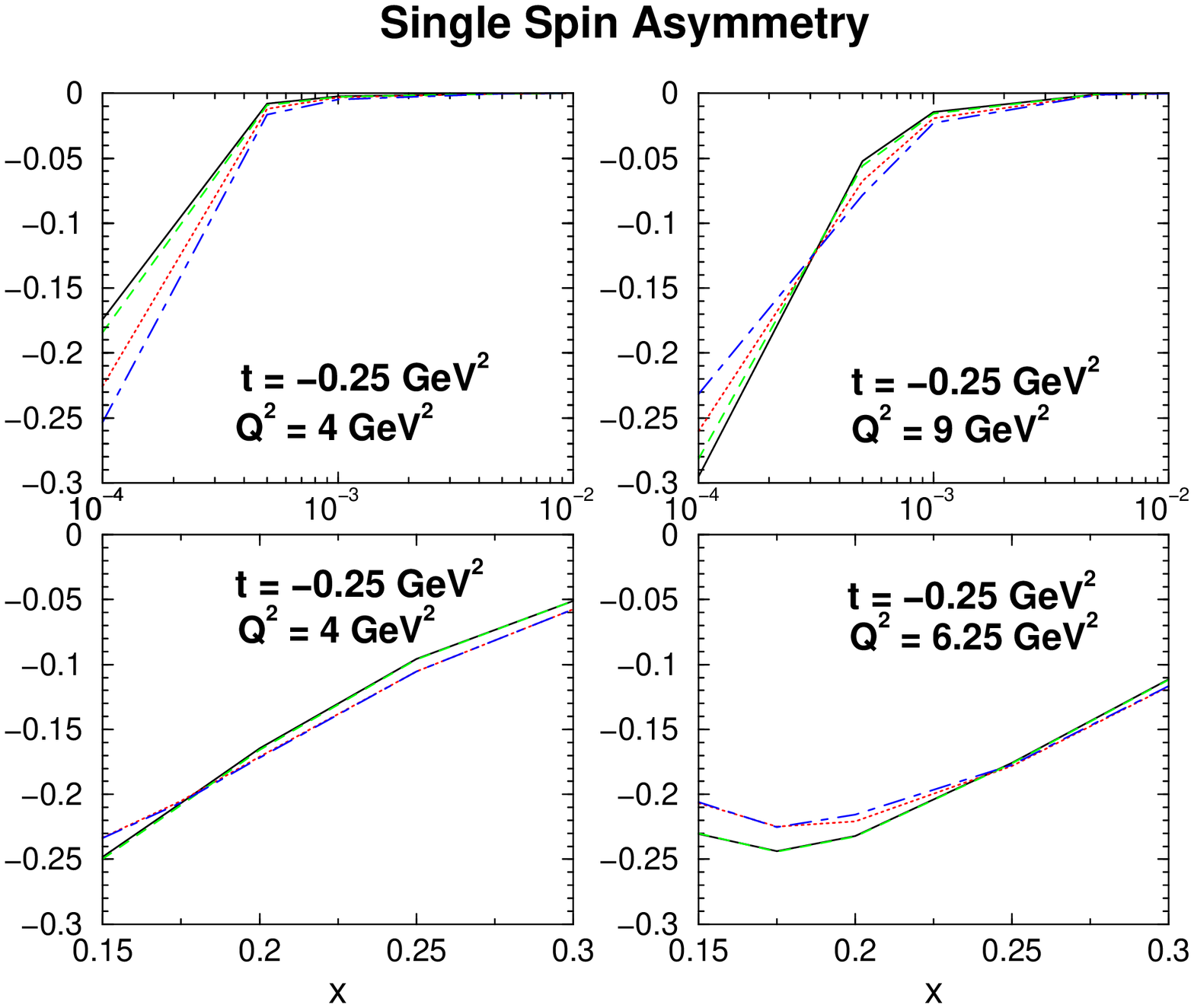,width=9.5cm,height=8.5cm}} 
\caption{The SSA in $x=\zeta$ for fixed 
  $t$ and $Q^2$. The solid (dotted) curve is the MRSA$^{'}$ set in LO (NLO) 
  and the dashed (dashed-dotted) curve is the GRV98 set in LO (NLO).} 
\label{ssa3} 
\end{figure}

\begin{figure} 
\centering 
\mbox{\epsfig{file=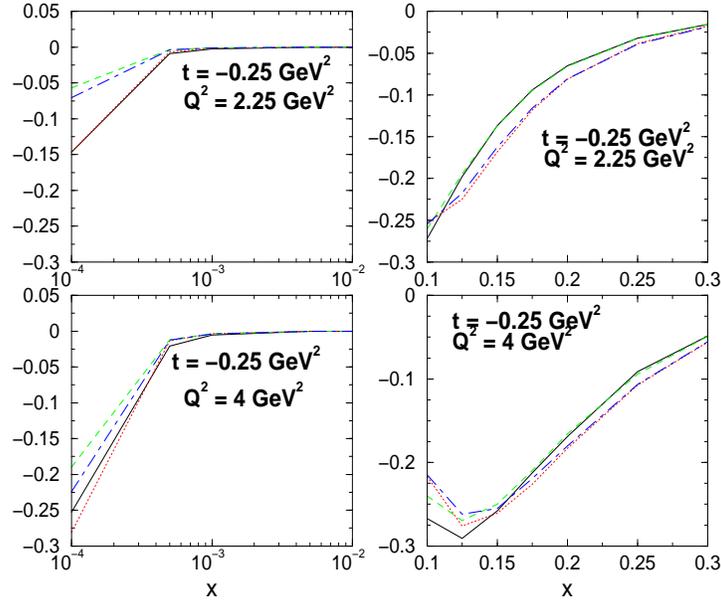,width=9.5cm,height=8.5cm}} 
\caption{The SSA as a function of $x=\zeta$ for fixed $t$ and $Q^2$. 
The solid (dotted) curve is the CTEQ5M set in LO (NLO) and the dashed 
(dashed-dotted) curve is the MRST99 set in LO (NLO).} 
\label{ssa4} 
\end{figure}

Figs.\ \ref{ssa3}, \ref{ssa4} show that the SSA drops steeply in $\zeta=x$ 
for fixed $Q^2$ and $t$, suggesting that the HERA experiments will only be able to measure the SSA  
in the small $\zeta$ regime ($\zeta \in [10^{-3}, 10^{-4}]$).
We would also like to point out that the differences for the different input sets becomes so small on 
the scale of the actual value of the SSA at larger $\zeta$ in HERA kinematics, that 
only very high statistics would be able to discriminate between 
them. Thus, for HERA, only a small $\zeta$ measurement would gives a reasonable discrimination 
between the various inputs.
For HERMES kinematics, where for fixed $\zeta=x, Q^2$ one of course probes a 
different $y = Q^2/xS$ value, the SSA again becomes sizeable.
Unfortunately, in this high $\zeta=x$ regime, the SSA is proportional to a linear combination 
of imaginary parts of ${\cal H}_1, {\tilde{{\cal H}}_1}, {\cal E}_1$ 
rather than just $Im~{\cal H}_1$ as at small $\zeta$ (see second term of 
eq.(30) of \cite{bemu1}).
        
\begin{figure} 
\centering 
\mbox{\epsfig{file=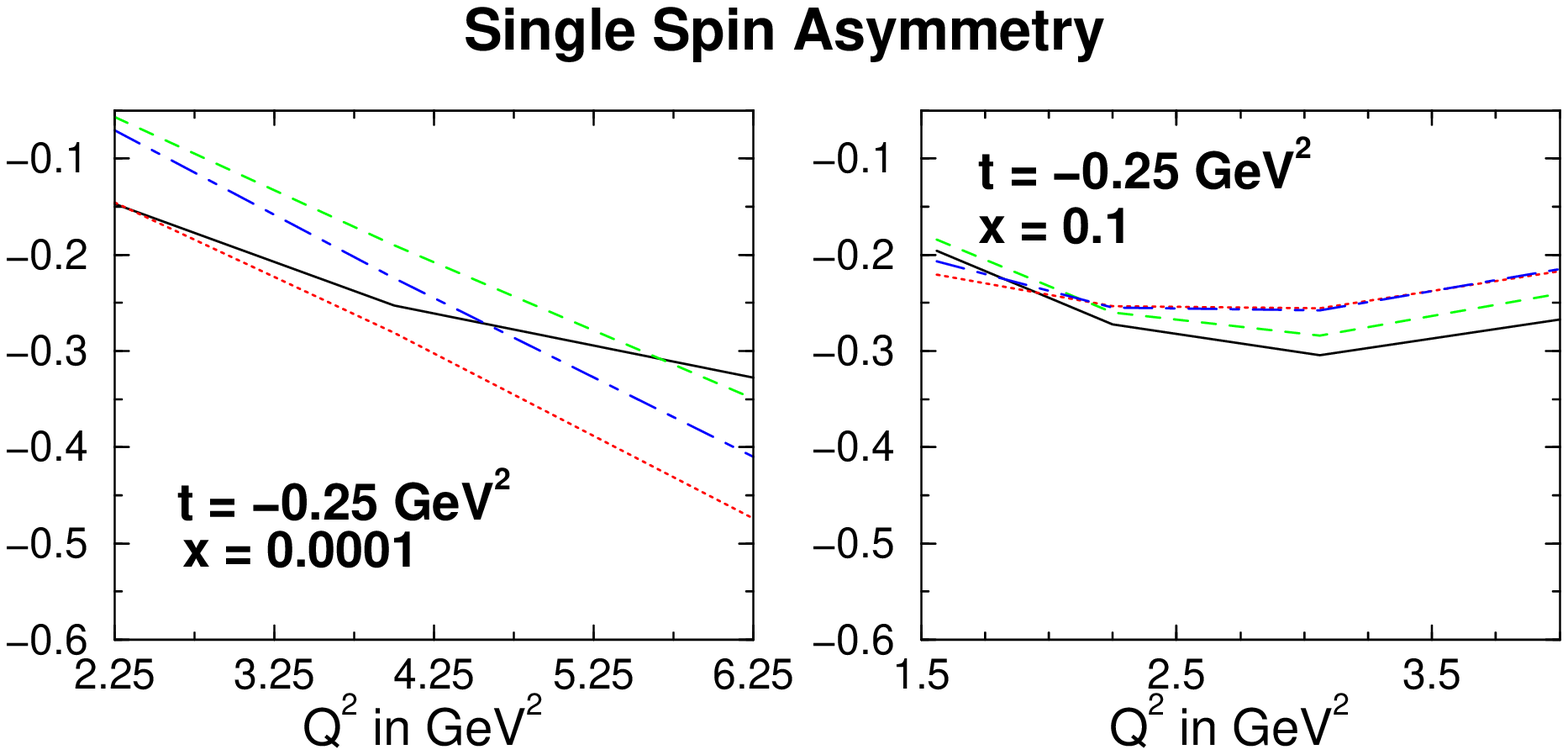,width=9.5cm,height=6.5cm}} 
\caption{The SSA as a function of $Q^2$, for fixed 
  $t$ and $x=\zeta$. The solid (dotted) curve is the CTEQ5M set in LO (NLO)  
  and the dashed (dashed-dotted) curve is the MRST99 set in LO (NLO).} 
\label{ssa5} 
\end{figure}

Finally, in Fig.\ \ref{ssa5} we plot the $Q^2$-dependence for fixed $\zeta=x$ and $t$.
For small $\zeta$ we observe that the magnitude of the SSA increases with 
$Q^2$, for both distributions. At large $\zeta$, both MRST99 and CTEQ5M have very similar $Q^2$ behavior and the NLO corrections 
are small. 
Note that at small $\zeta$ the NLO corrections appear to grow in $Q^2$ which at first sight looks 
strange. This is due to the fact that the BH process gains prominence relative to the DVCS process, 
so any differences between calculations of $Im {\cal H}_1$ for the interference term in the 
numerator become enhanced. However, the NLO corrections remain moderate between
$15-30\%$.

If one switches now to a longitudinally polarized proton target as 
available at HERMES, but not currently planned for HERA, and an 
unpolarized lepton probe, one can form the unpolarized single spin 
asymmetry UPLT which is directly sensitive to the imaginary part of a 
combination of DVCS amplitudes. Furthermore, for the small $\zeta$ regime within 
HERA kinematics, we find on inspection of the first term (proportional to $\Lambda \sin \phi$) 
of eq.(31) of \cite{bemu1} that the UPLT at small $\zeta$ is directly 
proportional to the imaginary part of the polarized amplitude ${\cal {\tilde H}}_1$ and that 
the other amplitudes like the numerically large ${\cal H}_1$ are suppressed by 
$x$. Hence, even though $Im~{\cal {\tilde H}}_1$ is about a factor 
of one thousand smaller than $Im~{\cal H}$ at $x=10^{-4}$, the suppression 
factor of $x$ means that $Im~{\cal H}_1$ constitutes only a $10\%$ correction. 
Hence the UPLT is mainly sensitive to $Im~{\cal {\tilde H}}_1$ at small $x$. 
Since this amplitude is numerically small, we found the UPLT asymmetry itself to be very small, at small $x=\zeta$, and thus virtually impossible to measure. 
Therefore, we do not show plots for HERA kinematics but rather only for HERMES
kinematics. 

\begin{figure} 
\centering 
\vskip+0.5cm 
\mbox{\epsfig{file=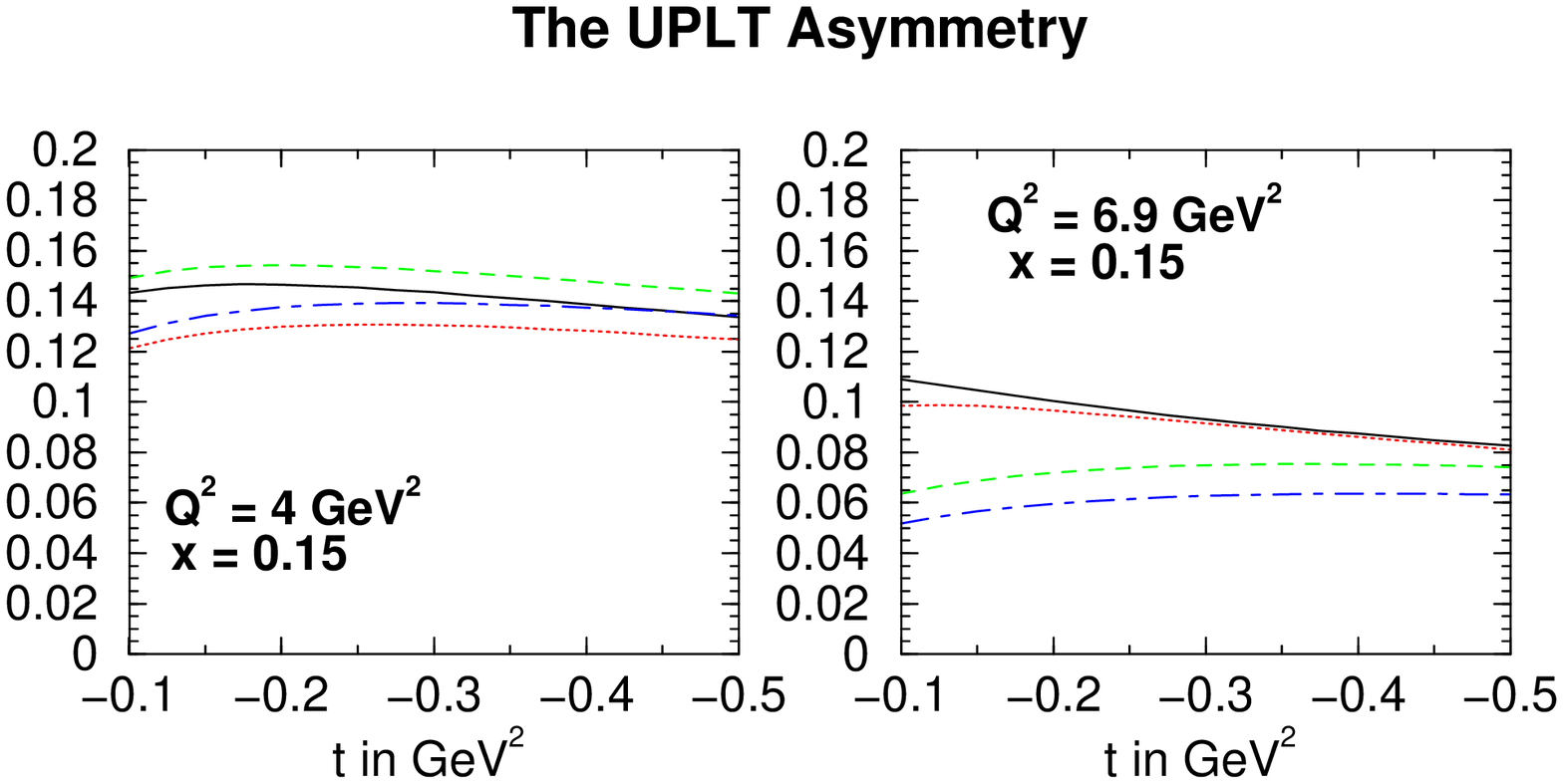,width=9.5cm,height=6.5cm}} 
\vskip+1.25cm 
\caption{The UPLT as a function of $t$, for fixed 
  $x$ and $Q^2$. The solid (dotted) curve is the MRSA$^{'}$ set in LO (NLO) 
  and the dashed (dashed-dotted) curve is the GRV98 set in LO (NLO).} 
\label{uplt1} 
\end{figure} 
 
\begin{figure} 
\centering 
\mbox{\epsfig{file=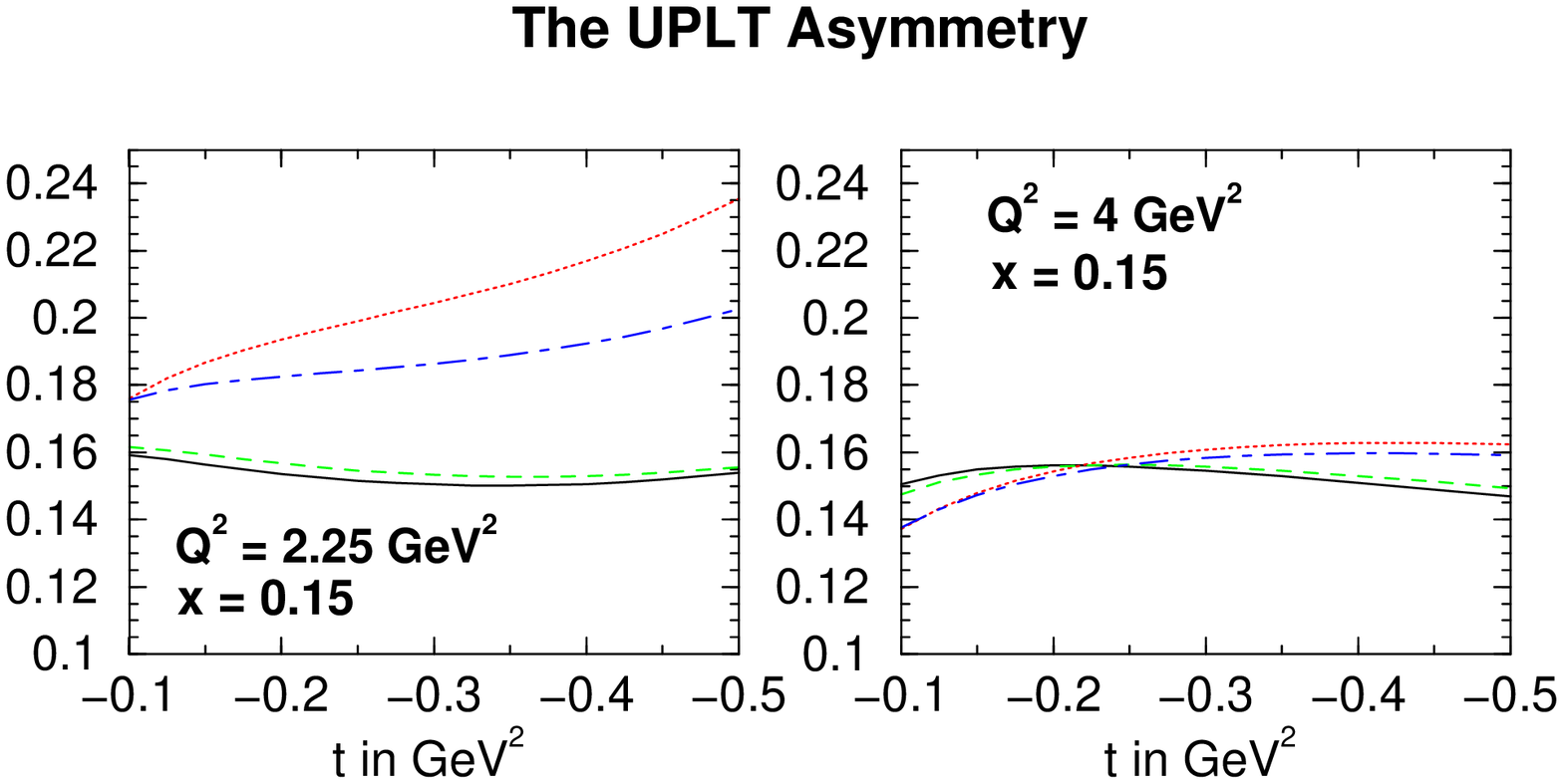,width=9.5cm,height=6.5cm}} 
\caption{The UPLT as a function of $t$, for fixed 
  $x$ and $Q^2$. The solid (dotted) curve is the CTEQ5M set in LO (NLO) 
  and the dashed (dashed-dotted) curve is the MRST99 set in LO (NLO).} 
\label{uplt2} 
\end{figure} 

\begin{figure} 
\centering 
\mbox{\epsfig{file=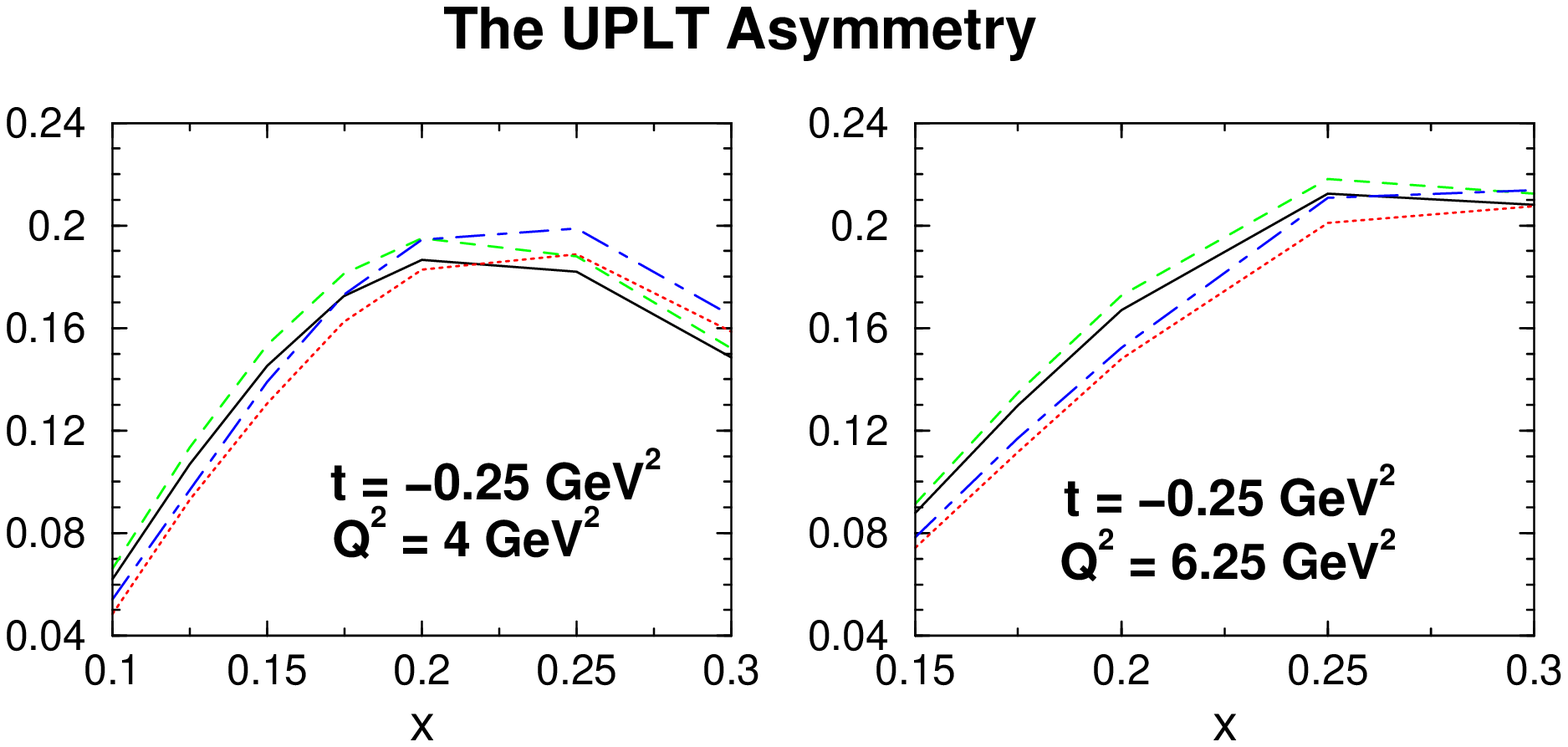,width=9.5cm,height=6.5cm}} 
\caption{The UPLT as a function of $x=\zeta$, for fixed 
  $t$ and $Q^2$. The solid (dotted) curve is the MRSA$^{'}$ set in LO (NLO) 
  and the dashed (dashed-dotted) curve is the GRV98 set in LO (NLO).} 
\label{uplt3} 
\end{figure}

\begin{figure} 
\centering 
\mbox{\epsfig{file=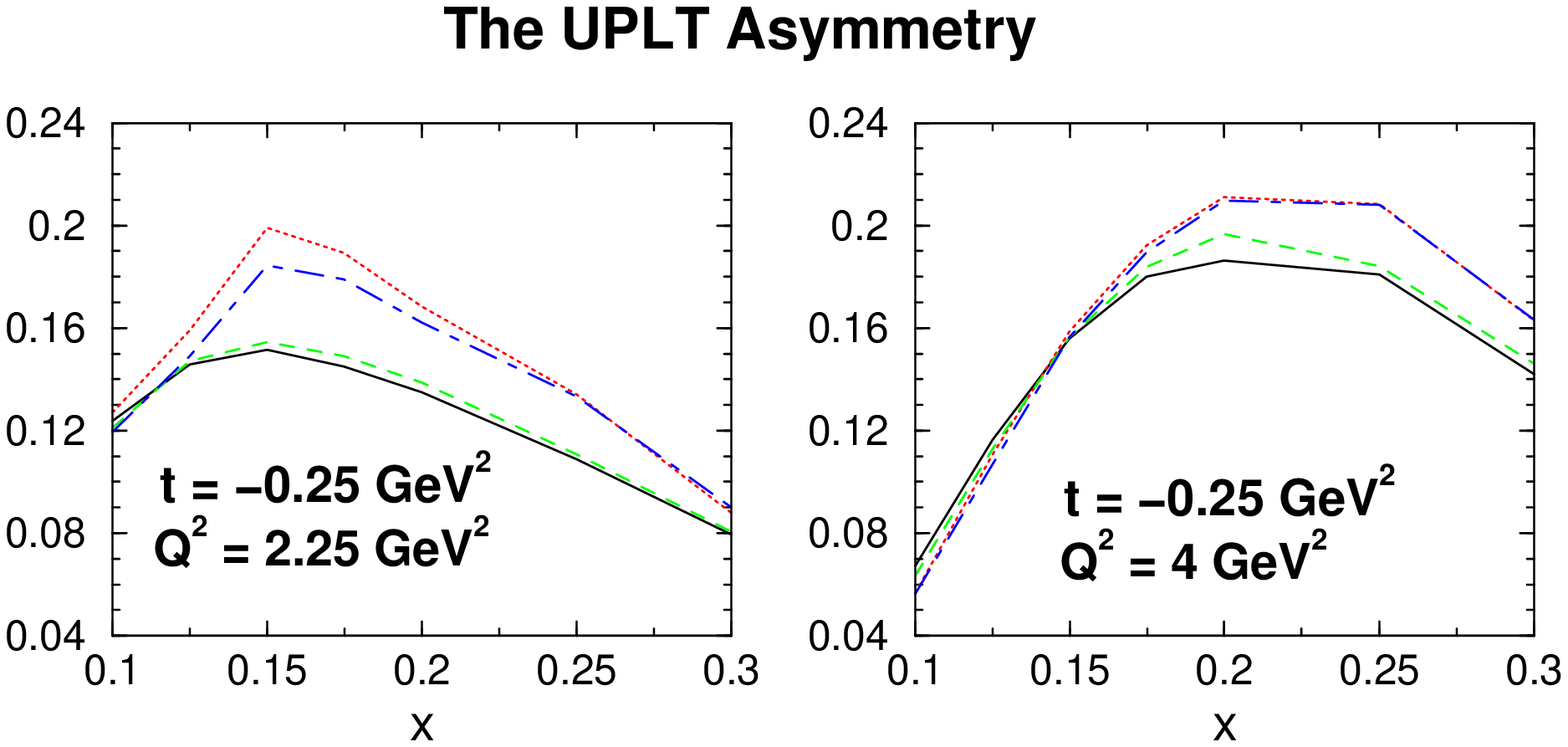,width=9.5cm,height=6.5cm}} 
\caption{The UPLT as a function of $x=\zeta$, for fixed 
  $t$ and $Q^2$. The solid (dotted) curve is the CTEQ5M set in LO (NLO) 
  and the dashed (dashed-dotted) curve is the MRST99 set in LO (NLO).} 
\label{uplt4} 
\end{figure}

\begin{figure} 
\centering 
\mbox{\epsfig{file=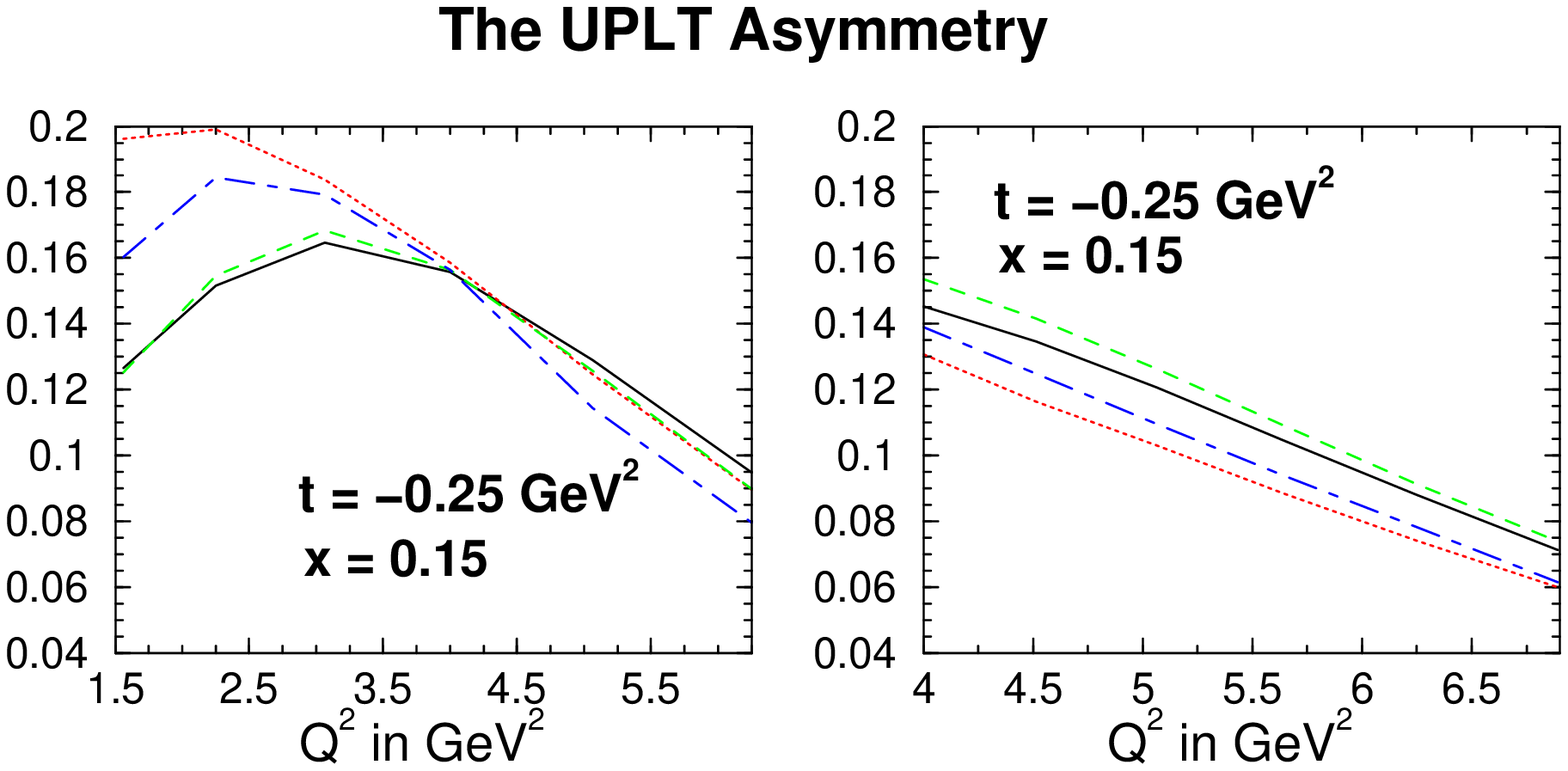,width=9.5cm,height=6.5cm}} 
\caption{The UPLT as a function of $Q^2$, for fixed 
  $t$ and $x=\zeta$. In the left figure, the solid (dotted) curve is the 
CTEQ5M set in LO (NLO) and the dashed (dashed-dotted) curve is the MRST99 set 
in LO (NLO), and in the right figure, the solid (dotted) curve is the 
MRSA' set in LO (NLO) and the dashed (dashed-dotted) curve is the GRV98 set 
in LO (NLO).} 
\label{uplt5} 
\end{figure}

In Figs.\ \ref{uplt1} and \ref{uplt2}, we show the UPLT 
as a function of $t$ for fixed $Q^2$ and $\zeta=x=0.15$, i.e. in HERMES kinematics, where 
$Im~{\cal H}$ and other amplitudes contribute significantly. Hence, 
despite the fact that this asymmetry should be measurable at HERMES, the 
results would only be useful in the context of a program of asymmetry 
measurements that would enable the individual DVCS amplitudes to be isolated. 
Again a rather flat behavior is observed in $t$. 
We observe that all sets agree well with one another and their NLO corrections 
are small, in terms of percentages. 
In Figs.\ \ref{uplt3}, \ref{uplt4} we plot the UPLT as a function of $\zeta=x$ at fixed $t,Q^2$. Note the good agreement of all sets in the 
$Q^2=4~\mbox{GeV}^2$ figures. This agreement does not bode well for the usefulness of the UPLT to discriminate between different input models. 
The NLO effects are found to be generally small. 
Finally, in Fig.\ \ref{uplt5} we show that the $Q^2$-behavior of the UPLT asymmetry, is rather complicated (for the same reasons as in the case of the SSA in HERMES kinematics). Note the close agreement of all sets for the $Q^2>4~\mbox{GeV}^2$ behavior and the smallness of the NLO corrections.

\subsection{The azimuthal angle asymmetry (AAA)} 
 
We now discuss the (unpolarized) azimuthal angle asymmetry, defined in eq.
(\ref{aaadef}), which directly probes the real part of DVCS amplitudes and 
only $Re~{\cal H}_1$ at small $\zeta=x$ where it dominates the other 
amplitudes (cf. the term proportional to $\cos \phi$ in eq.(30) of \cite{bemu1}). 
Again we would like to point out that subtracting the BH contribution is 
necessary since it does not cancel in the asymmetry due to the 
$\phi$-dependence of the propagators ${\cal P}_1$ and ${\cal P}_2$.

\begin{figure} 
\centering 
\mbox{\epsfig{file=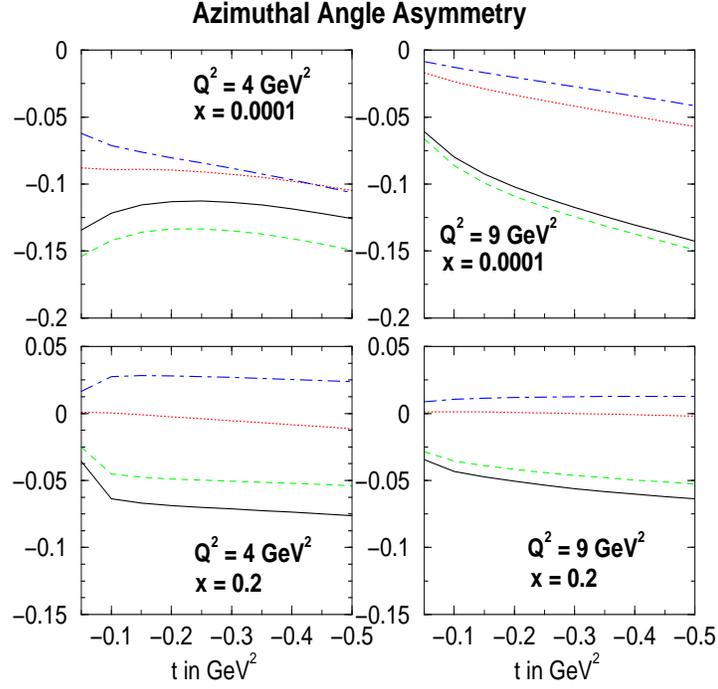,width=9.5cm,height=9.5cm}} 
\caption{The AAA in $t$ for fixed 
  $x=\zeta$ and $Q^2$. The solid (dotted) curve is the MRSA$^{'}$ set in LO (NLO) 
  and the dashed (dashed-dotted) curve is the GRV98 set in LO (NLO).} 
\label{aaa1} 
\end{figure}

\begin{figure} 
\centering 
\mbox{\epsfig{file=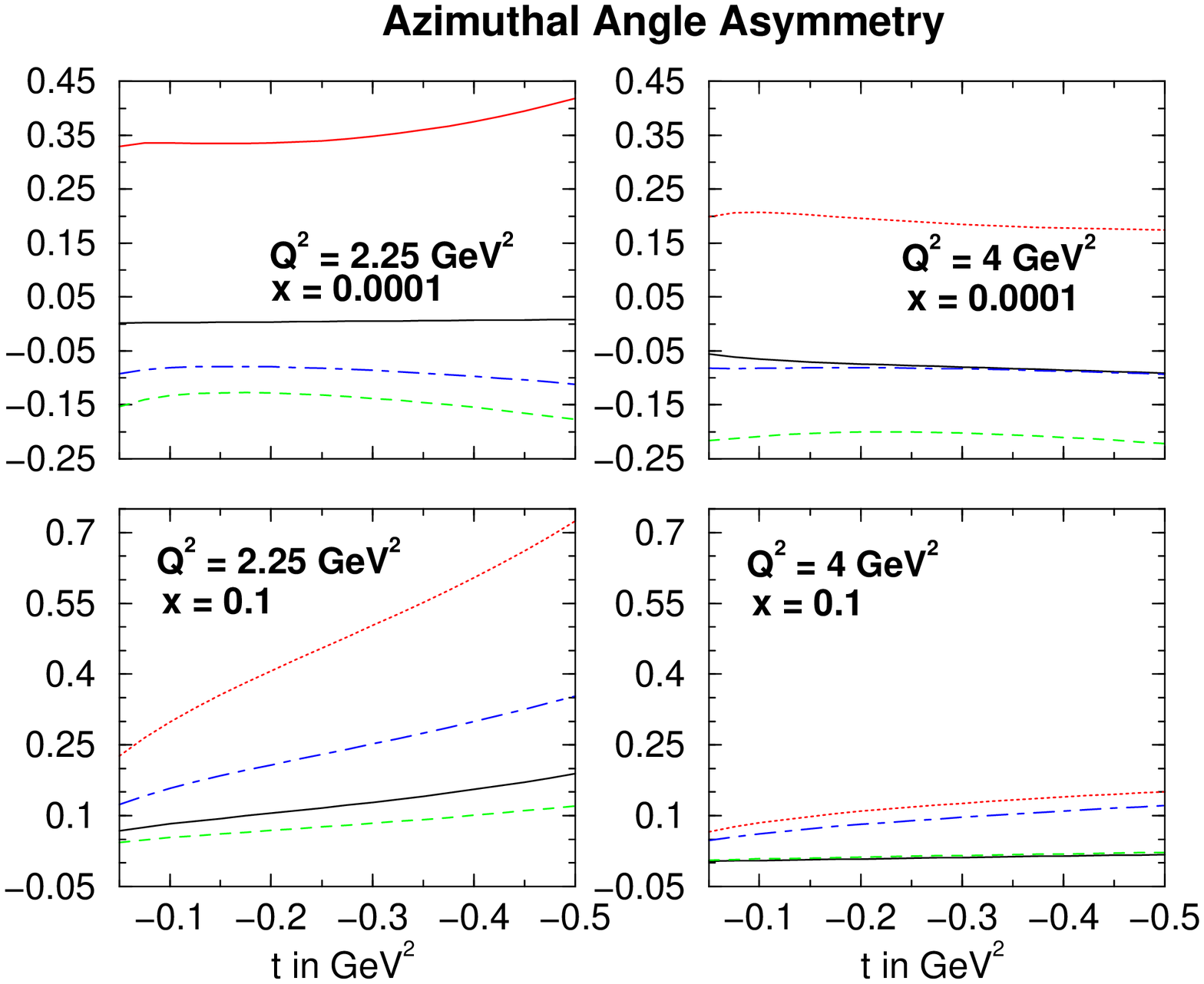,width=9.5cm,height=9.5cm}} 
\caption{The AAA in $t$ for fixed 
  $x=\zeta$ and $Q^2$. The solid (dotted) curve is the CTEQ5M set in LO (NLO) 
  and the dashed (dashed-dotted) curve is the MRST99 set in LO (NLO).} 
\label{aaa2} 
\end{figure}

In general, Figs.\ \ref{aaa1}, \ref{aaa2} reveal a rather flat behavior of the 
AAA in $t$, for fixed $\zeta, Q^2$. We note that the spread of predictions 
coming from different inputs is rather large, indicating a strong sensitivity 
of the AAA to the input distributions and to the order in perturbation theory.
The NLO corrections are very large, simply because the gluon enters at NLO for
the first time, with a relative minus sign compared to the quarks 
in the real parts of the amplitudes.
The wide spread in results indicates that the AAA is a 
highly sensitive discriminator between different input models, even those 
which have, up until now, agreed very well with one another. Hence, measuring 
the AAA both at HERA and HERMES with high precision is imperative for 
constraining the GPDs via a global fit. 
Note also that this strong sensitivity to the details of the shape and size of
the GPD both in the ERBL and DGLAP region, as anticipated by the results in 
\cite{frmc2,frmc3}, indicates that an extraction of the GPDs with 
reasonable precision may be possible.

\begin{figure} 
\centering 
\mbox{\epsfig{file=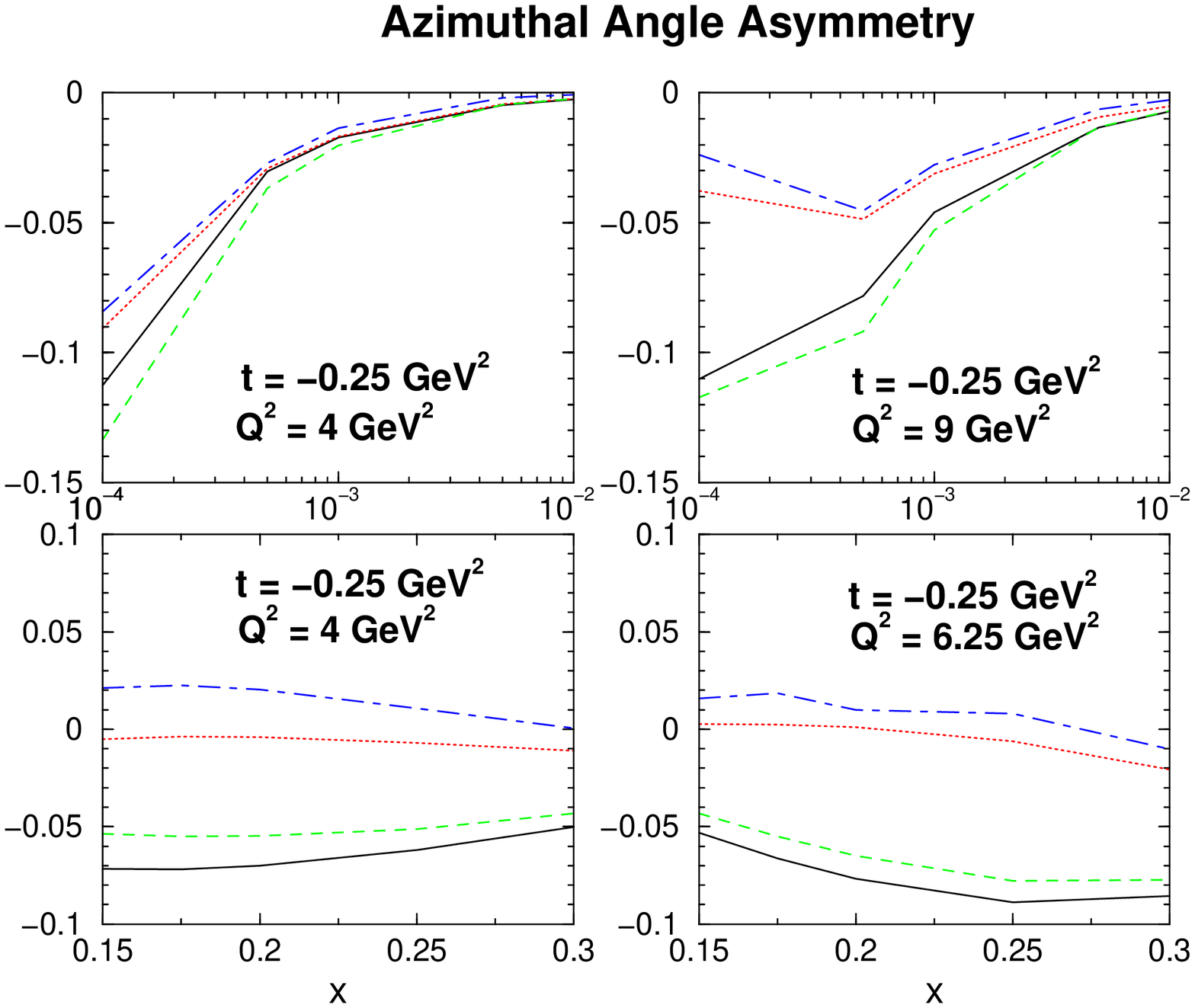,width=9.5cm,height=9.5cm}} 
\caption{The AAA in $x=\zeta$ for fixed $t$ and $Q^2$. The solid (dotted) curve is the MRSA$^{'}$ set in LO (NLO) 
  and the dashed (dashed-dotted) curve is the GRV98 set in LO (NLO).} 
\label{aaa3} 
\end{figure} 
 
\begin{figure} 
\centering 
\mbox{\epsfig{file=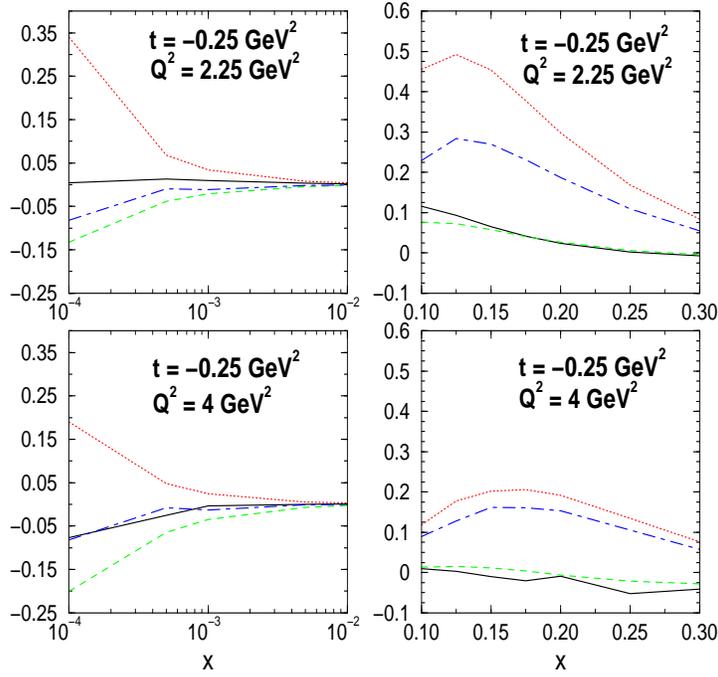,width=9.5cm,height=9.5cm}} 
\caption{The AAA in $x=\zeta$ for fixed $t$ and $Q^2$. The solid (dotted) curve is the CTEQ5M set in LO (NLO) 
  and the dashed (dashed-dotted) curve is the MRST99 set in LO (NLO).} 
\label{aaa4} 
\end{figure} 

Figs.\ \ref{aaa3} and \ref{aaa4} illustrate the strong decrease of the AAA as 
$\zeta$ decreases within HERA kinematics (i.e. small $\zeta$), for fixed 
$Q^2,t$. However, as expected from dispersion relations, this behavior at 
small $\zeta$ of AAA, which is sensitive to $Re {\cal H}_1$, is naturally not 
as steep as that of SSA,  which is sensitive to $Im {\cal H}_1$. Note again 
the kink at $Q^2 = 9~\mbox{GeV}^2$ when going from a BH dominated region to a 
DVCS dominated one, i.e. from high to low $y$.

The $\zeta$-dependence of the AAA in the valence region probed at HERMES 
(where it has already measured the SSA), shows very large NLO effects, 
consistent within all sets, which make the AAA large and thus measurable at 
HERMES. The large overall size and spread of predictions is encouraging for 
measurements of the AAA at large $x$ since even here the discriminating power 
between GPD models is very good.
 
\begin{figure} 
\centering 
\mbox{\epsfig{file=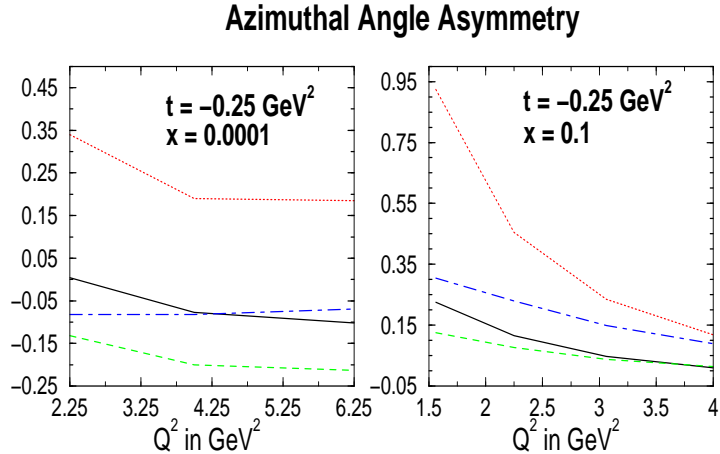,width=9.5cm,height=6.5cm}} 
\caption{The AAA in $Q^2$ for fixed $t$ and $x=\zeta$. The solid (dotted) curve is the CTEQ5M set in LO (NLO) 
  and the dashed (dashed-dotted) curve is the MRST99 set in LO (NLO).} 
\label{aaa5} 
\end{figure} 

Finally we show the $Q^2$-dependence of the AAA in Fig.\ \ref{aaa5} (again only
for MRST99 and CTEQ5M, due to their lower input scale). For small $\zeta=x$ we
see a fairly flat behavior of the AAA with $Q^2$ with the NLO to LO changes 
staying fairly constant in $Q^2$. At large $\zeta=x$, the predictions for the 
two sets quickly approach one another in both LO and NLO as $Q^2$ increases. 
Also the NLO to LO change decreases as $Q^2$ increases in line with the 
argument that BH starts dominating at large $Q^2$ due to the associated 
increase in $y$. 
 
\subsection{The charge asymmetries (CA,CADSFL)} 
 
Other asymmetries which measure the real part of DVCS amplitudes are 
the charge and charge double spin flip asymmetries (see eqs.(\ref{cadef}, \ref{cadsfldef})). These asymmetries require the measurement of DVCS with both 
the positron and the electron and, in the case of the CADSFL, a longitudinally
polarized probe and target at the same time. 
Thus they are more difficult to measure than the AAA or other observables. 
The CA isolates the same combination of real parts of DVCS amplitudes as the 
AAA (i.e. the first term in eq.(30) of \cite{bemu1}, which is proportional to $\cos \phi$), and would therefore serve as a useful complementary measurement. 

The CADSFL isolates the combination of real parts of DVCS amplitudes given in 
the second term of eq.(31) of \cite{bemu1}. The UPLT asymmetry isolates the 
imaginary part of the same linear combination.
Thus a combined measurement of both CADSFL and UPLT at small $x$ reveals 
information about the real and 
imaginary parts of the polarized helicity non-flip DVCS amplitude 
${\cal {\tilde H}}_1$.

\begin{figure} 
\centering 
\mbox{\epsfig{file=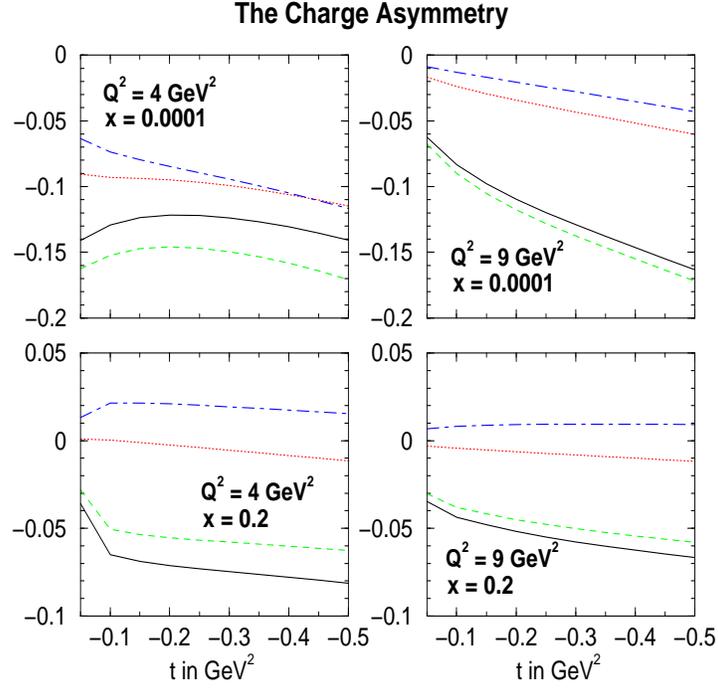,width=9.5cm,height=9.5cm}} 
\vskip+1.0cm 
\caption{The CA in $t$ for fixed $x=\zeta$ and $Q^2$. The solid (dotted) curve is the MRSA$^{'}$ set in LO (NLO) 
  and the dashed (dashed-dotted) curve is the GRV98 set in LO (NLO).} 
\label{ca1} 
\end{figure}

\begin{figure} 
\centering 
\mbox{\epsfig{file=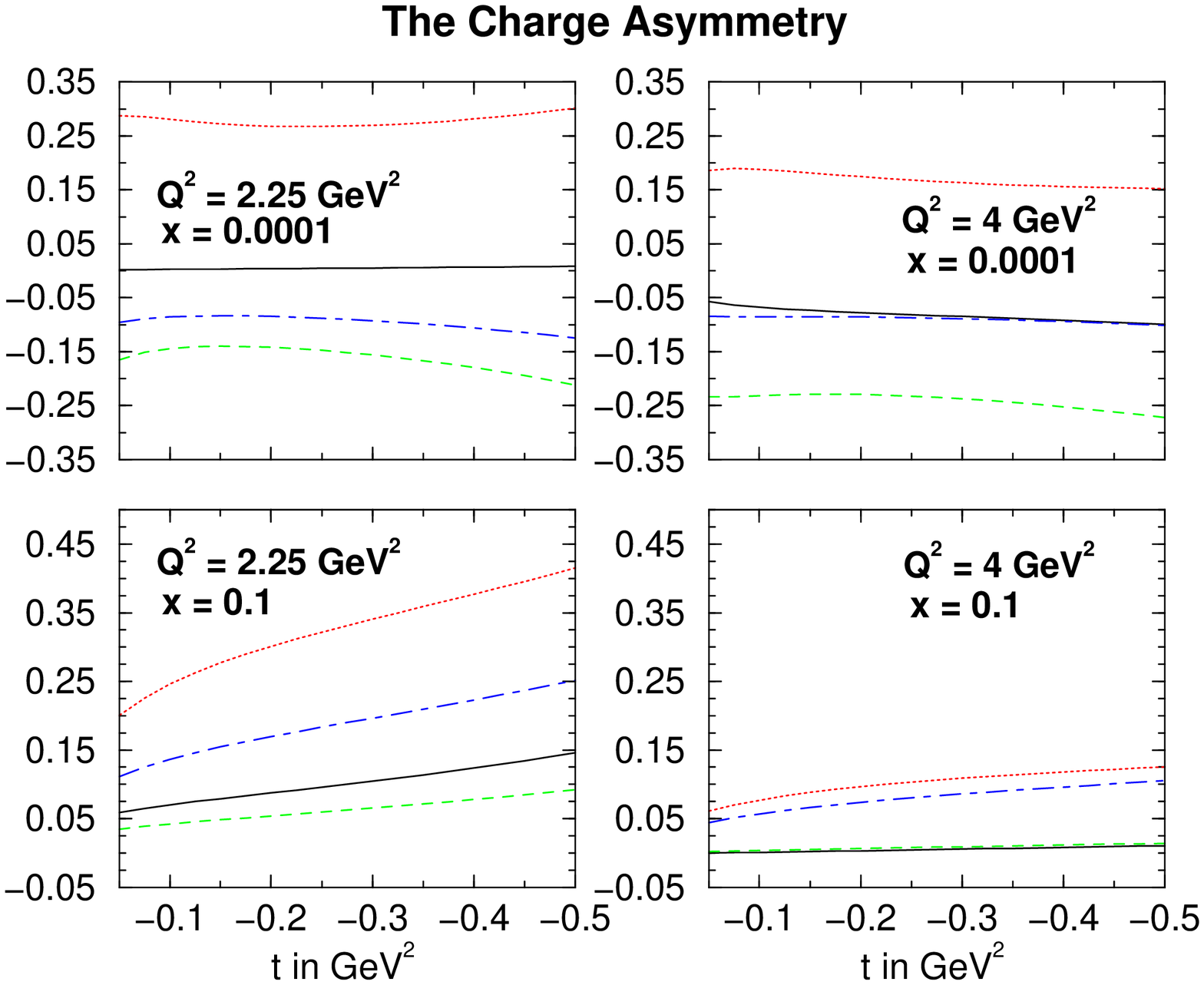,width=9.5cm,height=9.5cm}} 
\vskip+1.0cm 
\caption{The CA in $t$ for fixed $x= \zeta$ and $Q^2$. 
The solid (dotted) curve is the CTEQ5M set in LO (NLO) and the dashed 
(dashed-dotted) curve is the MRST99 set in LO (NLO).} 
\label{ca2} 
\end{figure}

\begin{figure} 
\centering 
\mbox{\epsfig{file=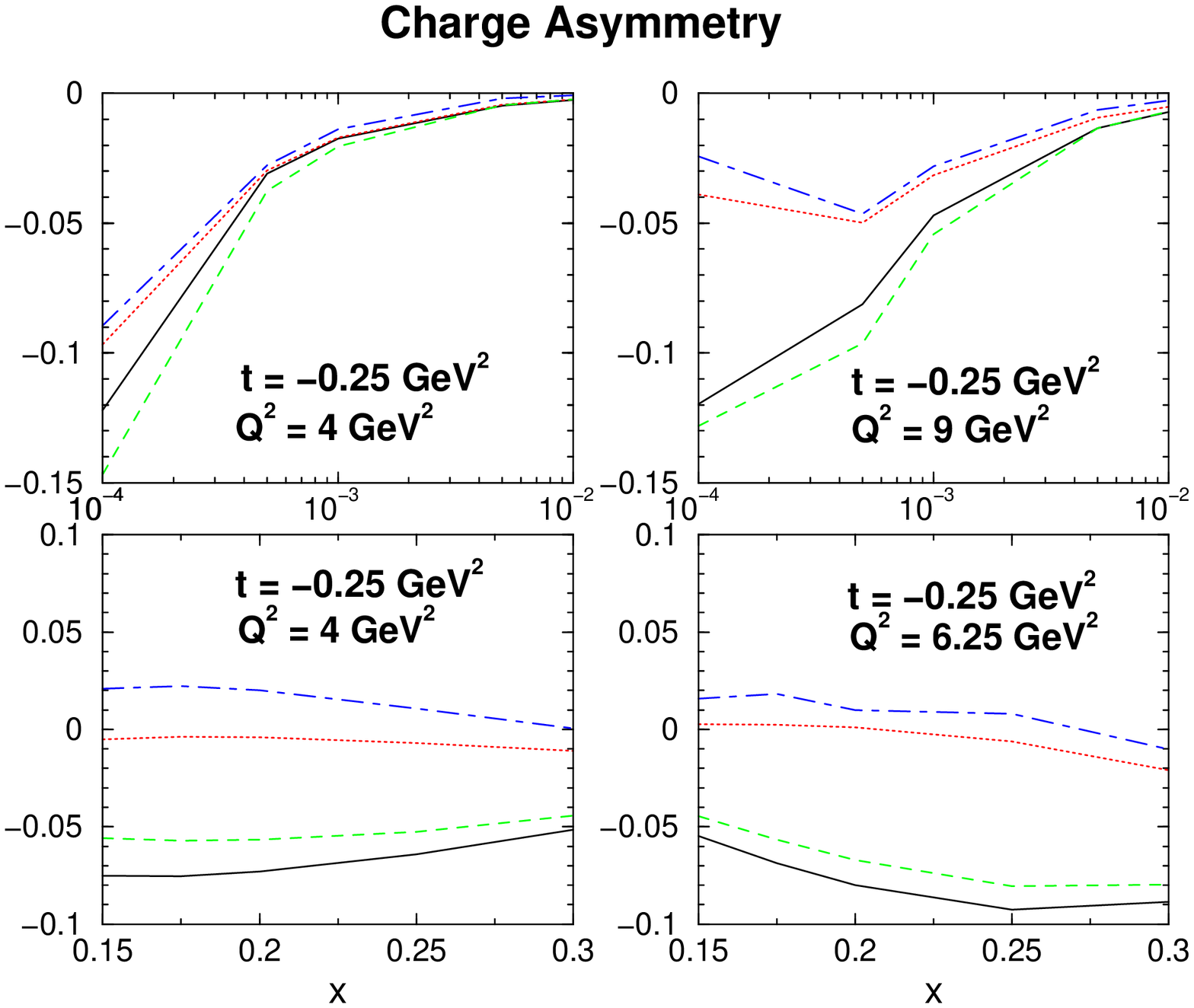,width=10.5cm,height=9.5cm}} 
\caption{The CA in $x=\zeta$ for fixed $t$ and $Q^2$. 
The solid (dotted) curve is the MRSA$^{'}$ set in LO (NLO) and the 
dashed (dashed-dotted) curve is the GRV98 set in LO (NLO).} 
\label{ca3} 
\end{figure}

\begin{figure} 
\centering 
\mbox{\epsfig{file=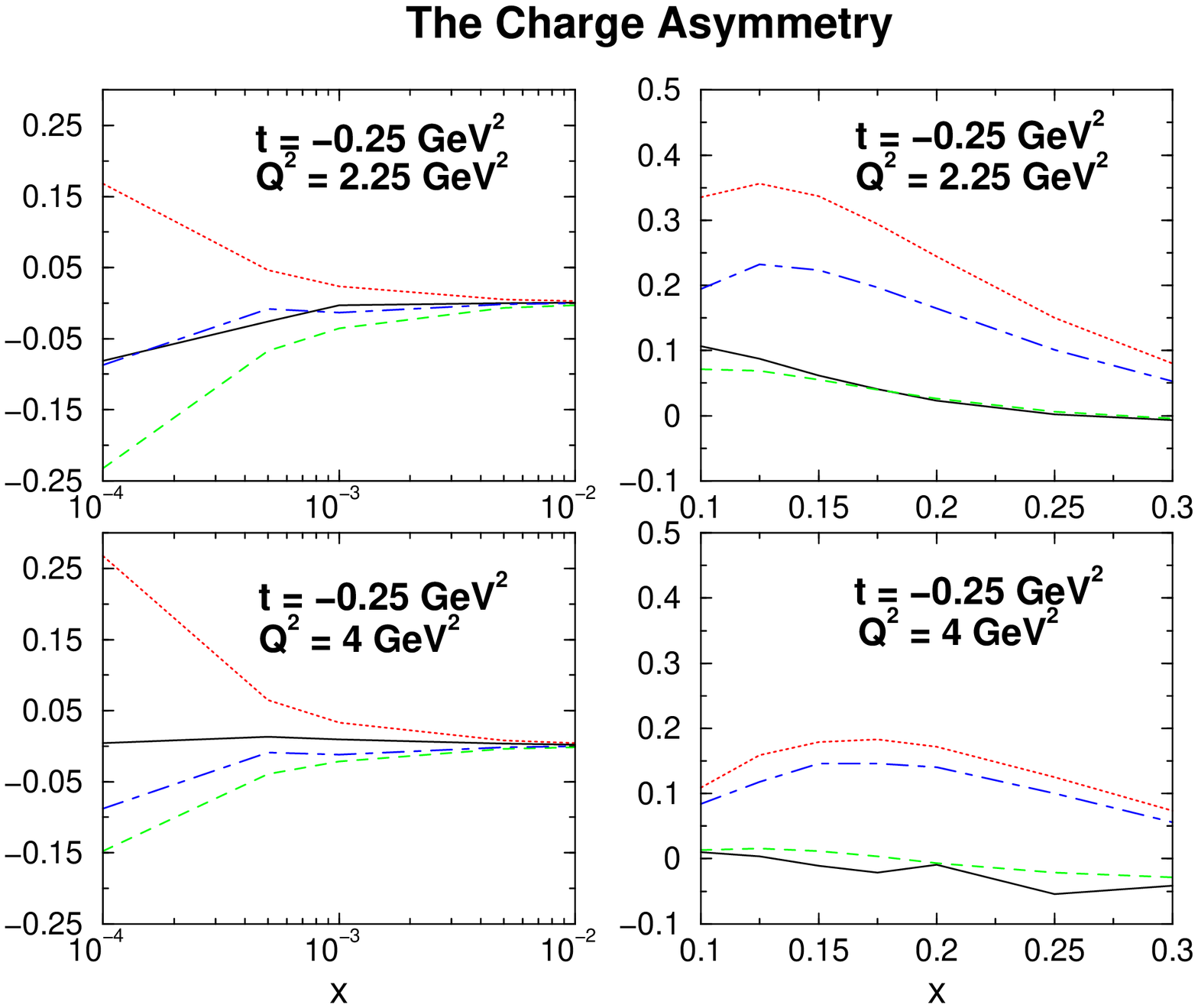,width=10.5cm,height=9.5cm}} 
\caption{The CA in $x=\zeta$ for fixed 
  $t$ and $Q^2$. The solid (dotted) curve is the CTEQ5M set in LO (NLO) 
  and the dashed (dashed-dotted) curve is the MRST99 set in LO (NLO).} 
\label{ca4} 
\end{figure}

\begin{figure} 
\centering 
\mbox{\epsfig{file=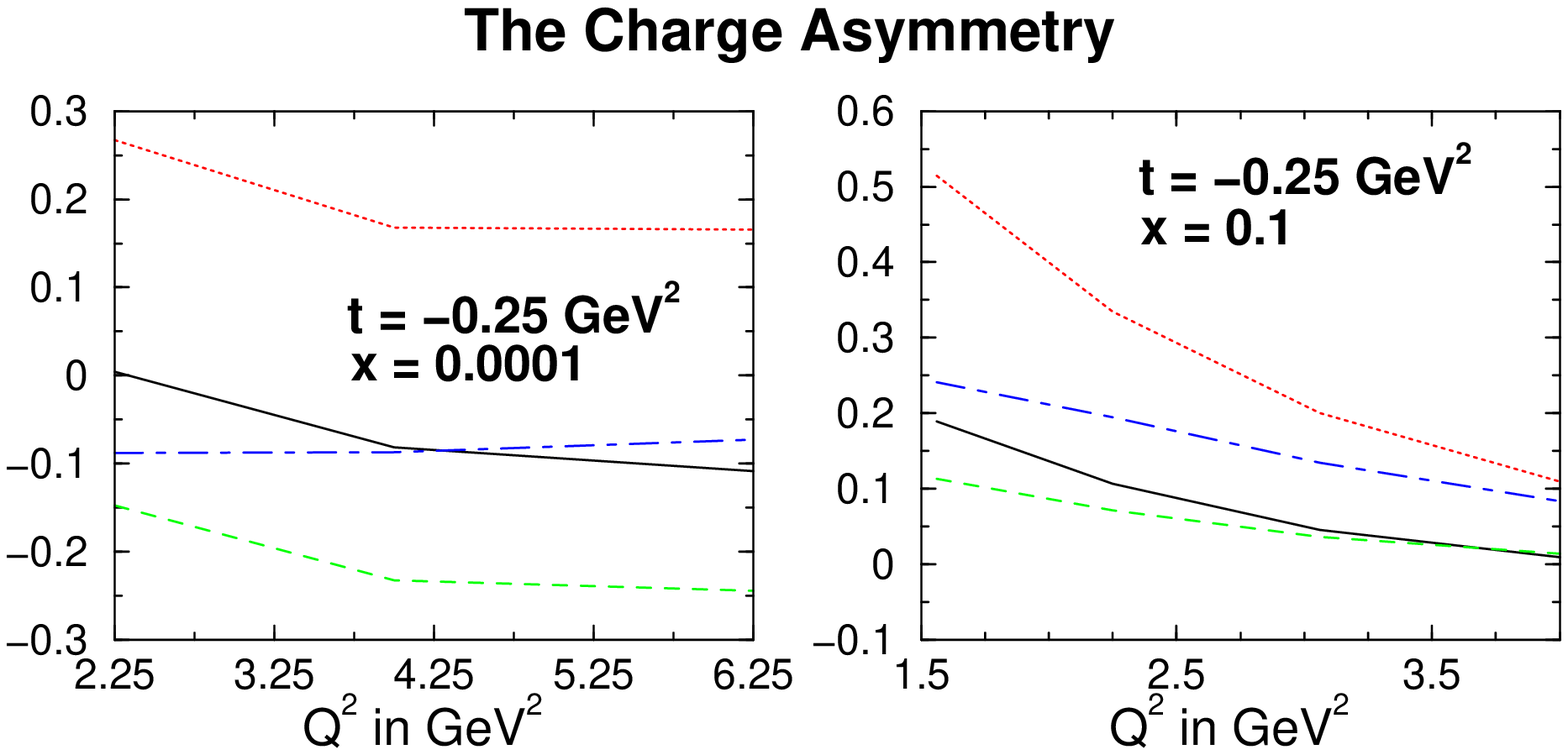,width=10.5cm,height=6.5cm}} 
\caption{The CA in $Q^2$ for fixed $t$ and $x=\zeta$. The solid (dotted) curve is 
the CTEQ5M set in LO (NLO) and the dashed (dashed-dotted) curve is the MRST99 set in LO (NLO).} 
\label{ca5} 
\end{figure}

We discuss the CA first. Figs.\ \ref{ca1}-\ref{ca5} reveal that the CA 
is very similar to the AAA in the BH dominated region, and is  
within $20\%$ of it in the DVCS dominated region for both LO and NLO. 
It can be easily seen that the $t,x$ and $Q^2$ behavior is very similar 
for both asymmetries. This behavior can be understood quite easily:
in contrast to the AAA case, for CA the interference term drops out in the normalization in eq. (\ref{cadef}) due to the sign change of the interference term 
in going from a positron to an electron probe. The numerator is the same for
both asymmetries. Hence, deviations between the AAA and the CA results show directly the influence of the interference term relative to the BH and DVCS 
contributions. The larger the deviation the more important the interference 
term. Hence, a precise measurement of the CA and the AAA will serve as a very 
good consistency check of the experimental analysis, since strong deviations 
between the CA and AAA are not expected. One caveat here is the
possible importance of higher twist contributions in the interference term.
 
Moving now to a longitudinally polarized probe and target, we will study 
the charge asymmetry in a double spin flip experiment, the CADSFL. 

\begin{figure} 
\centering 
\mbox{\epsfig{file=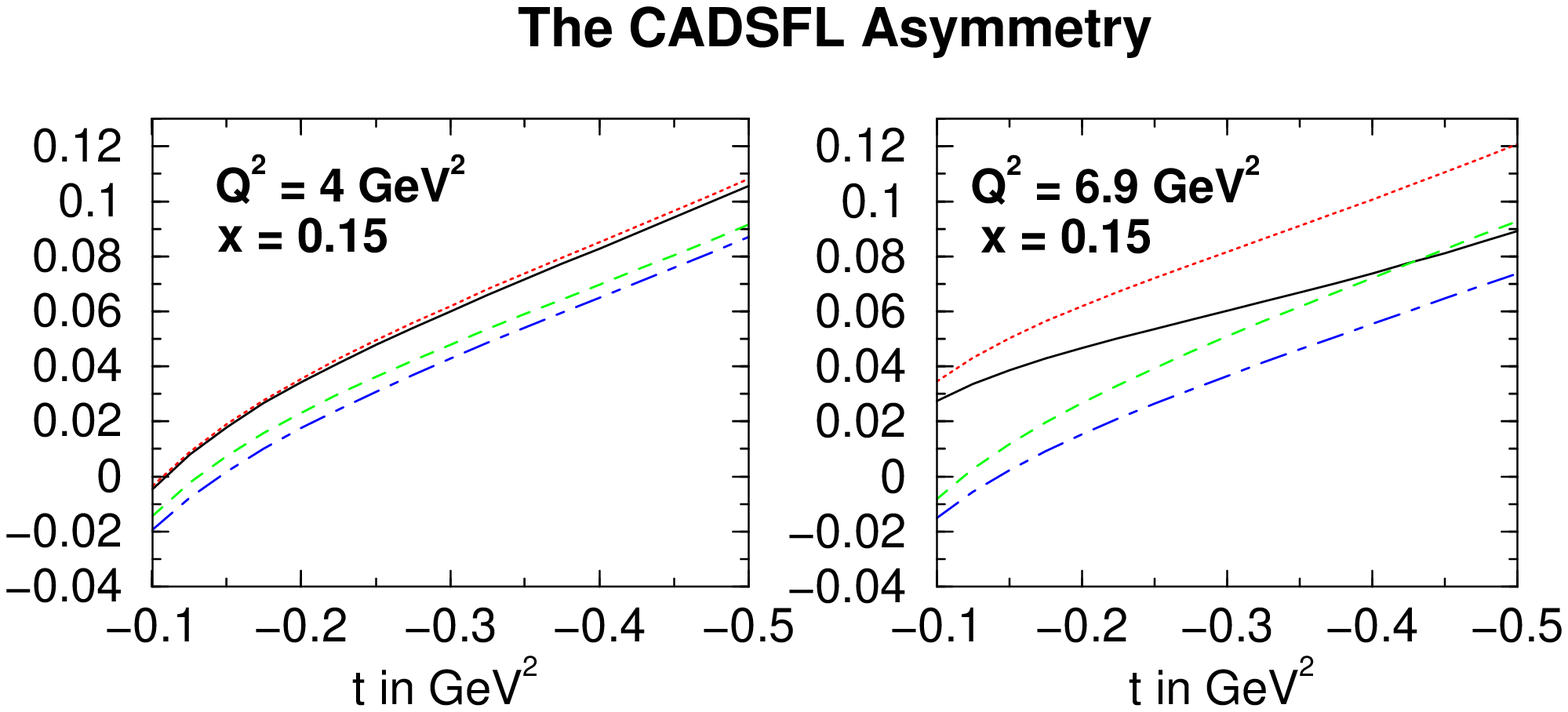,width=10.5cm,height=6.5cm}} 
\vskip+1.0cm 
\caption{The CADSFL in $t$ for fixed 
  $x=\zeta$ and $Q^2$. The solid (dotted) curve is the MRSA$^{'}$ set in LO (NLO) 
  and the dashed (dashed-dotted) curve is the GRV98 set in LO (NLO).} 
\label{cadsfl1} 
\end{figure} 

\begin{figure} 
\centering 
\mbox{\epsfig{file=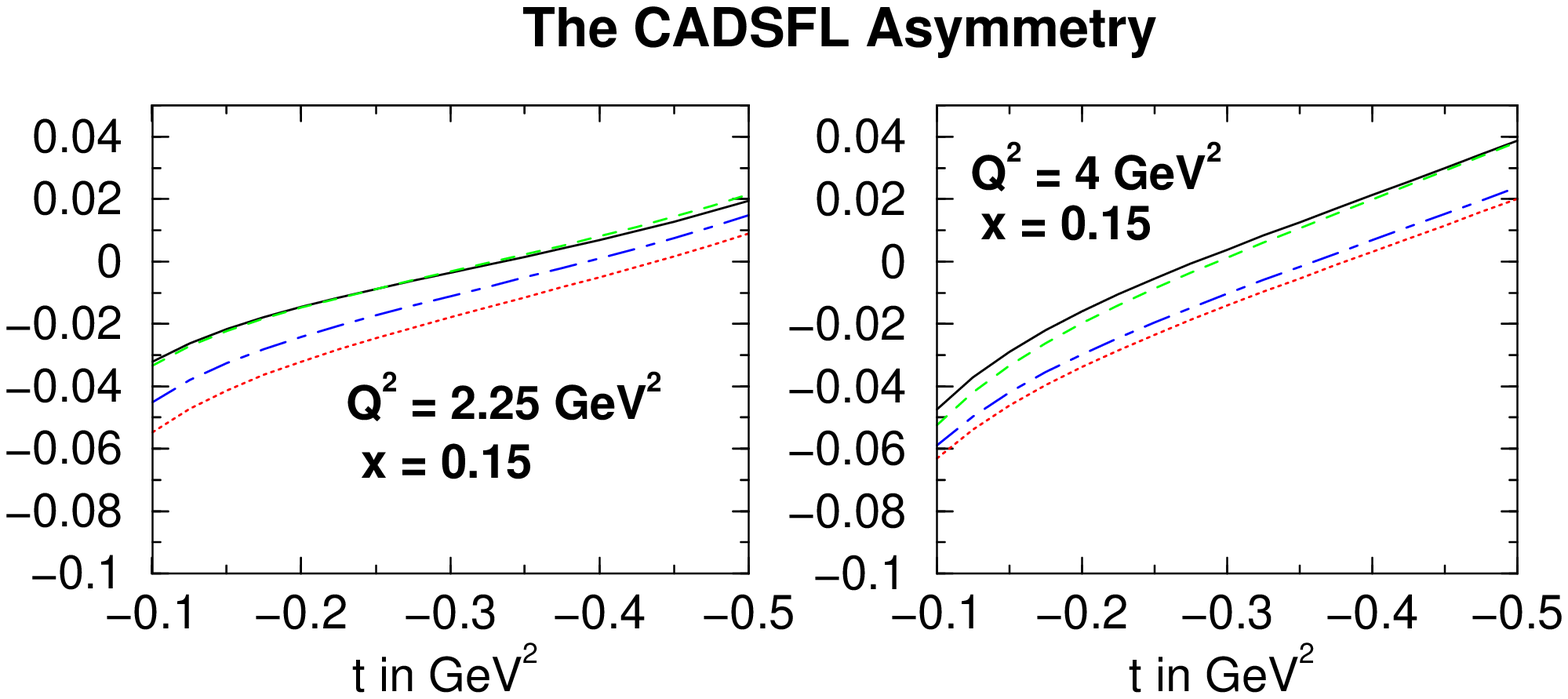,width=10.5cm,height=6.5cm}} 
\caption{The CADSFL in $t$ for fixed 
  $x=\zeta$ and $Q^2$. The solid (dotted) curve is the CTEQ5M set in LO (NLO) 
  and the dashed (dashed-dotted) curve is the MRST99 set in LO (NLO).} 
\label{cadsfl2} 
\end{figure}

\begin{figure} 
\centering 
\mbox{\epsfig{file=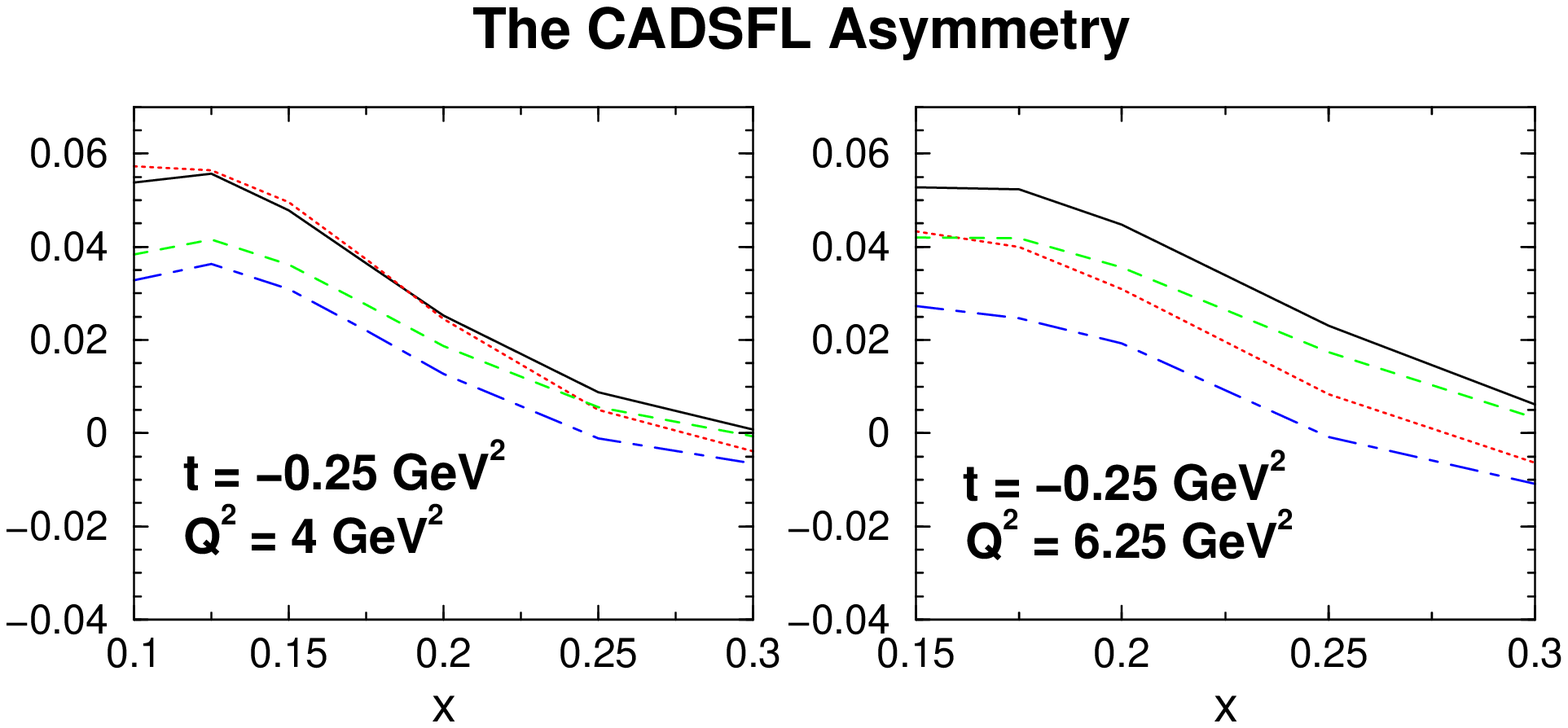,width=11.5cm,height=6.5cm}} 
\caption{The CADSFL in $x=\zeta$ for fixed 
  $t$ and $Q^2$. The solid (dotted) curve is the MRSA$^{'}$ set in LO (NLO) 
  and the dashed (dashed-dotted) curve is the GRV98 set in LO (NLO).} 
\label{cadsfl3} 
\end{figure}

\begin{figure} 
\centering 
\mbox{\epsfig{file=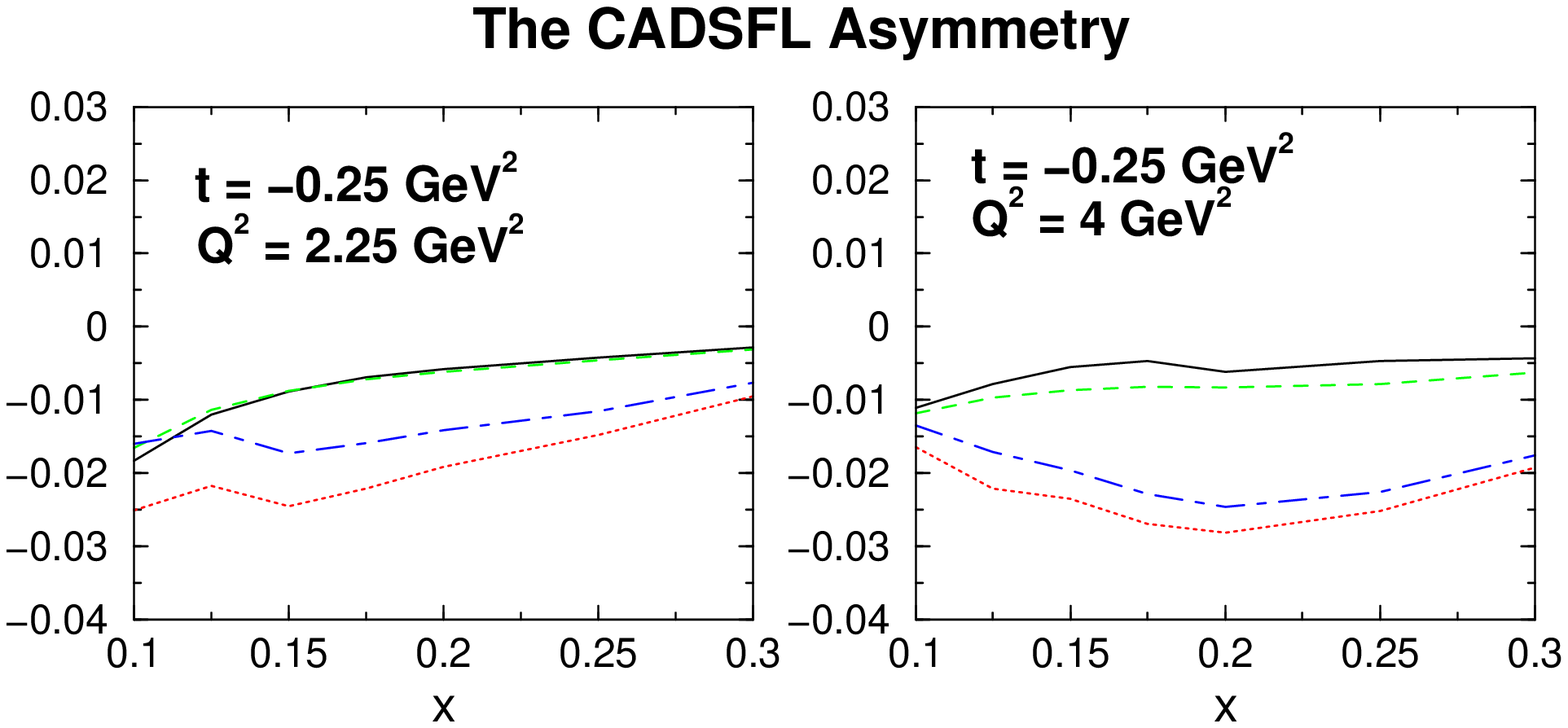,width=11.5cm,height=6.5cm}} 
\caption{The CADSFL in $x=\zeta$ for fixed 
  $t$ and $Q^2$. The solid (dotted) curve is the CTEQ5M set in LO (NLO) 
  and the dashed (dashed-dotted) curve is the MRST99 set in LO (NLO).} 
\label{cadsfl4} 
\end{figure}

For HERA kinematics the CADSFL turns out to be too small to be measured,  
i.e. consistent with zero for any practical purposes, and thus we will not show
plots for HERA kinematics. For HERMES kinematics, the asymmetry is of the order
of a few percent, so one might hope to be able to measure it.
In Figs.\ \ref{cadsfl1} and \ref{cadsfl2} we show the $t$-dependence for fixed $\zeta=x$ and 
$Q^2$, which turns out to be rather steep. In Figs.\ \ref{cadsfl3} 
and \ref{cadsfl4} we show the $\zeta$-dependence for fixed
$t,Q^2$. They indicate that although in percentage terms the spread of 
predictions is large the overall size of the CADSFL (only a few percent) 
would seem to indicate that exploiting this spread of predictions is 
impractical.

\begin{figure} 
\centering 
\vskip+1cm 
\mbox{\epsfig{file=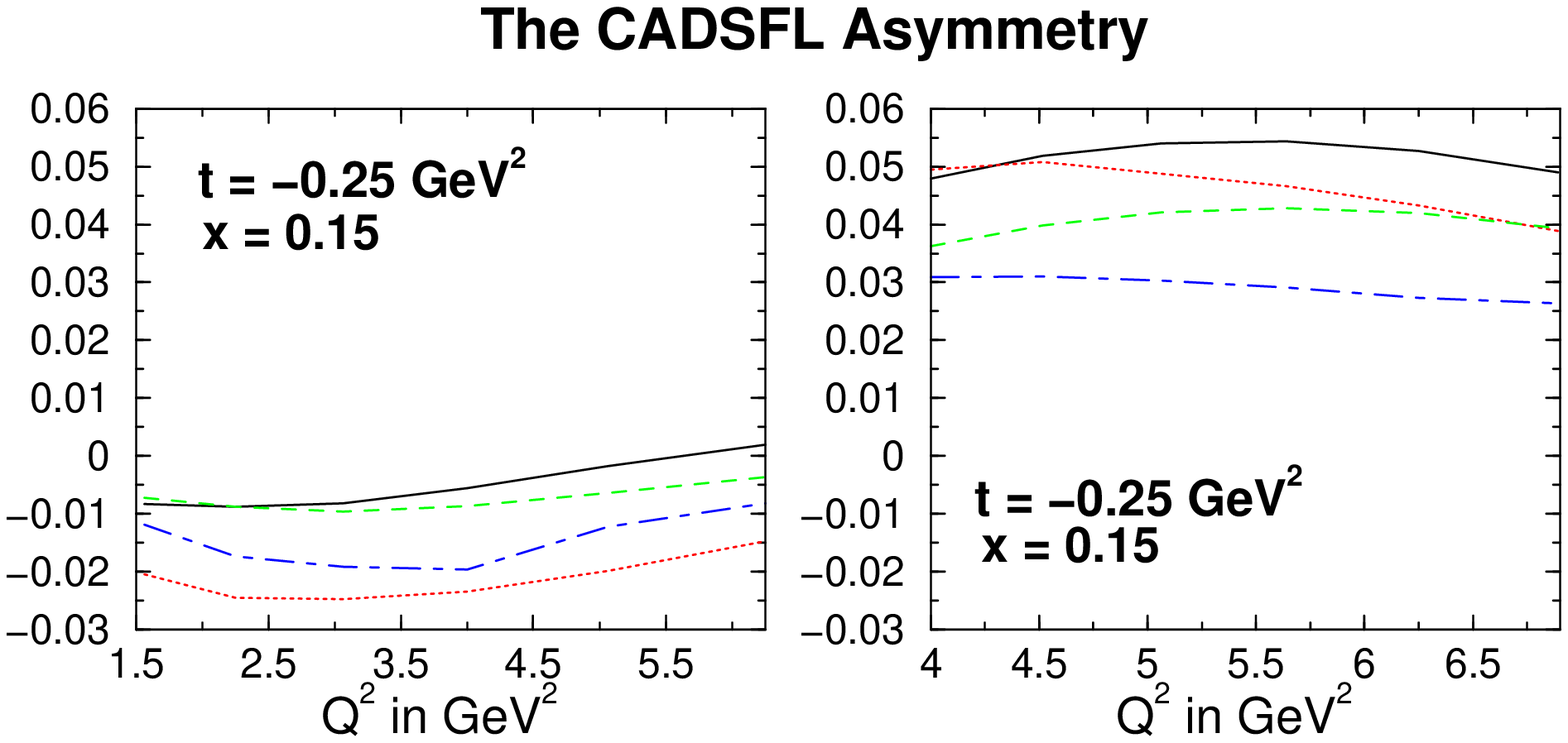,width=11.5cm,height=6.5cm}} 
\vskip+1.5cm 
\caption{The CADSFL in $Q^2$ for fixed $t$ and $x=\zeta$. In the left figure, 
the solid (dotted) curve is the CTEQ5M set in LO (NLO) and the dashed 
(dashed-dotted) curve is the MRST99 set in LO (NLO), and in the right figure, 
the solid (dotted) curve is the MRSA' set in LO (NLO) and the dashed 
(dashed-dotted) curve is the GRV98 set in LO (NLO).} 
\label{cadsfl5} 
\end{figure}

Finally we show the $Q^2$-behavior of CADSFL in Fig.\ \ref{cadsfl5}, for fixed 
$t,\zeta$. We see quite a moderate to large difference in going from LO to NLO.
Note the sign difference between the GRV98 and MRSA' sets on the one hand and 
the CTEQ5M and MRST99 sets on the other hand. This gives some hope to use this 
asymmetry as a discriminator between different model inputs. 
Generally speaking, due to its overall smallness, it is not as 
promising a candidate as the CA asymmetry as a good DVCS observable, 
even though it is directly sensitive to the real part of a 
polarized amplitude at small $\zeta$. This concludes our presentation of DVCS 
observables in LO and NLO. 

\subsection{Summary of optimal DVCS observables} 
 
We summarise the results of the previous subsections as follows:

\begin{itemize} 
\item The triple differential cross section is quite a good 
  discriminator between different GPD models, particularly at $Q^2$ values 
  close to the input scale (at higher $Q^2$ the BH process dominates). 
  We predict a steep rise with decreasing $\zeta$ reflecting the underlying 
  powerlike behavior in $\zeta$ of the DVCS amplitudes.
\item The SSA is quite sizable and well behaved in NLO for all the 
  input GPDs concerned, which makes it a very good candidate to be 
  measured both at HERA and HERMES. For small $\zeta$ (HERA kinematics) 
  we predict a strong increase in magnitude as $\zeta$ decreases (a feature shared by AAA and 
  CA) and good discriminating power between input models, especially at larger
  $Q^2$.       
\item The AAA and CA seem to be very good candidates for discriminating 
  between different input GPDs simply because they measure the real 
  part of DVCS amplitudes which are very sensitive to the underlying 
  details of the GPDs, in Radyushkin's ansatz, especially at NLO where the 
  gluon enters with a relative minus sign for the real part of its amplitude. 
  Practical measurements for both asymmetries 
  appear feasible for both HERMES and HERA kinematics. 
\item  The measurement of UPLT (CADSFL) appears to be practical only at 
  large $\zeta$ in HERMES kinematics, where it tests a linear combination of 
  imaginary (real)  parts of several DVCS amplitudes. Therefore these 
  measurements have very limited discriminating power between model GPDs. 
\end{itemize} 
 
In summary it seems that four out of six observables are large enough to make 
measurements at HERA and HERMES feasible, and which have good discriminating 
power between input scenarios. Thus the prospects of unraveling the details of 
the DVCS process experimentally 
are in principle very good. 
 
\section{Comparison with experiment and other calculations} 
\label{sec:dat} 

In this section we compare our LO and NLO results for each of our 
input sets with the available published experimental data from H1 and 
HERMES. 
The ZEUS collaboration announced the first measurement of DVCS at HERA \cite{zeus} (see also \cite{zeus2}), but has yet to publish a cross section. 
 
H1 recently published their data \cite{h1} on the measured DVCS 
cross section on both the lepton level and the photon level. They 
use the equivalent photon approach, which relates the pure DVCS cross 
section on the lepton level (with the interference neglected\footnote{This may be ignored at small $y$, typical of HERA kinematics, and after integration over $\phi$.} and the pure BH term subtracted) to the virtual-photon proton cross section: 
\begin{align} 
\frac{d^{2}\sigma (ep \to ep\gamma)}{dy dQ^2} = \Gamma~\sigma_{DVCS}(\gamma^*p\to 
\gamma p)~\qquad~\mbox{where}~~\qquad~\Gamma = \frac{\alpha_{e.m.} (1+(1-y)^2)}{2\pi 
  y Q^2}. 
\end{align} 
\noindent By integrating eq.\ (\ref{crossx}) over $t, \phi$, and changing variables from $x_{bj}$ to $y$, we can establish the formula for the photon-level cross section in terms of 
our amplitudes as 
\begin{align} 
\sigma_{DVCS}(\gamma^*p \to \gamma p) = \frac{\alpha^2x^2\pi}{Q^4{\cal B}}|{\cal 
  T}_{DVCS}|^2|_{t=0}, 
\end{align} 
where ${\cal B}$ stems from the $t$-integration and, within our model for the 
$t$-dependence, ${\cal B} \approx 6.5$~GeV$^{-2}$ 
(with a surprisingly small spread of about $1$~GeV$^{-2}$ !). 
We observed that for the kinematical region of the H1 data which is 
limited to small $x$, i.e. to small $\zeta$, we could write 
\begin{align} 
\int^{t_{\mbox{\small{tmin}}}}_{-\infty}~dt~|{\cal T}_{DVCS}|^2 \approx 
\frac{1}{{\cal B}}|{\cal T}_{DVCS}|^2|_{t=0} 
\end{align} 
as if we had assumed a global exponential dependence, $e^{Bt}$, as was used by H1 in their 
comparison to other \cite{ffs,dodo01} calculations (dropping all amplitudes except ${\cal H}$). 
On numerical inspection we found that indeed ${\cal H}$ clearly dominates all the other amplitudes 
for HERA kinematics. Therefore, in comparing to the 
data, we can safely neglect all other amplitudes (i.e. polarised and helicity-flip) 
in the DVCS square term. 

The same assumption was made in previous LO QCD \cite{ffs} and two-component 
dipole model \cite{dodo01,mss} calculations, which both reproduce the H1 data quite well.
 Since ours is a QCD calculation 
we point out the differences and similarities with \cite{ffs}. 
The latter was based on a LO input (CTEQ3L) with the assumption that the GPDs 
are equal to the PDFs at the input scale. The imaginary part of the 
unpolarized helicity non-flip amplitude was then computed using the aligned 
jet model and the imaginary part, rather than the GPD, was then evolved to 
higher $Q^2$. This is equivalent to what we did at LO\footnote{Note however 
that our LO calculation is based on NLO input PDFs, 
and a different ansatz for the GPDs, 
so we don't necessarily expect the results to agree closely.} since the imaginary part of the DVCS 
amplitude is simply $\pi$ times the GPD at $\zeta=x_{bj}$, legitimising 
the approach in \cite{ffs}. The DVCS triple differential cross section, at 
small $x$, was then computed by comparing the imaginary part of DVCS to DIS 
and thus introducing the structure function $F_2$ through the optical theorem.
In order to reconstruct the real part of the DVCS amplitude a dispersion 
relation approach was used which exploited the slope of $F_2$ in $\ln(1/x)$ at 
small $x$ (which was extracted from data to give 
$\eta = Re A_{\mbox{{\scriptsize DIS}}} / Im A_{\mbox{{\scriptsize DIS}}}$) 
together with the comparative factor 
$R = Im A_{\mbox{{\scriptsize DVCS}}} / Im A_{\mbox{{\scriptsize DIS}}}$.  
The specification was completed with 
the additional assumption of a global $t$-dependence, $e^{Bt}$. In contrast to
\cite{ffs}, we computed both parts of the DVCS amplitudes directly and used 
them in the unapproximated expressions for the DVCS triple differential cross 
section, which also contains the polarized as well as helicity flip amplitudes
(however these can be safely neglected in the cross section at small $\zeta$). 
Furthermore, we assumed a dipole type $t$-dependence which is close to the 
exponential behavior at small $\zeta$ and $t$, as well as being correct at 
larger $t$ (and small $\zeta$). 
 
\begin{center} 
\begin{figure}  
\mbox{\epsfig{file=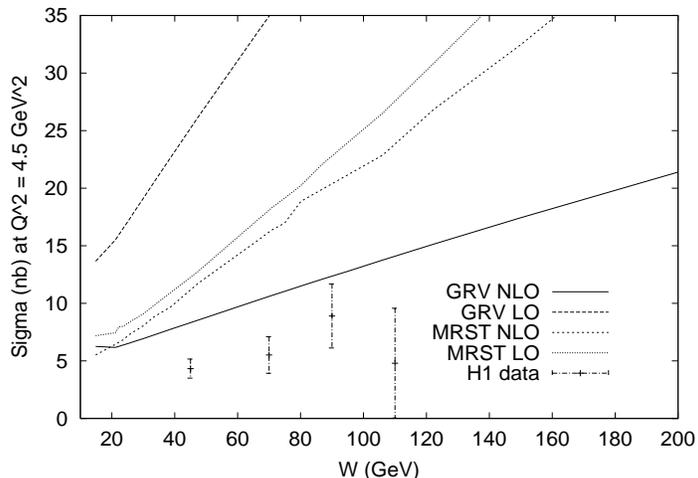,width=9.5cm,height=6.5cm}}
\caption{The photon level cross section $\sigma (\gamma^* P \to \gamma P)$ as a function of \ $W$ at fixed $Q^2 = 4.5$~GeV$^2$, for GRV98 and MRST99 at LO and NLO. Also shown are the recent H1 data, at $Q^2 = 4.5$~GeV$^2$, with systematic and statistical errors added in quadrature.}
\label{h1comw}
\end{figure} 
\end{center}

\begin{center} 
\begin{figure}  
\mbox{\epsfig{file=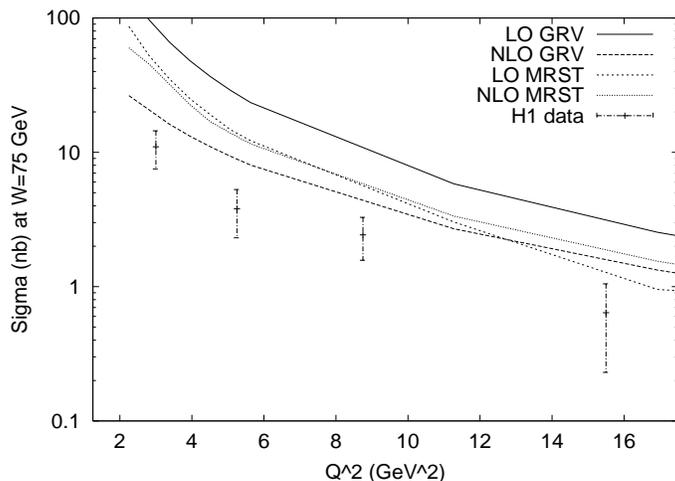,width=9.5cm,height=6.5cm}}
\caption{The photon level cross section $\sigma (\gamma^* P \to \gamma P)$ as a function of $Q^2$ at fixed $W = 75$~GeV, for GRV98 and MRST99 at LO and NLO. Also shown are the recent H1 data, at $W = 75$~GeV, with systematic and statistical errors added in quadrature.}
\label{h1comq}
\end{figure} 
\end{center}

In Fig.\ \ref{h1comw} we plot $\sigma_{DVCS}(\gamma^*p)$ at $Q^2 = 4.5$~GeV$^2$ 
in $W$ (this is of course similar to a plot in $1/x=W^2/Q^2$) for GRV and MRST 
input models at LO and NLO. Also shown 
is the H1 data, for which they quote $<\!Q^2\!> = 4.5$~GeV$^2$, with 
systematic and statistical error bars added in quadrature. In Fig.\ \ref{h1comq}, 
we plot $\sigma_{DVCS}(\gamma^*p)$ in $Q^2$ for fixed $W = 75$~GeV.
The most obvious observation to make is that all of the the curves lie well above the 
data. However they do have the correct shapes (rising with $W$ and falling rapidly with 
$Q^2$). 

In order to understand this discrepancy we re-examined Radyushkin's input model in detail. 
For the unpolarized gluon GPD in 
\cite{frmc2} we found a rather moderate enhancement of about $20\%$ of the GPD 
relative to the forward case at the point $X=\zeta$. However, for the quark singlet, 
in which $q$ rather than $xq$ is used in the double distributions, we found this enhancement to be as large as a factor four! Such a large enhancement stems from the fact that the quark singlet distributions from the chosen input GPDs are very singular 
in the small $x$ region ($q(x) \propto x^{-1-\lambda}$ ), and therefore lead to an integrable singularity in the double distribution at $x' = 0$ (the shape function for the quark singlet is $\propto x$ for $x \rightarrow 0$ and this reduces the degree of the singularity to an integrable 
one). This leads to a large enhancement close to $X=\zeta$ and a strong and unsatisfactory  
sensitivity to the extrapolation of the input PDFs to very small $x$, where they have 
not yet been measured (cf. eqs.(\ref{erblq},\ref{erblbq})). This is clearly a very unsatisfactory
behaviour from a physical point of view and thus we are led to conclude that 
Radyushkin's input model should be used only with non-singular inputs.

\begin{center} 
\begin{figure}  
\mbox{\epsfig{file=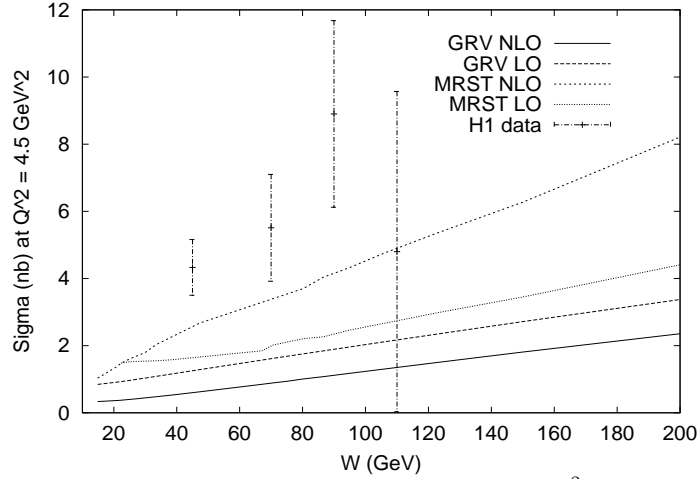,width=9.5cm,height=6.5cm}}
\caption{The photon level cross section $\sigma (\gamma^* P \to \gamma P)$ as a function of  $W$ at $Q^2 = 4.5$~GeV$^2$, for GRV98 and MRST99 at LO and NLO using the shifted ansatz.}
\label{h1comwmod}
\end{figure} 
\end{center}

\begin{center} 
\begin{figure}  
\mbox{\epsfig{file=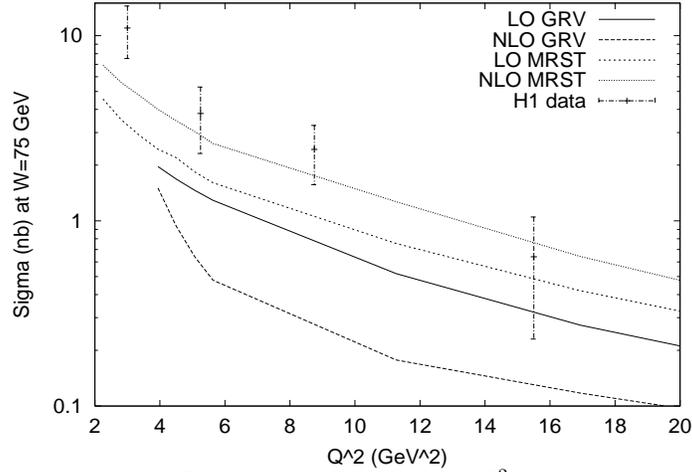,width=9.5cm,height=6.5cm}}
\caption{The photon level cross section $\sigma (\gamma^* P \to \gamma P)$ as a function of $Q^2$ at fixed $W = 75$~GeV, for GRV98 and MRST99 at LO and NLO, using the shifted ansatz.}
\label{h1comqmod}
\end{figure} 
\end{center}

To illustrate the strong sensitivity of the results to the particular choice of input GPD 
we make a very simple change to the input model, namely we shift the argument of the PDFs in eq.(14) of \cite{frmc1} from $x$ to  $x + \zeta$. This (admittedly rather 'ad hoc') modification, which we make purely for the purpose of demonstration, preserves the symmetries in the ERBL region and the forward limit, but can be expected to spoil the polynomiality properties, which are then restored\footnote{In \cite{bemu4}, in which the evolution is performed using moments, it is shown that expansion coefficients in an orthogonal basis allow only even powers of $\zeta$ in the expansion, in line with the symmetry properties of GPDs. Similarly, in our numerical solution, which is carried out in $x$ space rather than moment space, the non-polynomial pieces will be killed by the kernels under convolution (the kernels encode the GPD symmetries in their structure).} under evolution \cite{bemu4}.
In making this modification we ensure that the PDFs are never sampled below $x = \zeta$ 
(which removes the sensitivity to the very small $x$ region) and can be expected to 
reduce considerably the enhancement of the GPDs at $X=\zeta$. Figs.\ \ref{h1comwmod} and \ref{h1comqmod} show the very dramatic effect on the cross section. The theory curves for the modified ansatz now undershoot the experimental data by about a factor of two and appear to reproduce the shape in $W$ and $Q^2$ rather well. In summary we claim that this 
illustrates that the H1 data are already able to begin to constrain the input GPDs. 
It seems clear that if one is to use Radyushkin's ansatz at very small $\zeta$ the input 
PDFs should be non-singular. Alternatively, one must invent a new parameterization of the GPDs at the input scale that retains the required features (correct polynomiality and symmetry properties and faithful reproduction of the forward limit) without the problem of a strong sensitivity to the very small $x$ region and associated large enhancement of the GPDs relative to the PDFs at the point $X=\zeta$.
This might be achieved by choosing a non-singular double distribution at a very low input scale, with the small $X$ behavior generated by perturbative evolution.

We now turn to the SSA as measured by the HERMES collaboration \cite{herm}. 
They define the SSA, weighted with $\sin \phi$, as follows:
\begin{equation}
SSA = \frac{2 \int_{0}^{2 \pi} d \phi ~\sin \phi~(d \sigma^{\uparrow} - d \sigma^{\downarrow})}{
\int_{0}^{2 \pi} (d \sigma^{\uparrow} + d \sigma^{\downarrow})} \, .
\end{equation}
They quote two values: $SSA = -0.18 \pm 0.05 \pm 0.05$ and  $\langle SSA \rangle = -0.23 \pm 0.03 \pm 0.04$. The first, which assumes that their missing mass equals the proton mass, 
is quoted at the following average values: $<\!x\!> = 0.11, <\!Q^2\!> = 2.5$~GeV$^2$ and $<\!t\!> = -0.27$~GeV$^2$. The second (average) value is the SSA integrated over the missing mass at the same average values of $x, Q^2$ and $t$. At this point in $x, Q^2, t$ for the HERMES definition we find $SSA= -0.365$ for CTEQ5M and $SSA = -0.35$ for MRST99. 

Given the fact that the input models used do not describe the data at 
small $\zeta$, and that we find the same type of enhancement effect when 
comparing GPDs with forward PDFs at both small and large $\zeta$, the failure to 
describe the HERMES data is not surprising. 
In fact one needs higher statistics 
over a wide kinematic range in several DVCS observables to be able to begin to tune 
the input GPDs using a fitting method. Only then can one start discriminating between different 
choices for the PDFs in input models.  
Furthermore, for HERMES data one needs to know the normalizations of numerator and the 
denominator producing the measured asymmetry to help to constrain the normalization 
of the individual amplitudes (in order to be able to make any sensible statements 
about a comparison between theory and experiment). 
 
In comparing with other calculations we show qualitative agreement 
with \cite{ffs,dodo01,mss} and quantitative agreement with 
\cite{bemu1,bemu3} wherever we used the same input distributions (MRSA' and GS(A) only). 

\section{Conclusions} 
\label{sec:con} 
 
We have presented a detailed next-to-leading (NLO) QCD analysis of 
deeply virtual Compton scattering (DVCS) observables.  
We quantified the NLO corrections and established which observables 
have the best prospects to be measured accurately at HERA and HERMES 
(the triple differential cross section and the azimuthal angle (AAA), single spin (SSA) 
and charge (CA) asymmetries). We have demonstrated that
such measurements would have good discriminating power between different scenarios 
for the generalized parton distributions (GPDs), by examining four cases.
It turns out that AAA and CA, which test the real part of unpolarized DVCS 
amplitude at small $x$ are the most sensitive to the choice of GPD model.

We performed a comparison with presently available DVCS data and showed that within 
Radyushkin's ansatz for the GPDs, all of the models we examined overshoot the H1 data. 
This is due to their singular nature of the quark singlet at small $x$ which leads to 
a strong sensitivity to the PDFs in the extremely small $x$ region, where they have not 
yet been measured. We illustrated this strong sensitivity by artificially shifting 
the argument of the PDFs by $\zeta$, which led to a large reduction in the 
theoretical predictions for the photon-level cross section (which then undershoots the data). 
This illustrates an urgent need for improved input models, which do not rely on a singular 
double distribution and also for more, high precision data to help constrain them.

\section*{Acknowledgements} 
 
\label{sec:ack}  
  
A.\ F.\ was supported by the DFG under contract $\#$ FR 1524/1-1. M.\ M.\ was 
supported by PPARC. We gladly thank D.\ M{\"u}ller for helpful 
conversations and discussions throughout this project. 
We also thank R.~G.~Roberts for providing the input parameters of the 
`corrected' MRST parton set.

\end{document}